\documentclass[useAMS,usenatbib,a4paper, fleqn]{mnras}
\usepackage{ae,aecompl}
\title[Simulated and observed clusters]
{
The properties, origin and evolution of stellar clusters in galaxy simulations and observations
}
\author[Dobbs]
{C. L. Dobbs\thanks{E-mail:
dobbs@astro.ex.ac.uk}$^{1}$, A. Adamo$^2$, C. G. Few$^{1,3}$, D. Calzetti$^4$, D. A. Dale$^5$, B. G. Elmegreen$^6$, 
\newauthor A. S. Evans$^{7,8}$, D. A. Gouliermis$^{9,10}$, K. Grasha$^4$, E. K. Grebel$^{11}$, K. E. Johnson$^7$, 
\newauthor H. Kim$^{12,13}$, J. C. Lee$^{14}$, M. Messa$^2$, J. E. Ryon$^{15}$, L. J. Smith$^{13}$, D. A. Thilker$^{14}$, 
\newauthor L. Ubeda$^{14}$, B. Whitmore$^{14}$\\
$^1$ School of Physics and Astronomy, University of Exeter, Stocker Road, Exeter, EX4 4QL, UK \\
$^2$ Department of Astronomy, The Oskar Klein Centre, Stockholm University, AlbaNova University Centre, SE-106 91 Stockholm, Sweden \\
$^3$ E.A. Milne Centre for Astrophysics, Dept. of Physics \& Mathematics, University of Hull, Hull, HU6 7RX, UK \\
$^4$ Astronomy Department, University of Massachusetts, Amherst, MA 01003, USA \\
$^5$ Department of Physics \& Astronomy, University of Wyoming, Laramie WY, USA \\
$^6$  IBM Research Division, T.J. Watson Research Center, 1101 Kitchawan Road, Yorktown Heights, NY 10598 USA\\
$^7$ Dept. of Astronomy, University of Virginia, 530 McCormick Road, Charlottesville, VA 22904, USA \\
$^8$ National Radio Astronomy Observatory, 520 Edgemont Road, Charlottesville, VA 22903 USA \\
$^9$ Zentrum f\"ur Astronomie der Universit\"at Heidelberg, Institut f\"ur Theoretische Astrophysik, Albert-Ueberle-Str.2, 69120 Heidelberg, Germany \\
$^{10}$ Max Planck Institute for Astronomy, K{\"o}nigstuhl 17, 69117 Heidelberg, Germany \\
$^{11}$ Astronomisches Rechen-Institut, Zentrum f\"ur Astronomie der Universit\"at Heidelberg, M\"onchhofstr.\ 12--14, 69120 Heidelberg,
Germany \\
$^{12}$ Korea Astronomy and Space Science Institute, Daejeon, Korea \\
$^{13}$ Dept. of Astronomy, University of Texas at Austin, Austin, TX, USA \\
$^{14}$ Space Telescope Science Institute, 3700 San Martin Drive, Baltimore MD 21218, USA\\
$^{15}$ Dept. of Astronomy, University of Wisconsin--Madison, Madison, WI, USA\
}
\usepackage{amssymb}
\usepackage{amsmath}
\usepackage[pdftex]{graphicx}
\usepackage{epsfig}
\usepackage{multirow} 
\def\aap{{ A\&A}}

\def\apjl{{ApJL}}

\def\apjs{{ApJS}}

\begin{document}
\label{firstpage}
\date{\today}

\pagerange{\pageref{firstpage}--\pageref{lastpage}} \pubyear{2012}

\maketitle

\begin{abstract}
We investigate the properties and evolution of star particles in two simulations of isolated spiral galaxies, and two galaxies from cosmological simulations. Unlike previous numerical work, where typically each star particle represents one `cluster', for the isolated galaxies we are able to model features we term `clusters' with groups of particles.  We compute the spatial distribution of stars with different ages, and cluster mass distributions, comparing our findings with observations including the recent LEGUS survey. We find that spiral structure tends to be present in older (100s Myrs) stars and clusters in the simulations compared to the observations. This likely reflects differences in the numbers of stars or clusters, the strength of spiral arms, and whether the clusters are allowed to evolve. Where we model clusters with multiple particles, we are able to study their evolution. The evolution of simulated clusters tends to follow that of their natal gas clouds. Massive, dense, long-lived clouds host massive clusters, whilst short-lived clouds host smaller clusters which readily disperse. Most clusters appear to disperse fairly quickly, in basic agreement with observational findings. We note that embedded clusters may be less inclined to disperse in simulations in a galactic environment with continuous accretion of gas onto the clouds than isolated clouds and correspondingly, massive young clusters which are no longer associated with gas tend not to occur in the simulations. Caveats of our models include that the cluster densities are lower than realistic clusters, and the simplistic implementation of stellar feedback.
\end{abstract}


\section{Introduction}
The evolution and properties of stellar clusters is a huge area of interest in astronomy. To date most work in this area has been observational, looking at the age and mass functions, types and distributions of star clusters in nearby galaxies (see e.g. recent reviews by \citealt{Adamo2015} and \citealt{Longmore2014}). There has been relatively little input in this area from galactic numerical simulations. Here we use galaxy simulations to study the properties of clusters (which are represented by one or many star particles depending on resolution) and compare with observations. Where clusters are represented by many star particles, we also study their evolution including in relation to their natal molecular clouds.

Clusters evolve both in the context of their formation within a molecular cloud (i.e. at the embedded stage), and in terms of their dissociation from the gas of their natal cloud. Observational measures of cluster behaviour typically focus on the age and mass distributions of clusters. Cluster age distributions can be used to predict cluster evolution, since obviously a sharply declining distribution will suggest that clusters disperse much quicker than those with distributions with a flat profile. Recent observations show a constant power law profile followed by a clear downturn around 100 \citep{Silva-Villa2014} or 200 Myr \citep{Baumgardt2013}  in the cluster age distributions in nearby galaxies. This appears to  indicate that a significant population of clusters disperse on timescales of around 100 Myr. There is some discrepancy between results, with some groups finding steeper mass distributions \citep{Chandar2006,Chandar2014} compared to others \citep{Gieles2007a,Silva-Villa2014} even for the same galaxy. There does however appear to be a genuine environmental dependence, some galaxies (typically more quiescent galaxies such as the SMC) having flatter distributions, with less indication of cluster disruption, than others \citep{Adamo2015}. 

There are a number of possible mechanisms whereby clusters may disperse, or become unbound over time (see e.g. \citealt{Adamo2015}) and only a brief outline is included here. The first mechanism is so called `infant mortality', whereby the loss of gas (e.g. by stellar feedback processes) from a stellar cluster still associated with a molecular cloud changes the gravitational potential experienced by the cluster \citep{Lada1984}. Other processes include tidal effects within the cloud \citep{Elmegreen2010,Kruijssen2011, Kruijssen2012}. Other mechanisms are relevant instead to the evolution of the cluster independent of, or after the gas of the natal molecular cloud has disappeared. Again tidal effects in the galaxy may cause a cluster to become unbound, whilst the interactions of clusters with other molecular clouds may also play a role (e.g. \citealt{Spitzer1958}). Finally, N-body effects, and longer term stellar evolution may also have an effect on the cluster evolution. 

Another question regarding cluster evolution is the spatial evolution of the stars. Young stars are preferentially formed in the spiral arms of galaxies, but as clusters evolve they may tend to be less associated with a spiral pattern. Again, it is of interest over what timescale this occurs; the distribution of stars of different ages may vary according to the nature of the spiral arms in galaxies, whether they are density waves, short-lived transient arms, or arms induced by a bar or tidal perturbation (\citealt{Dobbs2009,Chandar2011}, Chandar al. 2016 in prep.). The distribution of stars is also often described as hierarchical (e.g. \citealt{Larson1995, Elmegreen1997, Bonnell2003, Elmegreen2014, Goul2015}), at least for young stars in clusters, often thought to be a consequence of the turbulent nature of the ISM. \citet{Grasha2015} recently  find that in NGC 628 this hierarchical nature tends to disappear for clusters older than 40 Myr. Other studies also indicate that hierarchical structure reduces over timescales of $\sim 70 $ Myr \citep{Gieles2008,Goul2015}, though longer timescales are found for the LMC \citep{Bastian2009}.

It is not currently computationally feasible to fully model cluster evolution in a galaxy. Even the evolution of isolated clusters is difficult to follow, with fully resolved studies of cluster formation and evolution limited to only small clusters \citep{Moeckel2010,Moeckel2012}, and these still require input regarding how much and how quickly gas disperses when a cluster forms. Although not able to follow individual stars, \citet{Fujii2016}, model more massive clusters, again combining hydrodynamic and N-body physics, and assuming the removal of gas after stars form. They find that the cluster evolution is largely dependent on the initial conditions; lower mass clouds tend to form open clusters, higher mass clouds form dense massive clusters, and lower density high mass clouds form `leaky clusters' \citep{Pfalzner2009}, clumpy clusters which more gradually evolve into less dense clusters as gas disperses. On galaxy scales, one approach is to model a cluster using an N-body code, subject to a galactic potential \citep{Baumgardt2003,Hurley2008,Renaud2013,Renaud2015,Rossi2015}, which allows the study of galactic tides, galaxy collisions and other such large scale processes on the cluster. \citet{Kruijssen2011} perform an N-body$+$SPH model of galaxies, allowing for the consistent formation of clusters in high gas density regions, and including a sub-grid model for stellar cluster evolution including mass loss from stellar evolution, two-body relaxation, and tidal shocks. The resolution of these models however is such that a single particle represents a full cluster, and the ISM and GMCs are not well resolved. In addition to simulations on galaxy scales or smaller, nowadays cosmological simulations have sufficient resolution in order for a single particle to represent a stellar cluster and at least give an indication of the spatial distribution of star clusters, although they do not follow cluster evolution.

Previous numerical work has also considered the evolution of GMCs, which may well be linked with the evolution of star clusters. \citet{Dobbs2013} showed that the evolution of GMCs is highly complex, with multiple cloud mergers (see also \citealt{Dobbs2015}) leading to the formation of GMCs, and likewise the dispersal of GMCs into multiple clumps. \citet{Dobbs2013} find typical lifetimes of 10 Myr (as measured by the time over which GMCs retain at least half their mass), with longer lifetimes for the most massive clusters. GMCs are dispersed by feedback and shear. \citet{Hopkins2012} find lifetimes typically of a few 10s of Myrs. Observational estimates find lifetimes of 20-30 Myr \citep{Kawamura2009,Meidt2015}, although \citet{Whitmore2014} estimate that gas is expelled form clusters within $\sim$ 10 Myr. \citet{Dobbs2014} also investigate stellar age ranges in GMCs. In particular, they note that long-lived inter-arm clouds typically contain a larger age spread compared to GMCs in the spiral arms, which tend to predominantly contain young stars.

In this paper, we primarily focus on isolated galaxy simulations, which have sufficient resolution to follow `clusters' as groups containing multiple stars. We note throughout that although we term these features `clusters' they may range from massive dense clusters to unbound associations. We also utilise cosmological simulations to compare the spatial distribution of clusters in quite different simulations, over longer time periods and with more realistically induced spiral arms. We consider a number of properties of the clusters, including spatial distribution, mass distributions and rough estimates of the age distributions. We also follow the more detailed evolution of the clusters in the isolated galaxy simulations. We do not include stellar evolution (other than stellar feedback) of the clusters, or mass loss other than the result of star particles dispersing, and likewise cannot model two-body relaxation. Instead the evolution of clusters in the isolated galaxy simulations is just driven by the immediate gas structure dynamics, gas and stellar gravity and, in relation to these, stellar feedback from the star particles.

\section{Details of Simulations}
We use four simulations, two of isolated galaxies and two cosmological simulations. The isolated and cosmological simulations have different properties, and advantages and disadvantages.  The isolated galaxy simulations provide much higher resolution (clusters consist of multiple star particles, as each particle is $\sim$ 300 M$_{\odot}$), but have the disadvantage that the simulations are limited to timescales of only a few 100 Myr.  Conversely the cosmological simulations have much worse resolution, and each star particle will represent an individual cluster, but these clusters can be followed over Gyr timescales.
The isolated galaxies also use an imposed spiral potential, which has the advantage that the spiral pattern is clearly visible, and we know where the spiral arms will be, but the spiral arms arise much more self consistently in the cosmological simulations (likely through both gravitational instabilities in the stars and interactions). Below we provide a brief outline of the simulations, but note that the details of the simulations have largely been described in previous literature.

\subsection{Isolated galaxy simulations}
The isolated galaxy calculations performed here are essentially the same as the spiral galaxy calculation described in \citet{Dobbs2014}. These calculations are performed with the smoothed particle hydrodynamics code sphNG \citep{Benz1990,Bate1995}. We model a gas disc with an imposed galactic potential. The potential is logarithmic, based on a model of a dark matter halo, and produces a flat rotation curve with maximum velocity $\sim$ 220 km s$^{-1}$ (and a galaxy mass of $\sim 2 \times 10^{41}$ kg at our maximum radius of 10 kpc) \citep{Binney1987}. To this we also add \textbf{a 2} armed spiral spiral potential \citep{Cox2002}, which constitutes a perturbation of a few per cent. There is no bulge component of the galaxy. The surface density of the gas is 8 M$_{\odot}$ pc$^{-2}$. The gas has a uniform distribution and extends to a radius of 10 kpc. The gas has an initial scale height of 400 pc, but settles into vertical equilibrium on timescales of 10s of Myrs. We use 8 million particles, giving a mass per particle of 312.5 M$_{\odot}$. Heating and cooling of the gas is added according to the chemical model of \citet{Glover2007}. Self gravity of the gas and stars is also included, with adaptive gravitational softening following \citet{PM2007}.  Star particles tend to be lower density than the gas particles, thus softening predominantly effects dense gas regions rather than star particles, which have wider separations. 

Stellar feedback, nominally in the form of supernovae feedback is added once a gas particle becomes sufficiently dense (exceeding 500 cm$^{-3}$), and the surrounding gas (over a radius of $\sim$ 15 pc, a distance based on the typical smoothing lengths of the particles at our resolution and which ensures we consider 10's of particles) is converging and gravitationally bound. The stellar feedback (inserted in the gas) consists of both thermal and kinetic energy, inserted in the form of a Sedov solution \citep{Dobbs2011new}, again over a radius of $\sim$ 15 pc. We associate an efficiency with the feedback inserted (see Table~1), as at our resolution, only a small fraction of the gas mass will form stars. Similar to \citet{Dobbs2014}, we first run a simulation for a period of time before implementing star particles. We take a simulation that has already run for a period of 200 Myr, with all the aforementioned physics and a feedback efficiency of 5\%, and use the end point of this simulation as initial conditions (these are also the same initial conditions as used in \citealt{Dobbs2014}). This means that the gas is already distributed with a clear spiral pattern,  and dense clouds and imminent regions of star formation are preferentially situated along the spiral arms. Moreover the gas distribution and properties no longer reflect the initial conditions (see Section 4.1). We ignore this first 200 Myr of evolution (during which we do not include star particles), and denote the end point of this first stage as a time of 0 Myr.
\begin{table*}
\begin{tabular}{c|cc|c|c|c|c|c}
 \hline 
 Isolated/ &  Simulation & (Star) particle & Feedback & Star Formation & Length of  \\
Cosmological & & mass (M$_{\odot}$) & efficiency (\%) & efficiency (\%) & simulation (Gyr) \\
 \hline
Isolated & Low feedback & 312 & 5 & 10 & 0.2 \\ 
Isolated & High feedback & 312 & 20 & 10 & 0.2 \\
Cosmological & Selene & $2.3\times10^4$ & $\gtrsim10$ & 1 & 13.7  \\ 
Cosmological & Castor & $2.4\times10^5$ & 10 & 2 & 13.7 \\  
 \hline
\end{tabular}
\caption{List of calculations performed, including the efficiencies for feedback and star formation, and the mass of an individual star particle. The isolated simulations model a 10 kpc radius galaxy with an SPH code, whilst the cosmological simulations use RAMSES to model the formation and evolution of galaxies in a cosmological context. We note that the efficiency parameters are not defined the same way in the cosmological and isolated simulations, and RAMSES-CH has no exact equivalent values to the isolated galaxy simulations (hence also why there is no fixed value for the feedback efficiency for Selene). We refer the reader to previous work \citep{Few2012} for more details.}\label{tab:simtable1}
\end{table*}

From this point, we run two simulations, with different levels of feedback each for 200 Myr. Then when star formation occurs, the densest particle (this will be the particle which is identified as $> 500$ cm$^{-3}$) is converted from a gas particle to a star particle. Stellar feedback occurs instantaneously, i.e. at the same instant the gas particle is converted to a star particle. Like \citet{Dobbs2014}, the star particles interact via gravity, but not hydrodynamic forces (they also do not accrete gas so are distinct from sink particles). The star particles do not experience stellar feedback, in the sense that their velocities are not changed when feedback is added. 
We make a small difference in the current work, in that we separate the efficiencies for stellar feedback, and star formation (whereas in \citealt{Dobbs2011new} they are assumed to be the same). Choosing the star formation efficiency shown in Table~1 means that star particles have the same mass as gas particles (given our choice of IMF, and resolution, see \citealt{Dobbs2014}) which is usually recommended in SPH, although in retrospect the relatively small differences in particle mass would probably not have been significant. The upshot of this approach is that we run two different simulations, which include different levels of feedback, but roughly the same amount of star formation for each star formation event. The main reason for testing two different levels of feedback is to compare the evolution and properties of the clusters when the gas is more or less readily dispersed by feedback. A summary of the calculations is shown in Table~1.

\subsection{Cosmological simulations}
Two cosmological simulations are used here, the first is Castor from \citet{Few2012} and is simulated using \textsc{ramses}. Of the galaxies presented in \citet{Few2012b}, Castor is chosen because it exhibits the clearest spiral structure and thus provides the best comparison with the isolated galaxy models. We also use the initial conditions for Selene from \citet{Few2012} but in this case the galaxy is simulated using \textsc{ramses-ch} \citep{Few2012b,Few2014}, at greater resolution, and with different feedback parameters (see Table 1 for details). Castor and Selene are evolved in a cosmological context using the ``zoom-in" method that allows the total simulated volume to be $20^3$~h$^{-3}$~Mpc$^3$ while the peak resolution is 436~pc (Castor) and 218~pc (Selene).

Castor is a member of a small galaxy group similar to The Local Group with a bar and clear spiral structure. The supernovae feedback in this model is not instantaneous but is delayed by 10 Myr from the moment of star formation and is deposited as kinetic energy. The number of SN per unit mass is consistent with a \citet{Salpeter1955} initial mass function.  Details of the model used to create Castor can be found in \citet{Few2012}. Star formation takes place within gas that is denser than 0.1~n$_{\mathrm{H}}$~cm$^{-3}$ with an efficiency of 10 \%. To ensure that the Jeans' length is resolved by the grid, a polytropic equation of state is enforced which acts as a density dependent temperature floor such that $T_\mathrm{g}Õ = MAX[T_\mathrm{g},T_\mathrm{th}(n_\mathrm{g}/n_\mathrm{th})]$ where $T_{\mathrm{th}}$=2900~K and $n_{\mathrm{th}}$=0.1~n$_{\mathrm{H}}$~cm$^{-3}$.

Selene is an unbarred field galaxy which also has clear spiral structure. The initial conditions and local environment for Selene are described in \citet{Few2012} but the model used to create this realisation of those initial conditions is detailed in \citet{Few2014}. In the case of Selene, feedback from the stars occurs throughout the lifetime of each star particle and represents type-II supernovae (SNII), type-Ia supernovae (SNIa) and asymptotic giant branch stars. Feedback from SNII is imposed on the gas phase as kinetic energy and feedback from SNIa is thermal. The number of SN per unit stellar mass is a function of the initial mass function which for this model is taken from \citet{Salpeter1955}. This model employs `delayed cooling' as described in \citet{Agertz2013} and \citet{Teyssier2013}. Star formation takes place above a density threshold of 0.1~n$_{\mathrm{H}}$~cm$^{-3}$  and the efficiency of feedback is directly controlled by the initial mass function, allowed mass of SN progenitors and the binary fraction (see \citealt{Few2014} for details). The polytropic equation of state is parameterised by $T_{\mathrm{th}}$=188~K and $n_{\mathrm{th}}$=0.1~n$_\mathrm{H}$~cm$^{-3}$ which allows the gas to remain at lower temperatures than in the Castor run.

Generally we would expect the cosmological simulations to show less clear structure. This is because they do not use an imposed potential but is also driven in part by the resolution. The large volume of the simulation means that the physical resolution is lower than in the isolated simulations which means the gas does not reach such high densities in the spiral arms.

\subsection{Cluster determination}
For the isolated galaxy simulations, the resolution is sufficiently high that clusters, particularly massive clusters, will be represented by multiple star particles. By contrast, for the cosmological simulations one star particle will correspond to one cluster. For some of our results for the isolated galaxies, we group the star particles into clusters. To do this we adopt a friends of friends approach, as used in \citet{Dobbs2015} to find clouds, but with some differences for the stars. For clouds (which are only used in Section 4.3), the friends of friends algorithm first selects particles above a density of 50 cm$^{-3}$. Then for those particles, all those within a distance of $l=10$ pc from each other are grouped together into a cloud. 
We specify that clouds should contain at least 100 particles, so they are resolved (e.g. properties such as the velocity dispersion can be computed). For the stars, we simply apply the friends of friends algorithm to all the star particles and group together all those within distances of $l=20$ pc from each other into stellar clusters. Clusters are specified to contain at least 10 particles; with the stars the algorithm is used simply to group particles together rather than resolve cluster properties or dynamics. With such a large $l$, our clusters are evidently not as concentrated as real clusters. Our clusters have radii which are typically several pcs to a few 10s of pcs (see Section 4.4.3) which means they likely reflect both unbound clusters and compact clusters.  However we cannot resolve down to the formation of individual stars with sub parsec separations, rather the clusters reflect the resolution of our galactic simulations. About 15\% of the star particles are grouped into clusters in the simulations; the remainder are isolated.

\section{Observational data}
To compare our simulations with real observations of Young Stellar Cluster (YSC) populations in local spirals, we use three  recently compiled cluster catalogues. The M\,83 cluster catalogue has been obtained from two HST datasets that provide a large coverage of the inner 8 kpc (diameter) of the galaxy. The dataset, and catalogue is widely described in \citet{Silva-Villa2014} and \citet{Adamo2015b}. We refer the reader to these works for more details. The other two YSC catalogues have been obtained from the two spiral galaxies NGC\,628 and NGC\,1566. Both targets are part of the Legacy ExtraGalactic UV survey (LEGUS, \citealt{Calzetti2015}). In the case of NGC\,628 we use only the catalogue obtained from the central pointing which covers about 2 kpc in radius of the inner region. The HST pointing of NGC\,1566 covers the inner region for about 5 kpc in radius. The method used to produce the LEGUS cluster catalogues is described in Adamo et al. 2016, in prep. YSC physical properties in the three galaxies have been derived with similar techniques. We use integrated fluxes from multiband HST photometry which covers the NUV up to the NIR cluster spectrum.

\section{Results}

\subsection{Evolution of simulations}
We do not provide much discussion of the evolution of the simulations as they are discussed elsewhere \citep{Few2012,Ruiz-Lara2016,Dobbs2014}, but we do give a brief summary here. In the isolated galaxy simulations, the gas arranges into dense clouds, preferentially along the arms, by a combination of gravitational instabilities, thermal instabilities and cloud-cloud collisions. As they leave the arms, the clouds tend to be sheared into spurs. Overall the properties of the simulations do not undergo significant changes after the first 150 Myr (-50 Myr in our timeframe, as we have already run our simulations for 200 Myr before the results presented here), for example the thermal distribution of the ISM, and star formation rates \citep{Dobbs2011new}. The main departure models sometimes show from equilibrium is the build up of strongly bound clouds which are not easily dispersed by feedback and have very long lifetimes \citep{Dobbs2011new}. In our low feedback model, the number of long-lived massive clouds increases over the course of the initial 200 Myr and the 200 Myr presented here, but most particularly the last 100 Myr. In the high feedback model, such clouds do not readily occur, and the evolution of the disc is basically in equilibrium.

The two cosmological simulations used in this work (Castor and Selene) are evolved from redshift $z=50$ to the present day in a fully cosmological context. This means they are subject to galaxy mergers and satellite accretion consistent with $\Lambda$CDM cosmology which strongly influences the morphology and mass assembly of the stellar disc itself \citep{Ruiz-Lara2016}. The galaxies initially form from hierarchical merging of small condensed gas structures for a period that lasts around $\sim$3.2~Gyr. There is then a period of $\sim$4.4~Gyr (averaged over the entire sample of 19 galaxies) during which several large satellites may be accreted. This epoch ends with the last such accretion event at a lookback time of 8.0~Gyr for Castor and 6.0~Gyr for Selene. Finally the galaxy evolves secularly until the present day with only very minor mergers (satellites with masses less than 1\% that of the host galaxy) taking place. These simulations differ from the isolated runs in that the spiral structures are transient and have no explicit potential that enforces them. Furthermore, orbiting satellites can trigger and distort the spiral arms. Outright mergers may also temporarily disrupt the internal structure of the galaxy but we find that spiral arms quickly reassert themselves following most typical mergers.

\subsection{Spatial distribution of different age stars}

\subsubsection{Spatial distribution of different age stars: isolated galaxy simulations}
In this section we consider the spatial distribution of star particles and clusters of different ages in the isolated galaxy simulations. 
We first consider the spatial distribution of the star particles, where we make no attempt to group the stars together into clusters. This has the advantage that we can analyse the spatial distribution without any uncertainties about how the clusters are grouped together, or how their ages are assigned. In particular, as  we show in Section 4.4.1, there may be a considerable spread in the age of star particles in some clusters meaning there is no obvious particular cluster age. The disadvantage of this approach is that these results are less comparable with the observations where clusters are used. As well as being lower masses than observed clusters, the star particles do not evolve; they simply traverse the galaxy without changing mass, or being created or destroyed. Real clusters by comparison can clearly undergo disruption, and are expected to lose mass over time. 

\begin{figure}
\centerline{\includegraphics[scale=0.3, bb=200 100 400 550]{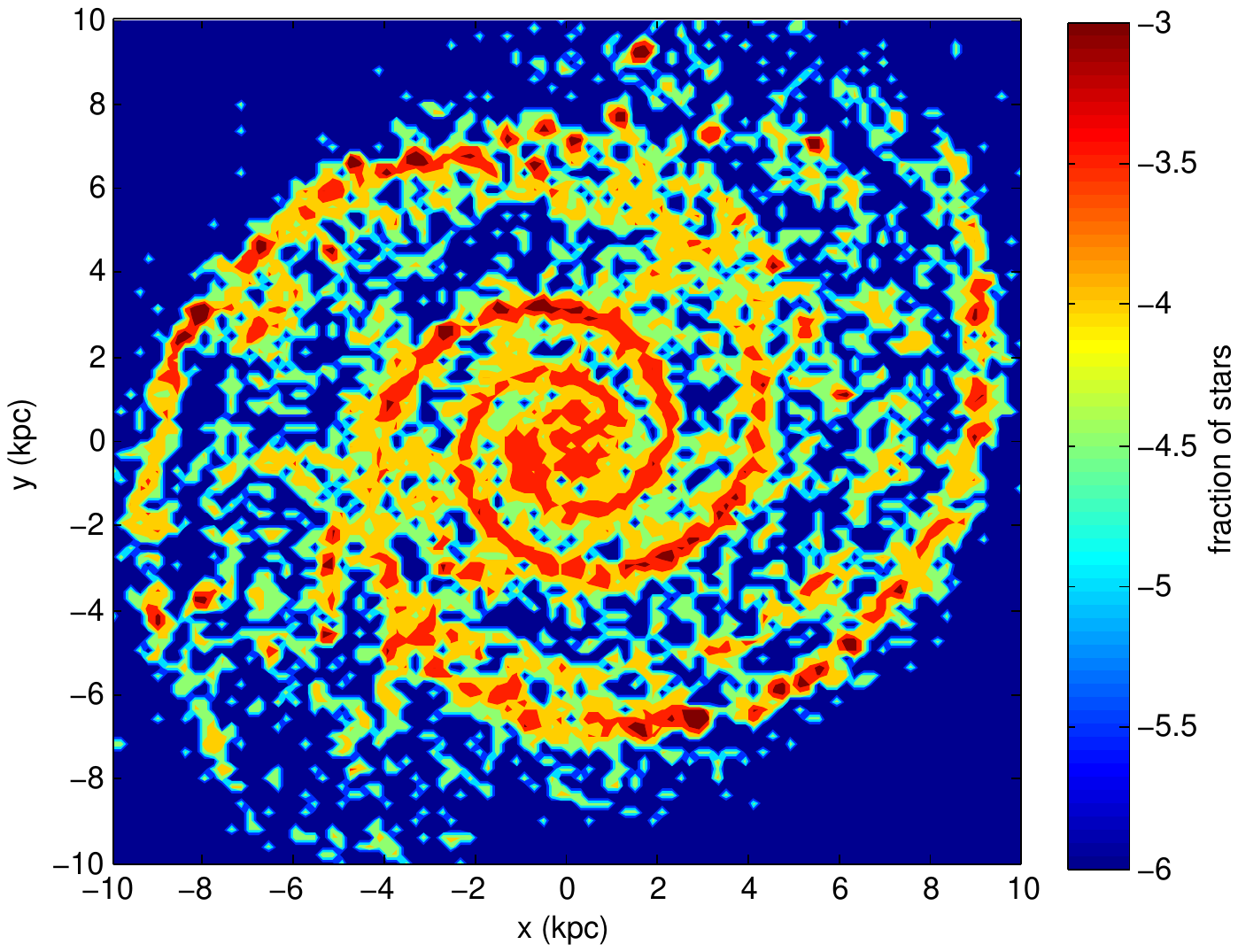}
\includegraphics[scale=0.3, bb=0 100 400 450]{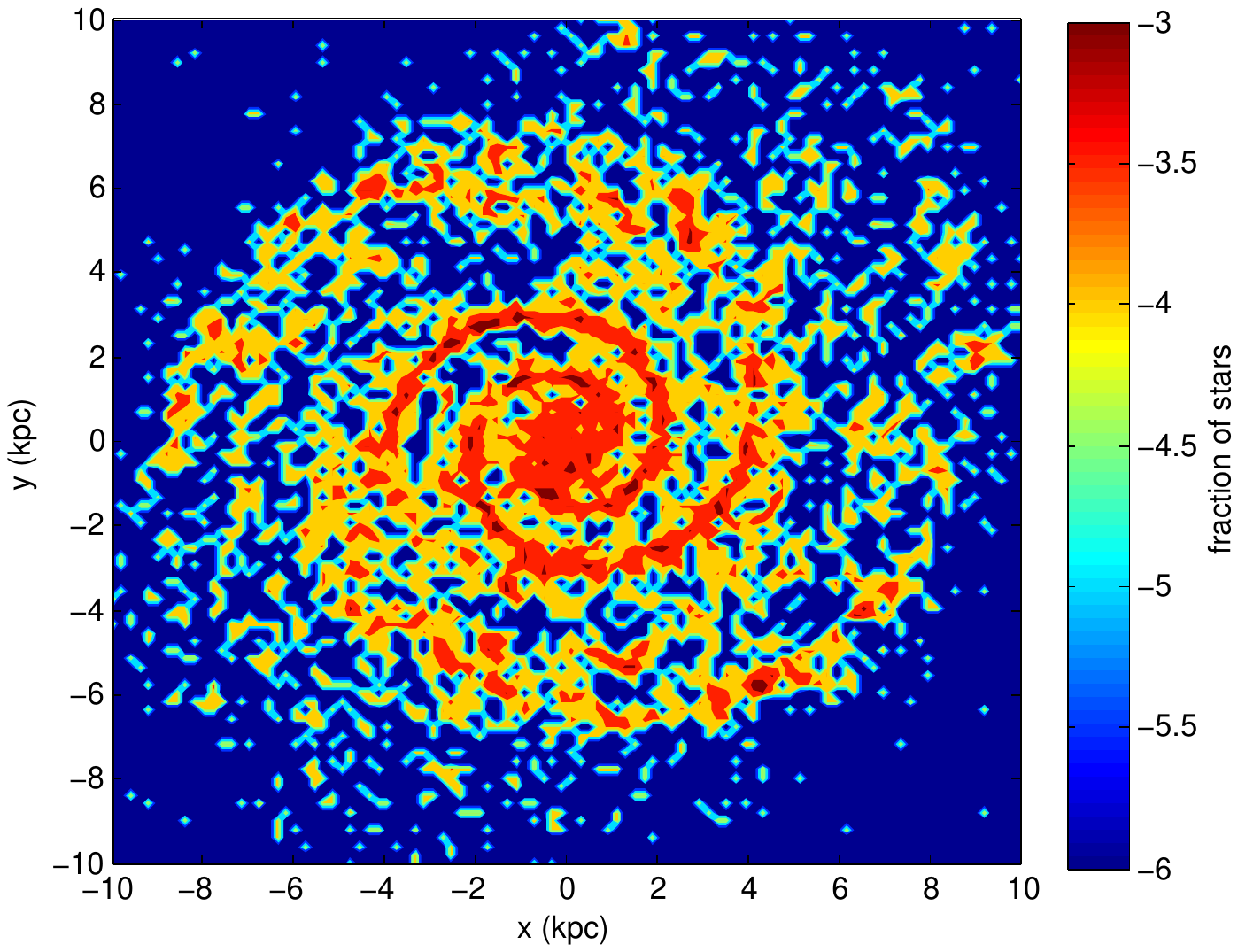}}
\centerline{\includegraphics[scale=0.3, bb=200 100 400 450]{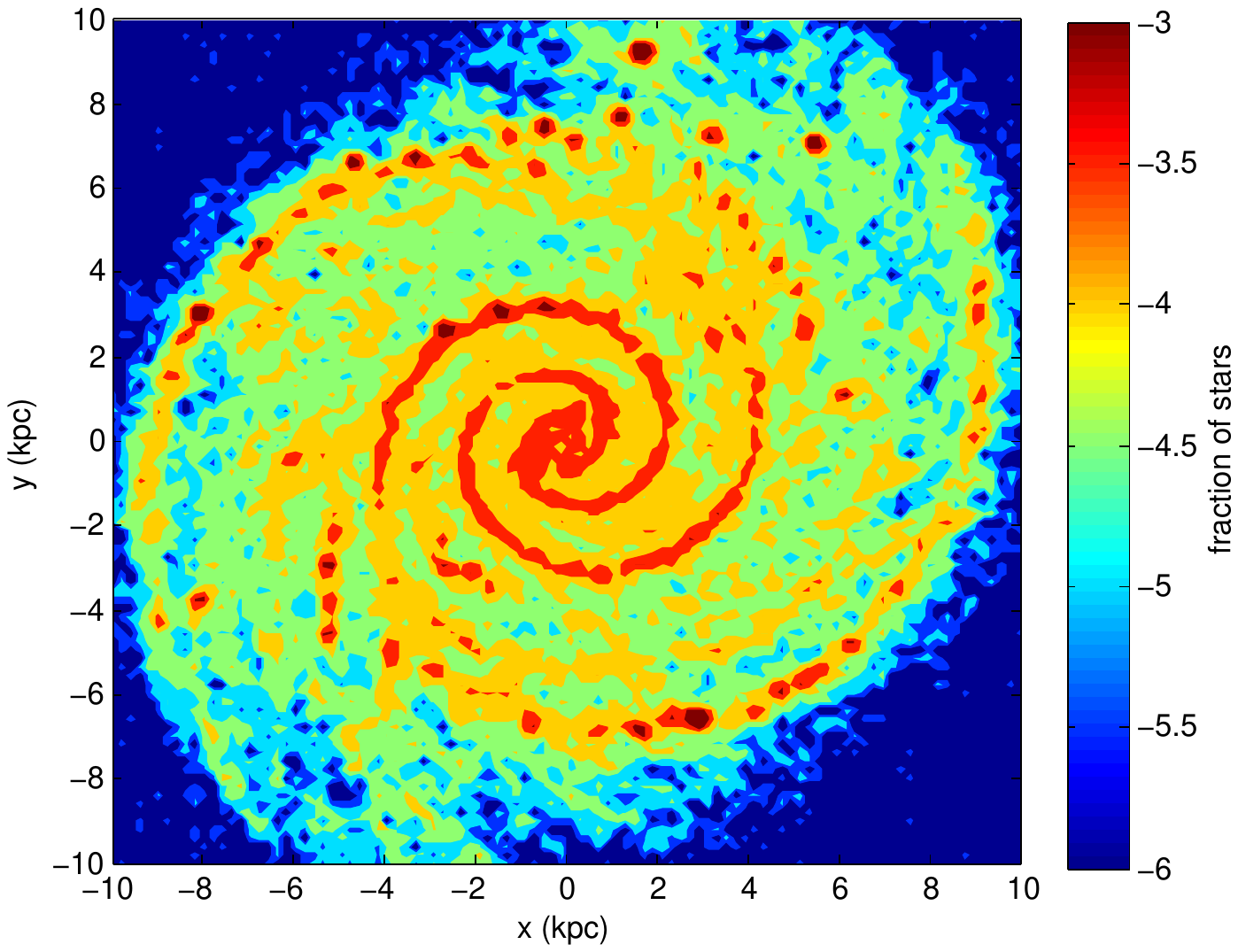}
\includegraphics[scale=0.3, bb=0 100 400 450]{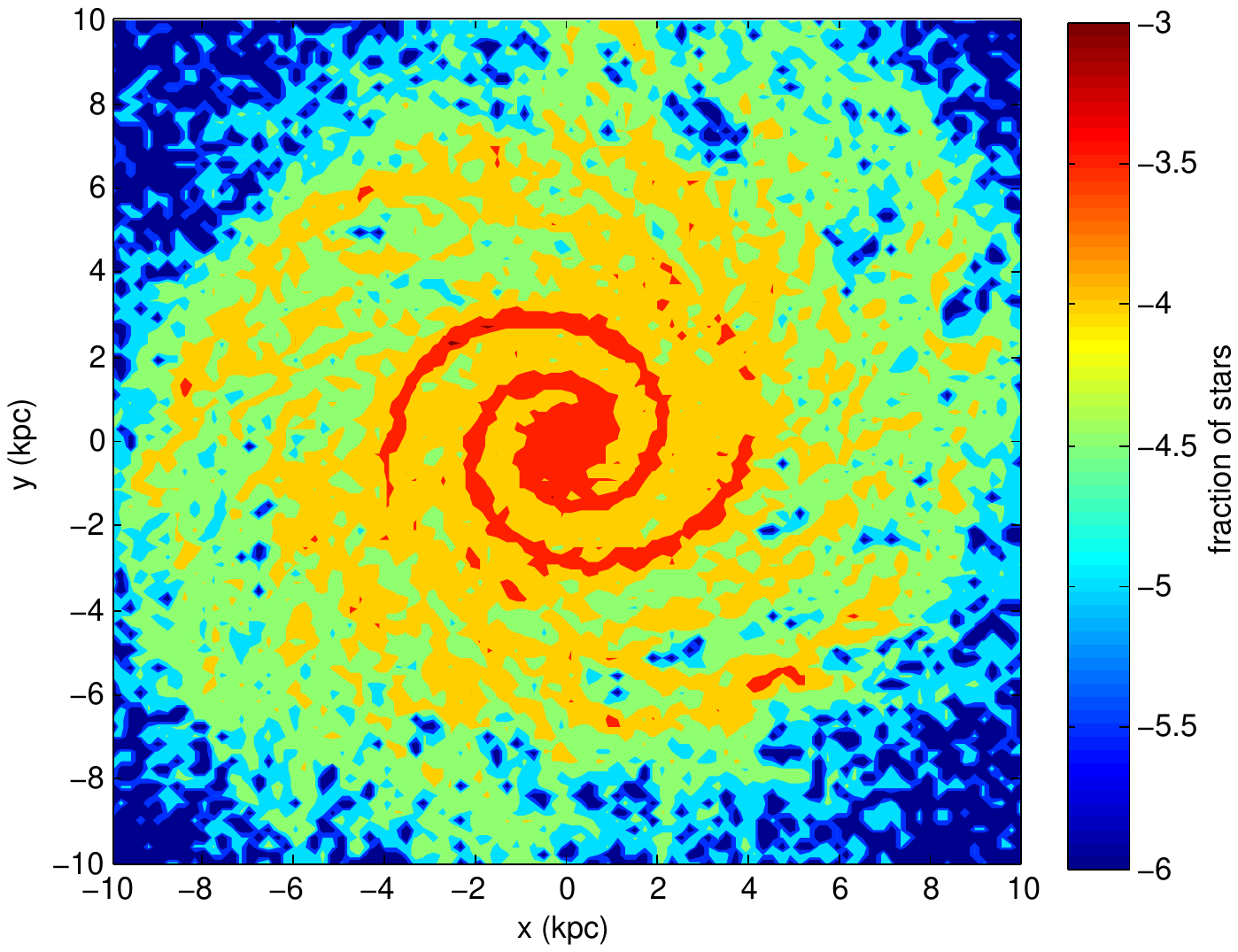}}
\centerline{\includegraphics[scale=0.3, bb=200 200 400 450]{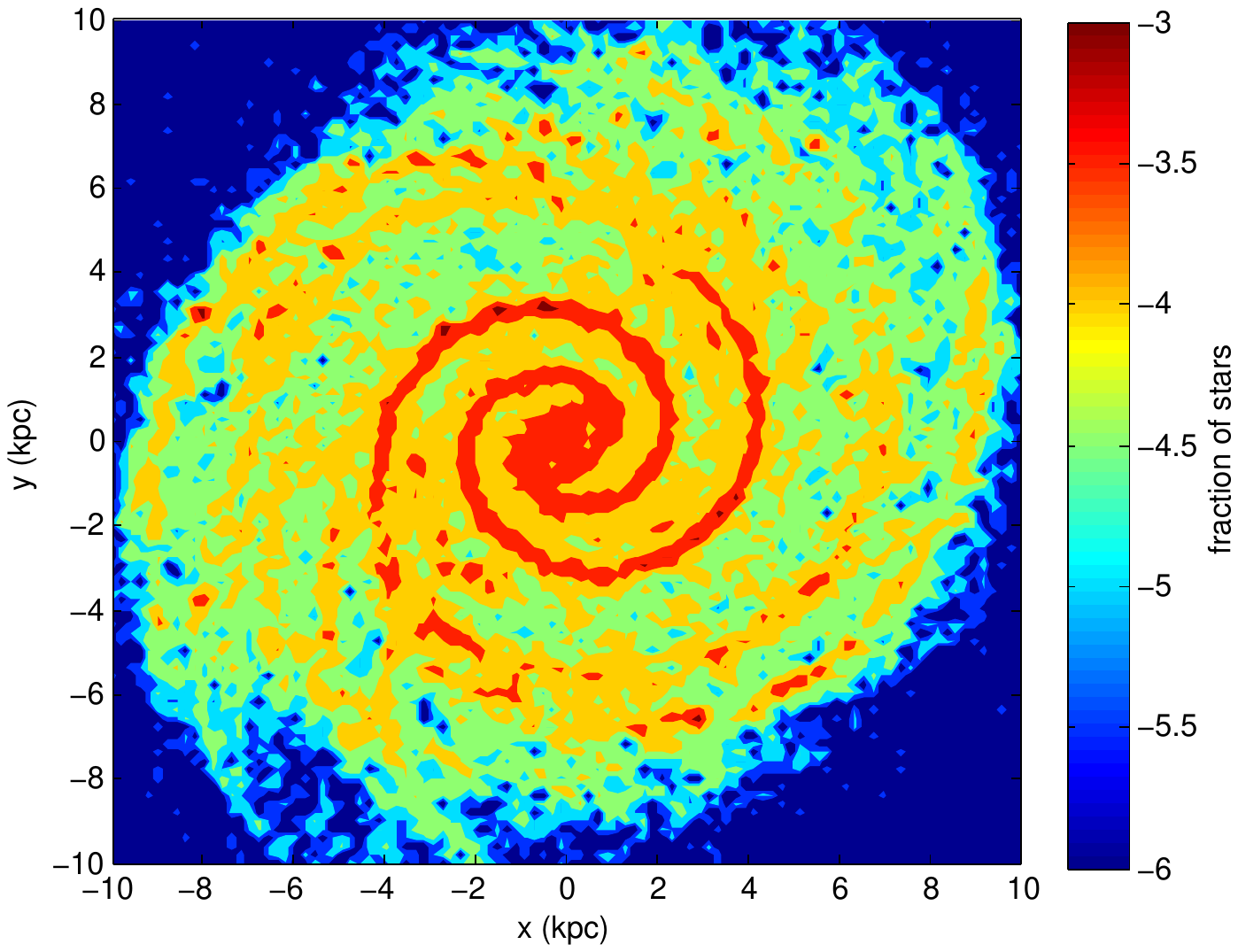}
\includegraphics[scale=0.3, bb=0 200 400 450]{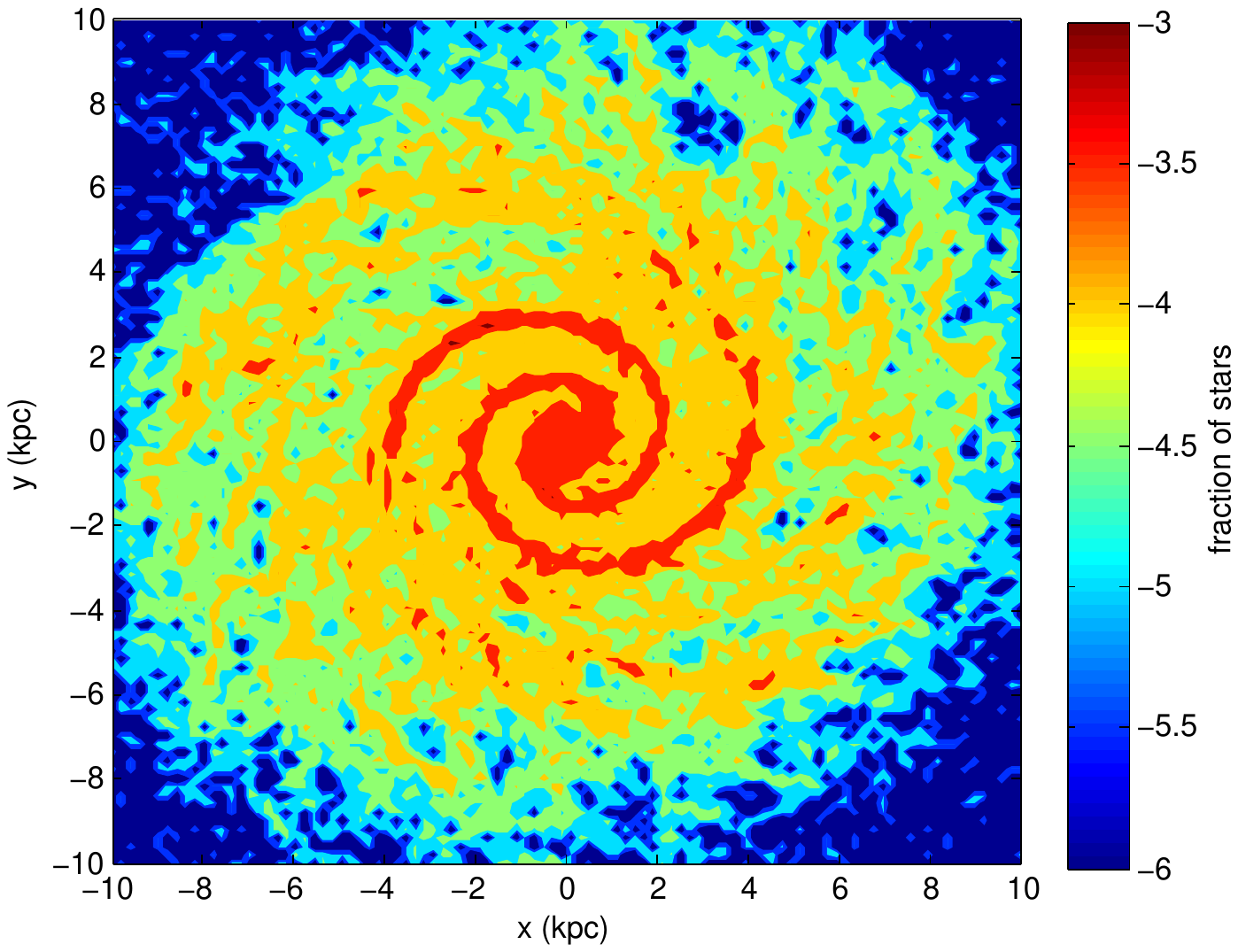}}
\caption{The normalised number density of stars is shown for the low feedback isolated galaxy model (left) and high feedback model (right). The number density is calculated for star particles of ages 0-10 Myr (top), 10-100 Myr (middle) and 100-200 Myr (lower panels). Spiral structure is clear for all stars, in all age ranges, though strongest in the youngest stars. The spiral arms are also stronger in the lower feedback model.}\label{fig:isostars}
\end{figure}

In most of our figures of the spatial distribution of clusters of different ages, we simply show points for the clusters. However when considering the star particles in the isolated galaxy models, there are many more star particles compared to clusters, either in the cosmological simulations, or the observations (where clusters are typically at least 1000 M$_{\odot}$). Since there are so many points, for easier viewing we instead produce contour plots of the number density of star particles of different ages (Figure~\ref{fig:isostars}). We group our ages into bins of 0-10 Myr, 10-100 Myr, and 100-200 Myr, which are comparable to the bins used for the analysis for the cosmological simulations and observations. To make the contour plots, we divide the galaxy into a 100 by 100 grid, whereby each grid cell is 200 pc $\times$ 200 pc. In each plot, we normalise the number density of stars by the total number of star particles in a given age bin. We show plots for the different age ranges and different feedback models in Figure~\ref{fig:isostars}. 

The star particles in all the age ranges show clear spiral structure, although with the higher feedback model, the spiral structure is weak in the outer parts of the galaxy. In the inner regions, the potential is strong enough that even when modelling test particles (i.e. particles not subject to gas pressure), a clear spiral pattern is seen (see e.g. initial conditions in \citealt{Dobbs2006}). In the inner regions, the stars spend relatively little time between the spiral arms. In the outer regions, the spiral structure is weaker, and there are more stars, and massive clusters (seen as the red circular patches) evident between the spiral arms. The low feedback model also shows a secondary feature in addition to the main spiral arms, which is associated with a branch occurring at the 4:1 resonance (see e.g. \citealt{Shu1973, Patsis1997,Few2016}. 

Figure~\ref{fig:isostars} shows quite a different distribution for the youngest stars (ages $<$ 10 Myr, top panels) compared to the older stars (lower and middle panels). In particular the youngest star particles are concentrated into patches with very low numbers of stars, or no stars in between. There is a high concentration of star particles in the arms, and a large difference between arm and inter arm regions. By contrast, even for the low feedback model, the star particles aged 10-100 Myr show a more even distribution, with a factor of 10 or less difference between the arm and inter arm region. There is relatively little difference between the ages of 10-100 and 100-200 Myr, particularly for the high feedback model. 

Figure~\ref{fig:isostars} also indicates some differences between the low (left) and high (right) feedback models. 
For the low feedback model, massive clusters are still obvious for ages 10-100 Myr, and to a lesser extent for ages 100-200 Myr, whereas in the high feedback model, massive clusters are not particularly evident. The reason for the difference between the models is that with the lower feedback, GMCs are less able to disperse. Statistically some GMCs will disperse, but others will remain bound for 10s if not 100s of Myrs. In these GMCs, the mass of gas and young clusters in the clouds simply builds up over time. At some point, all the gas may be turned into stars, but we do not reach this time in our simulation. In \citet{Dobbs2011new}, we also saw a couple of clouds which were not dispersed, whilst \citet{Tasker2015} similarly note a population of very long lived clouds. By comparison in the higher feedback model, the GMCs are fairly readily dispersed. 

We also compare the spatial distribution of `clusters' by grouping the star particles into clusters as described in Section~ 2.3. Whilst bearing more relation to observed clusters, this approach has the disadvantage that the clusters do not necessarily have well defined ages. Here we simply assign clusters to different age bins if there is a given overdensity of stars. So if the stars were distributed uniformly by age in a cluster, we would expect 5\% in the 0-10 Myr bin, 45\% in the 10-100 Myr bin, and 50\% in the 100-200 Myr age bin. We assign a cluster to a particular age bin if there are at least 50\% more stars in that bin compared to a uniform distribution (i.e. 7.5\% in the 0-10 Myr bin, 67.5\% in the 10-100 Myr bin, 75\% in the 100-200 Myr bin; if this occurs for more than two bins we choose the one with the highest overdensity).  We choose a 50\% overdensity to try and avoid having very few star particles in each bin, and note that we cannot take a 100\% higher overdensity or more, without starting to underpopulate the 100-200 Myr bin.   
\begin{figure*}
\centerline{\includegraphics[scale=0.3]{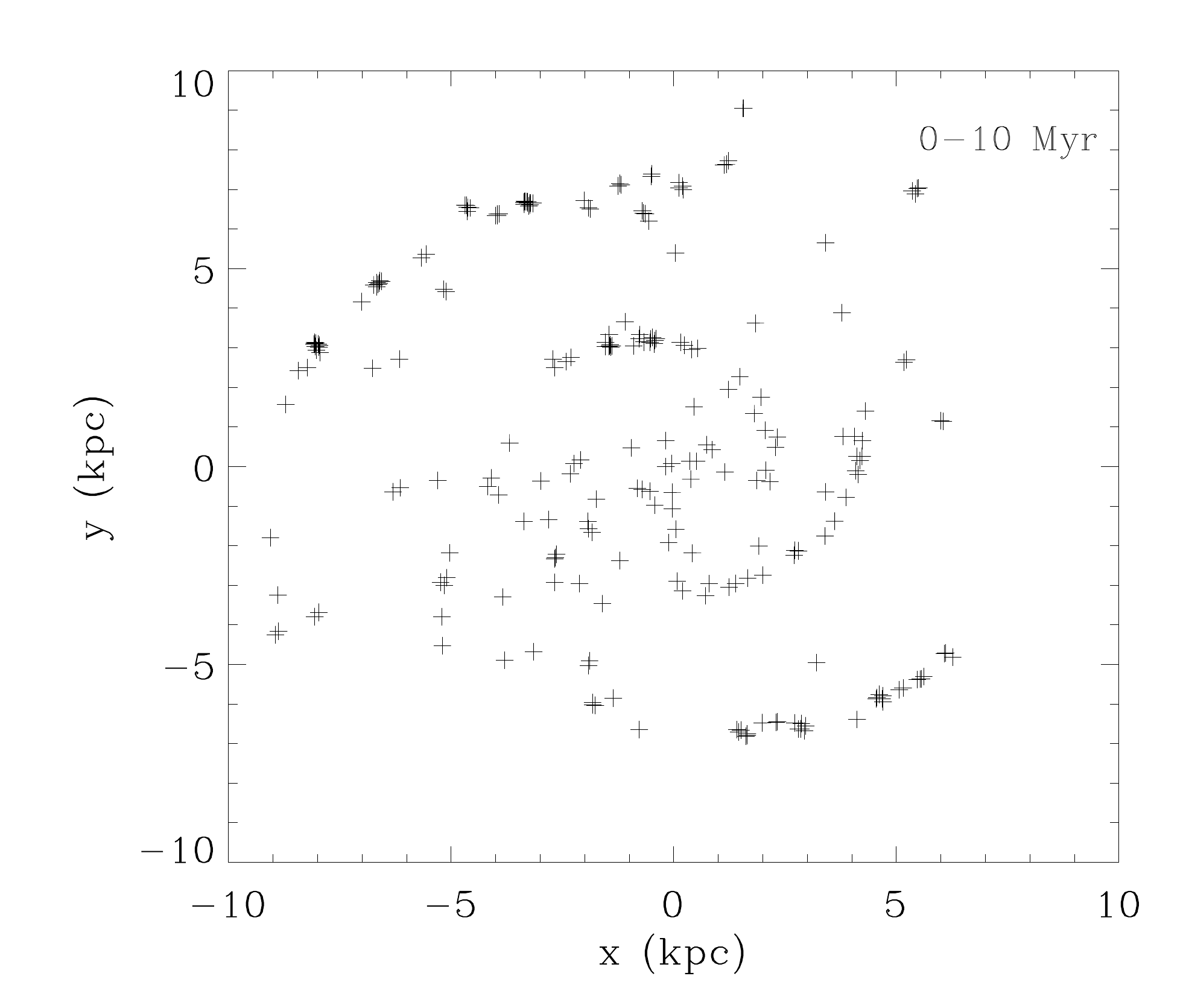}
\includegraphics[scale=0.3]{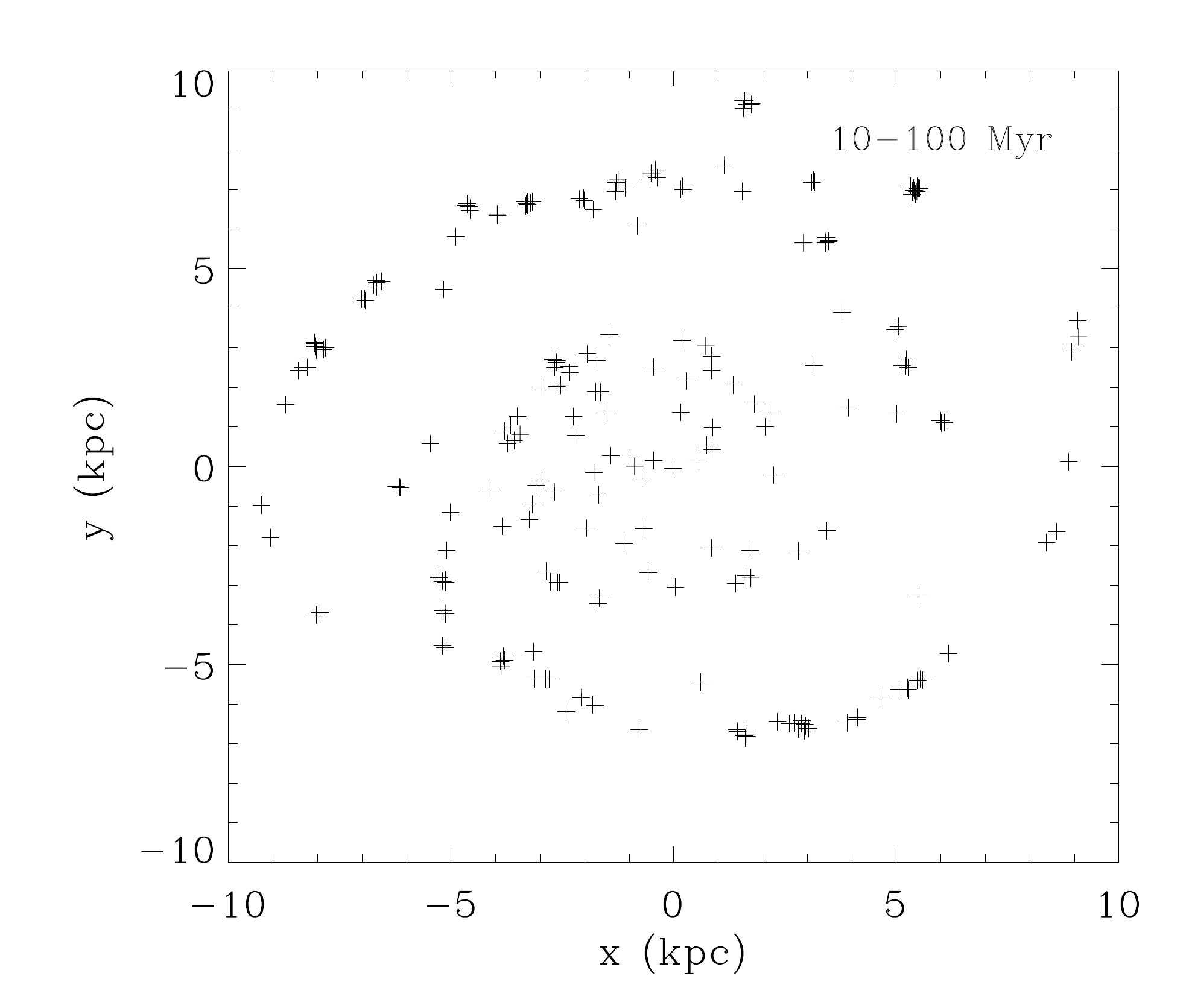}
\includegraphics[scale=0.3]{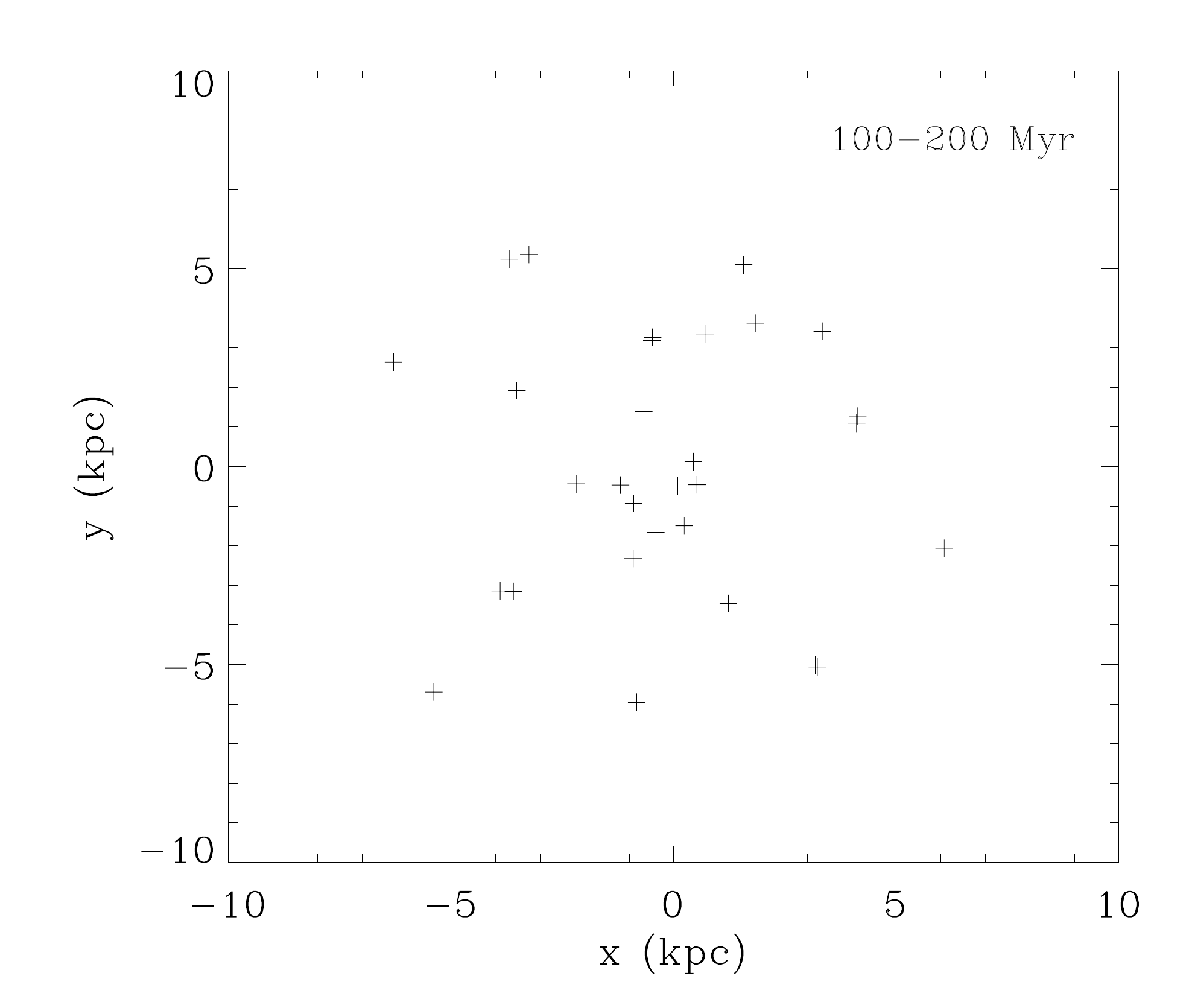}}
\caption{The distribution of clusters (see text) in different age bins is shown for the low feedback isolated galaxy simulation.}\label{fig:isoclusters}
\end{figure*}

Figure~\ref{fig:isoclusters} shows the distribution of `clusters' of different ages, determined as described above (now plotted simply as points) for the low feedback model. There are now far fewer points (and some clusters are not shown as they do not show an overdensity of stars of any age), which as we will see is more similar to the observational data. By grouping the star particles into clusters, we also now allow scope for cluster evolution (see  Section~4.4), and in particular clusters can disperse over time (though individual star particles survive indefinitely).  The clusters in the youngest bins (0-10 and 10-100 Myr) are still very concentrated in the arms. The main difference compared to taking the star particles is that there are relatively fewer clusters of older age. So there are approximately the same number of clusters in the 0-10 to the 10-100 Myr bin, and notably few clusters in  the 100-200 Myr bin. The scarcity of the clusters in the 100-200 Myr bin presumably reflects that the oldest clusters have dispersed, or become part of clusters now dominated by younger stars. As we will show in Section~4.4, both of these possibilities occur, although most clusters likely disperse as there are not so many massive clusters that have been in existence for at least 100 Myr. There are too few clusters in the 100-200 Myr bin to determine if there is spiral structure simply from Figure~\ref{fig:isoclusters}. Overplotting on an image of the galaxy simulation (not shown) indicates that they appear less concentrated ($\sim50$\% in arms and inter-arm regions respectively) in the spiral arms compared to the earlier bins. The higher feedback model shows similar behaviour, but also with fewer clusters in the 10-100 Myr range. 

We also tried using the mean or median age, using all of the clusters. 
Particularly when using the mean, the cluster ages are often larger because the age distributions tend to be skewed towards higher ages. Consequently we used the overdensity definition rather than the mean. When using the mean, there are few young clusters, and other clusters are generally given older ages, so consequently spiral structure tends to persist more in the older clusters.

\begin{figure}
\centerline{\includegraphics[scale=0.7]{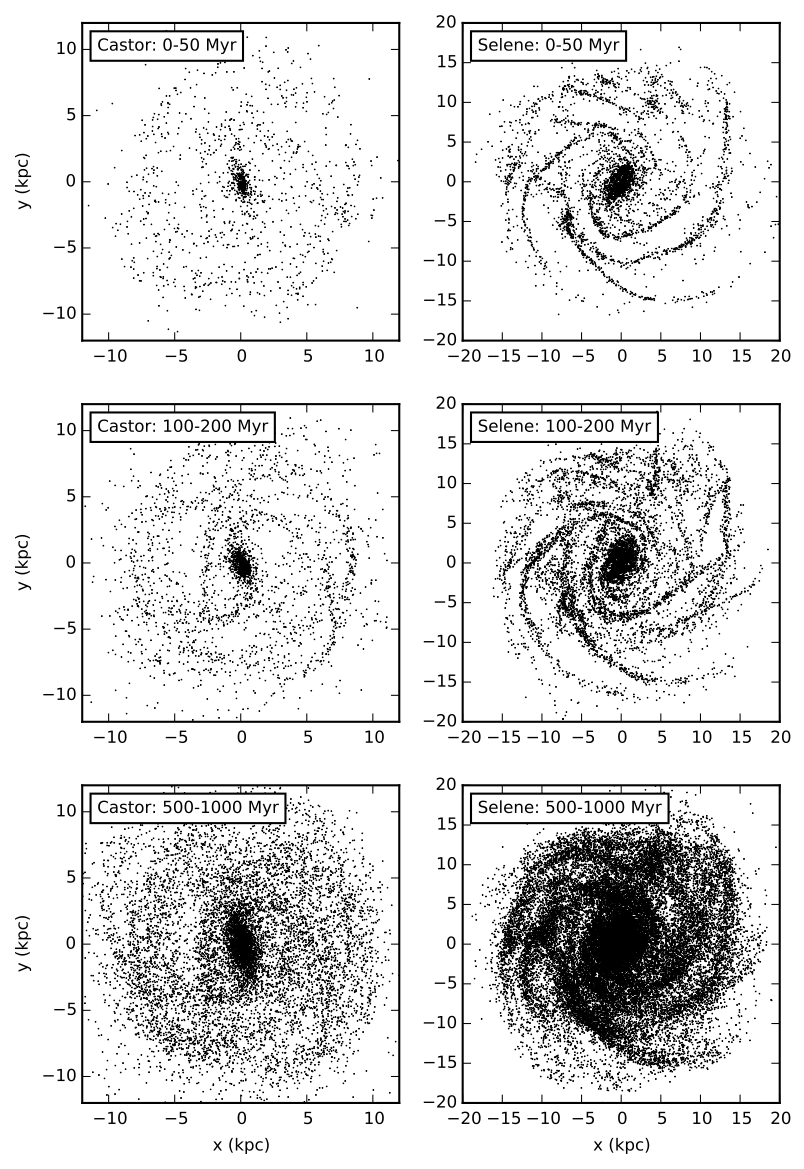}}
\caption{The distribution of stellar ages is shown for the cosmological simulations Castor and Selene. For Selene, with higher resolution, the spiral structure is much clearer in the stellar clusters. For both models, the oldest clusters are less confined to the spiral arms.}\label{fig:cosmoclusters}
\end{figure}

\subsubsection{Spatial distribution of different age stars: cosmological simulations}
We show the distribution of star clusters of different ages for the cosmological simulations Castor and Selene in Figure~\ref{fig:cosmoclusters}. We recall that the cosmological simulations have a more realistic development of spiral arms compared to the isolated galaxy simulations, and we note that the star particles (in this case clusters) survive indefinitely. There are far fewer star particles compared to the isolated galaxy simulations.  We also show a larger range of ages, taking 0-50 Myr, 100-200 Myr and 500-1000 Myr, reflecting the longer timescales involved.

Castor shows spiral structure in the stellar population, although it is not as clear as for the isolated galaxy simulations. Surprisingly the youngest clusters do not show particularly clear spiral structure, whereas for the isolated galaxies the spiral structure is strongest in the youngest stars, as would be expected. Slightly stronger spiral structure is seen in the clusters with ages 100-200 Myr. It is not clear why there is a difference in behaviour. This could simply reflect the low resolution of Castor compared to the isolated galaxy simulations, or a result of the time evolution of the arms (e.g. if the main arms are decaying then perhaps the strongest spiral patterns
in the stars are seen in slightly older stars when the spiral arms were stronger). Selene, which has higher resolution, shows more similar behaviour to the isolated galaxy simulations. For Selene, the clearest spiral structure is shown in the youngest stars. Like the isolated galaxy simulations, there is also still fairly clear spiral structure even in the older (100-200 Myr) stars. For the oldest stars (500-1000 Myr), beyond the time frame of the isolated galaxy simulations, the clusters show much more deviation from the spiral arms, both for Castor and Selene.

We also examined a number of other cosmological simulations, with both the lower resolution of Castor and higher resolution of Selene. We tended to see clear spiral patterns in the clusters only when the spiral arms were relatively strong, and the spiral structure tended to be clearer in the higher resolution simulations.

\subsubsection{Spatial distribution of different age stars: observations}
\begin{figure*}
\centerline{\includegraphics[scale=0.3]{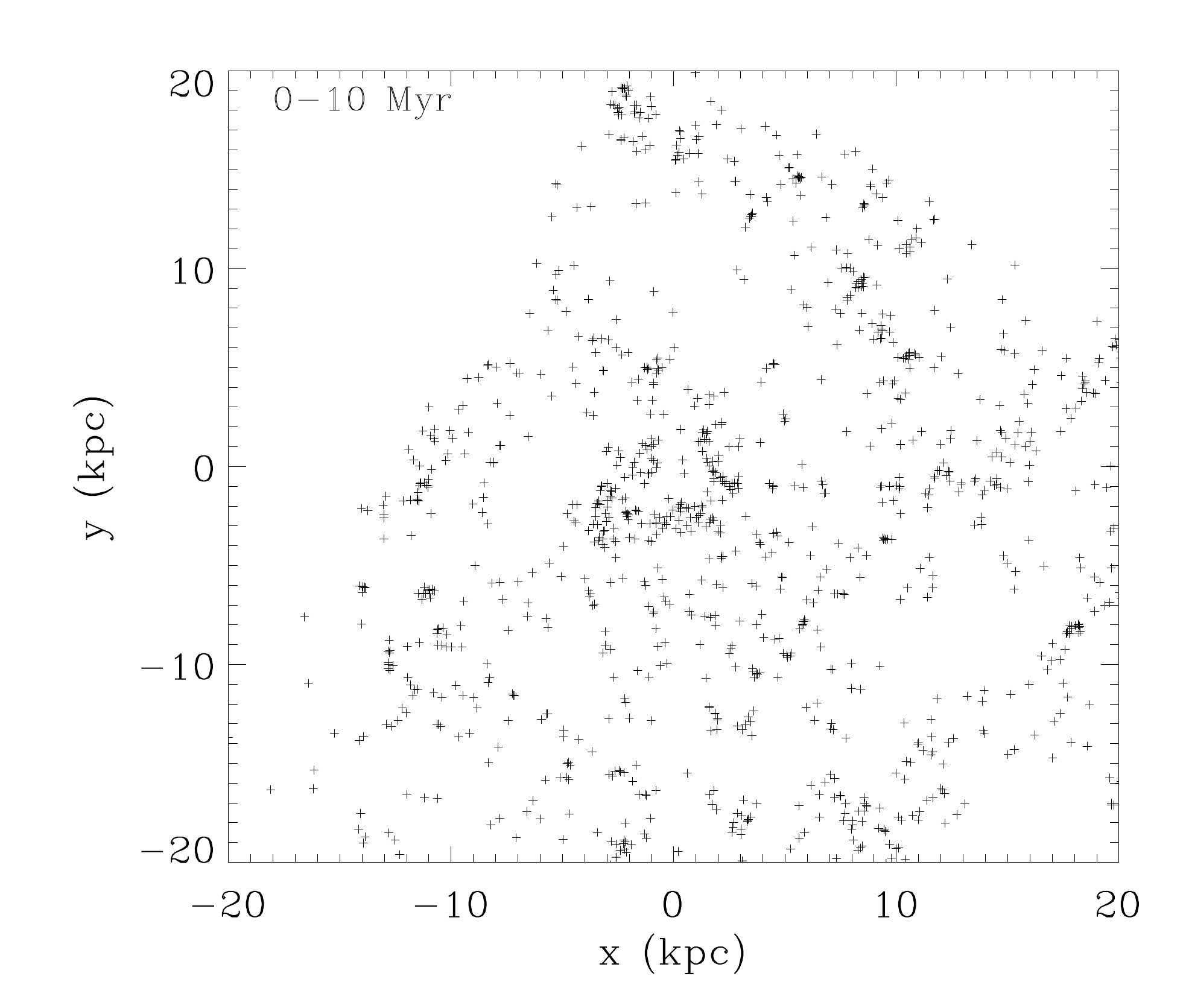}
\includegraphics[scale=0.3]{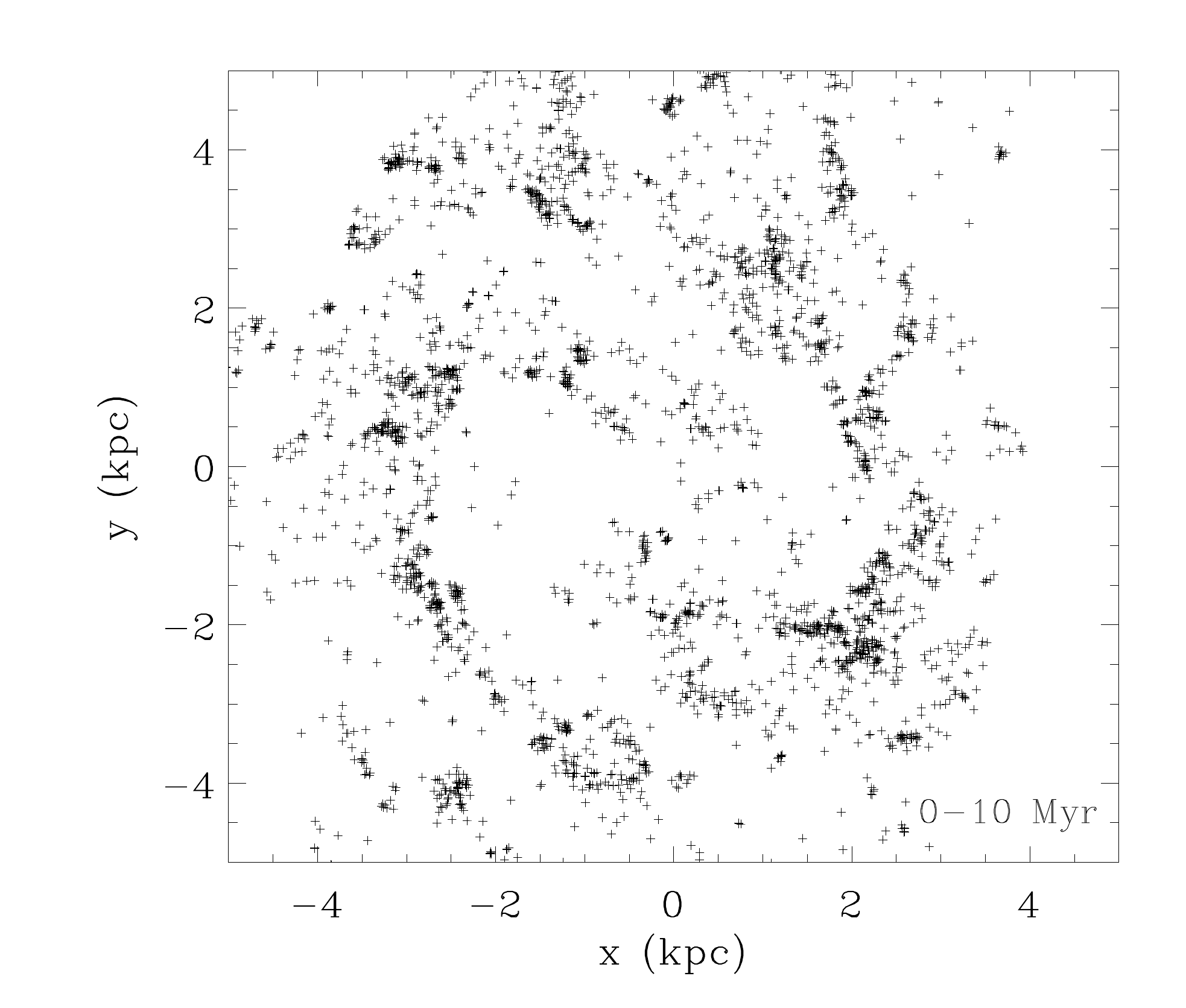}
\includegraphics[scale=0.3]{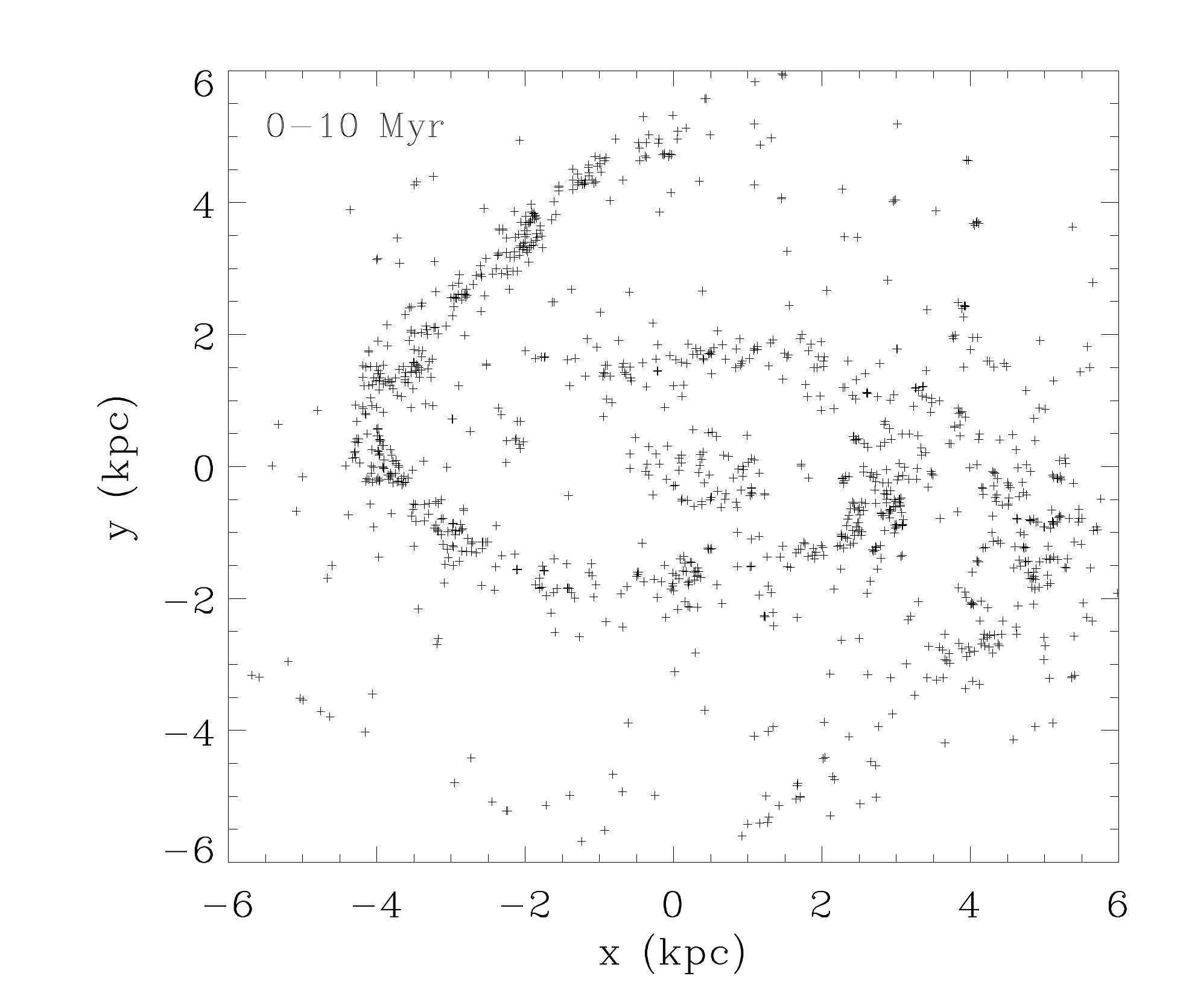}}
\centerline{\includegraphics[scale=0.3]{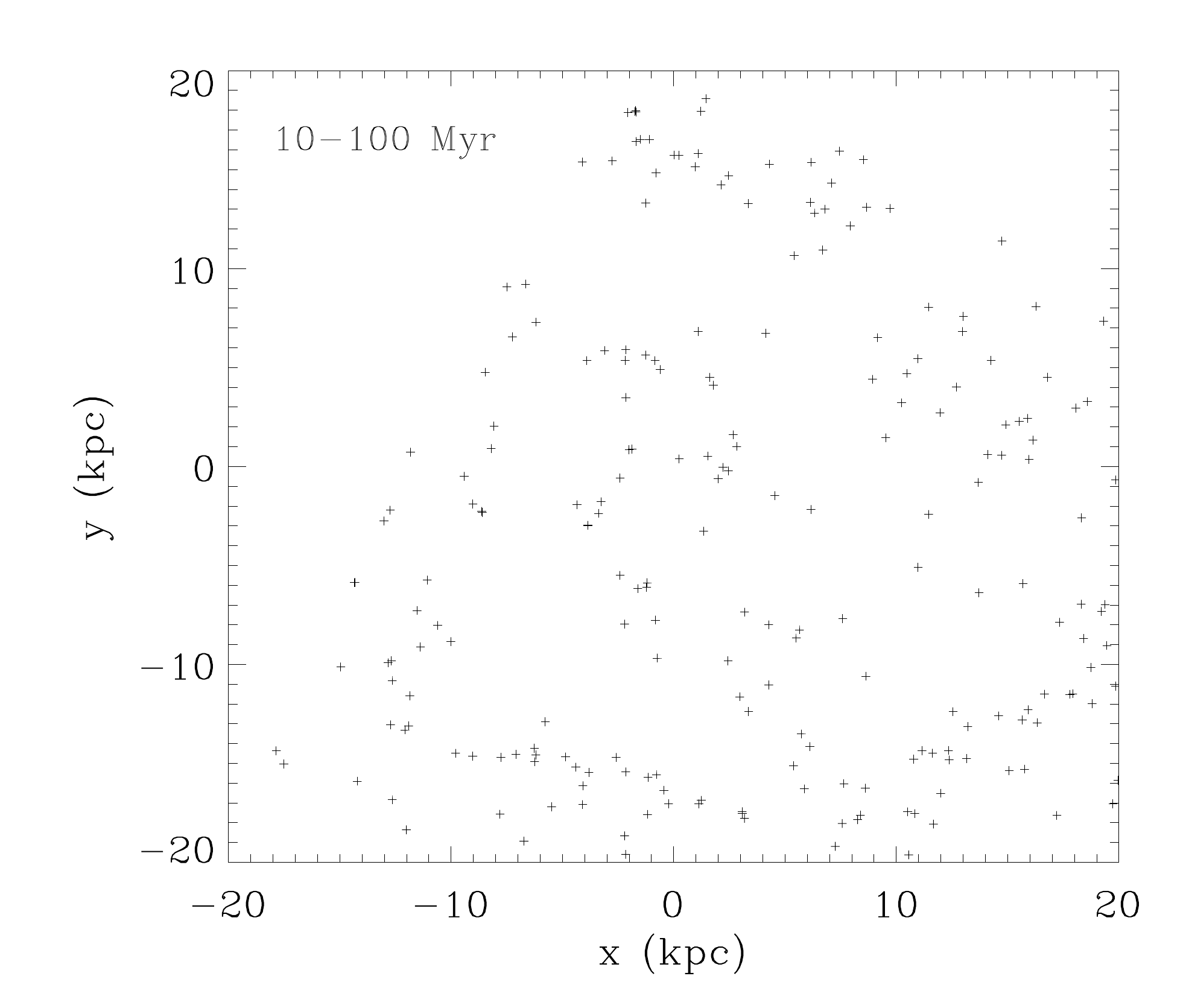}
\includegraphics[scale=0.3]{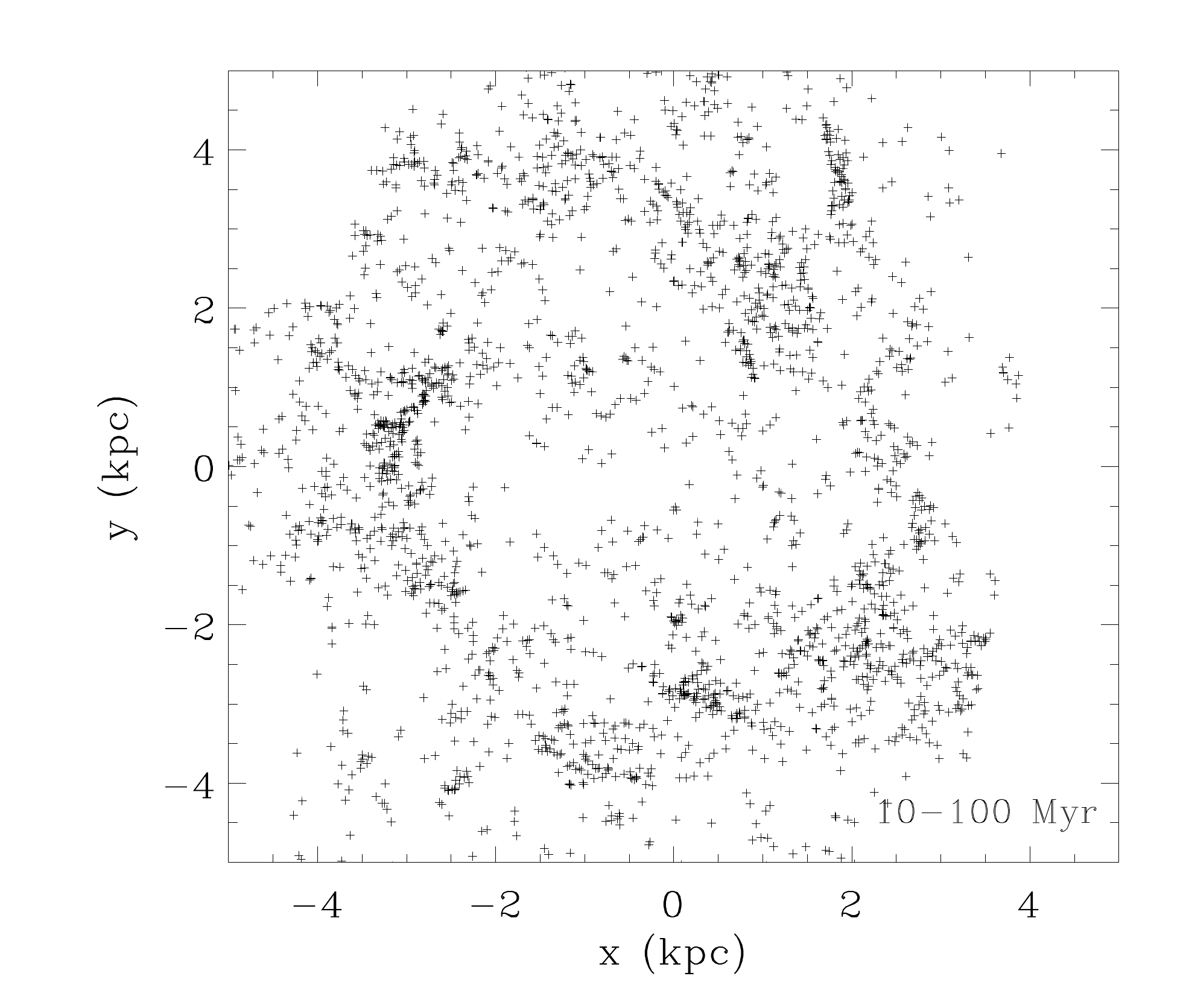}
\includegraphics[scale=0.3]{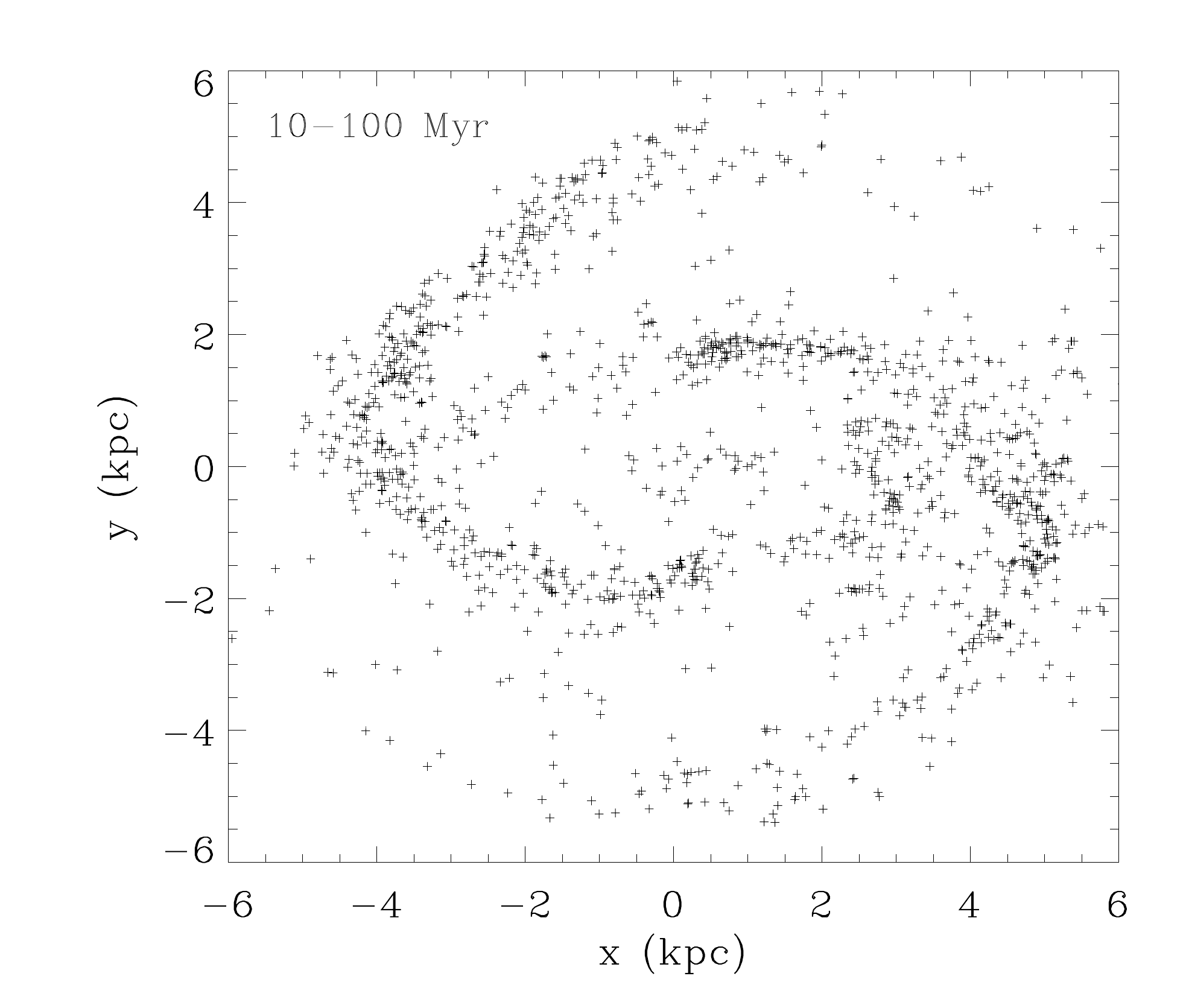}}
\centerline{\includegraphics[scale=0.3]{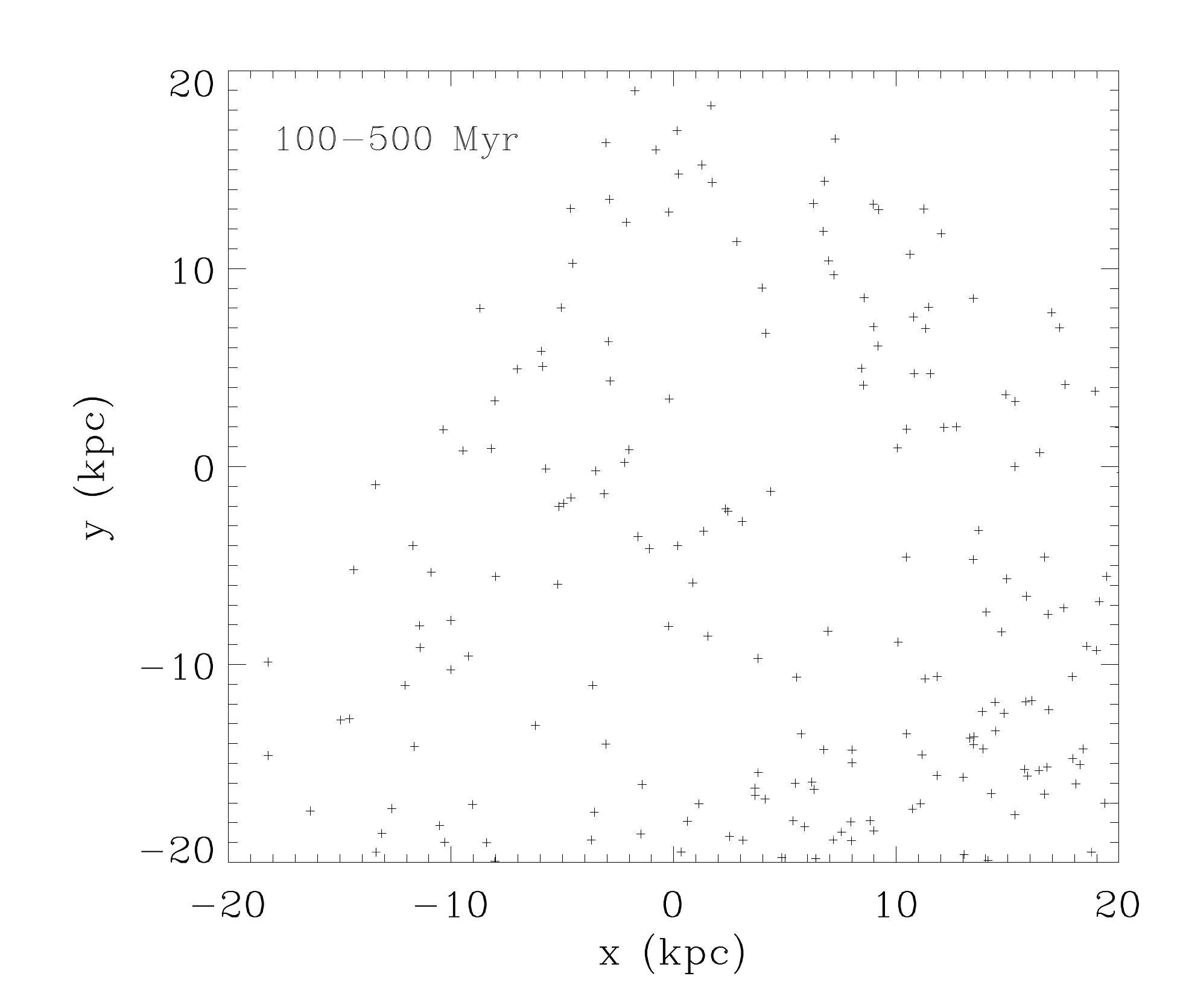}
\includegraphics[scale=0.3]{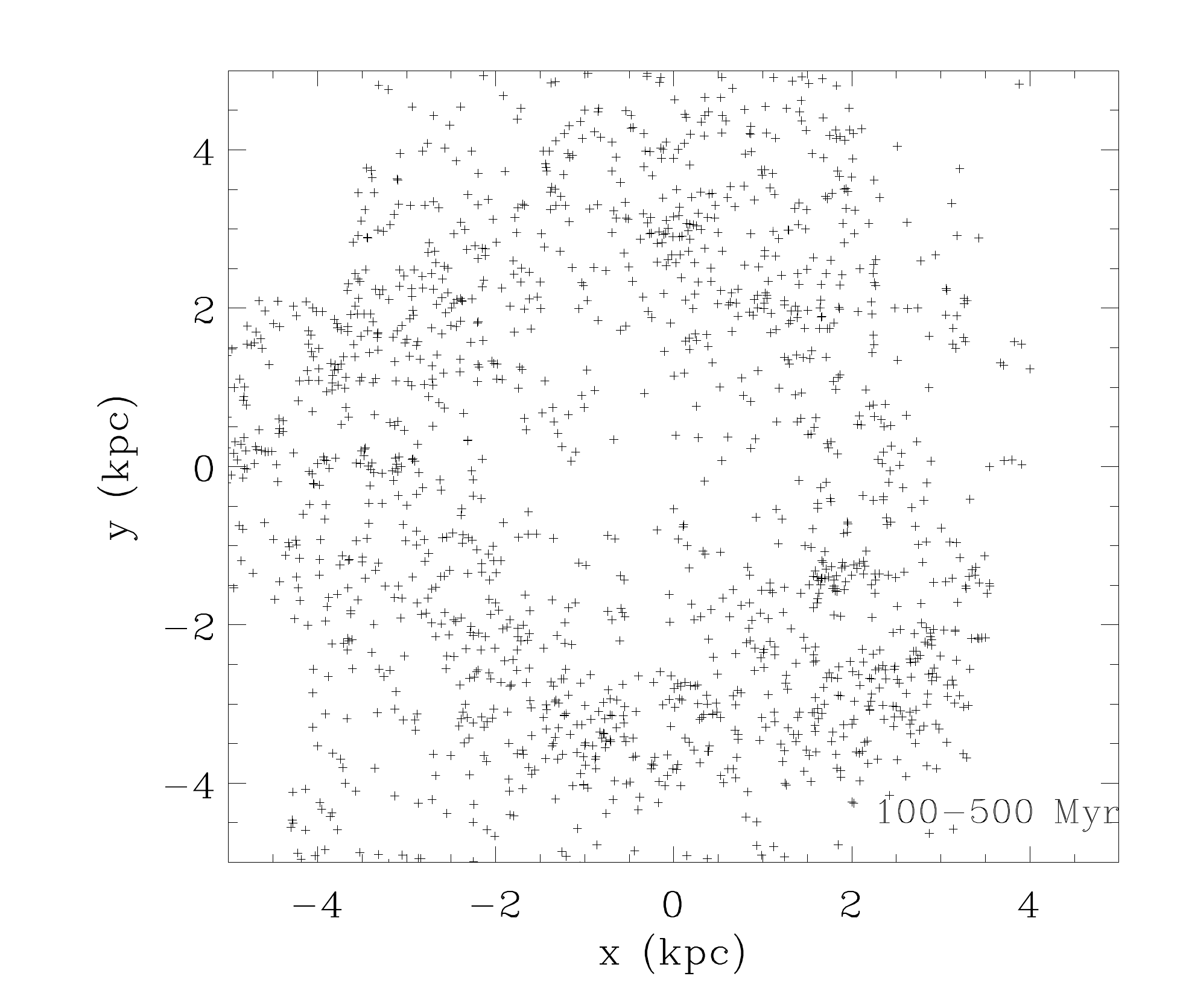}
\includegraphics[scale=0.3]{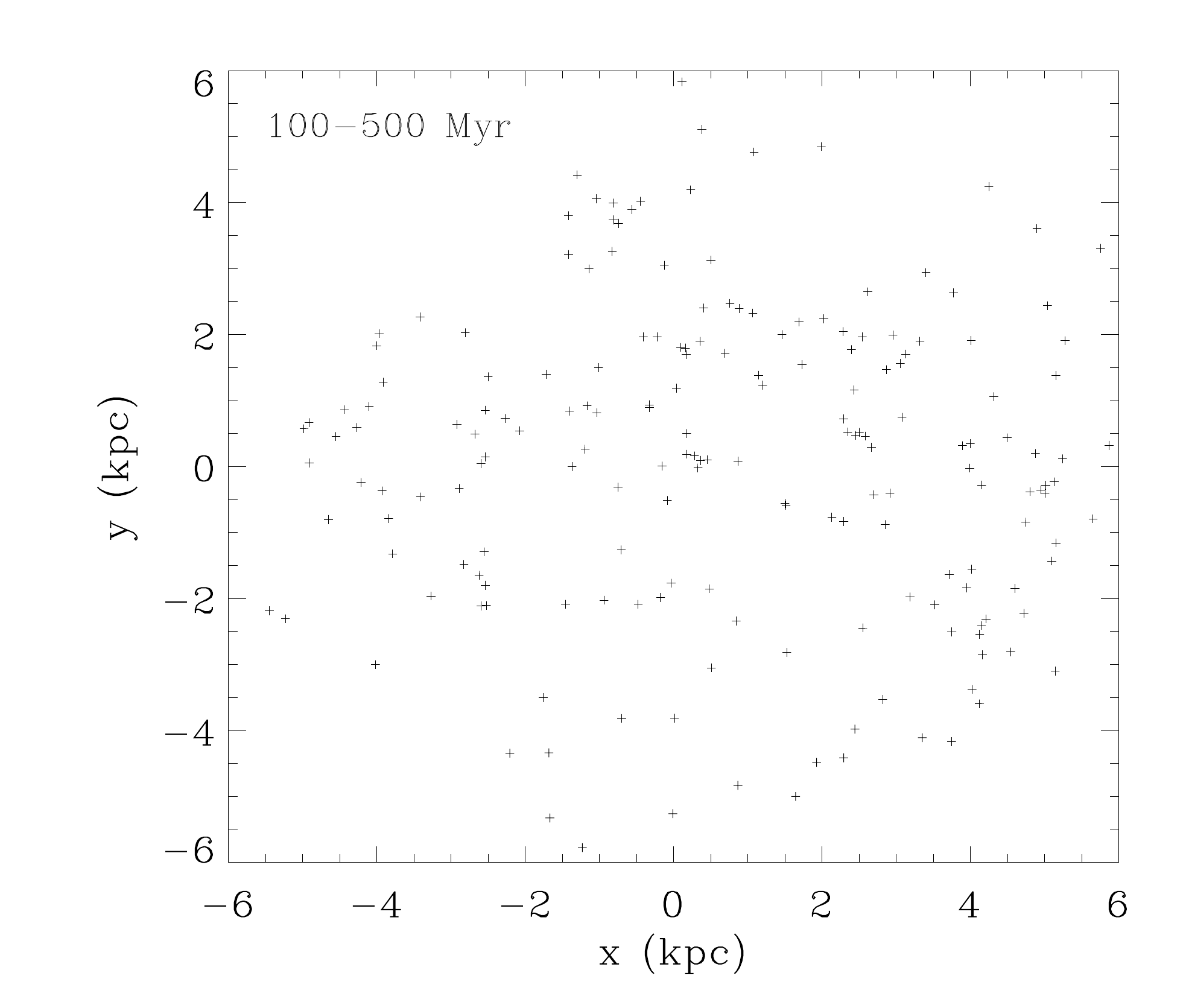}}
\caption{The spatial distribution is shown for stars of different ages for the galaxies NGC 628 (left), M83 (centre) and NGC 1566 (right). For M83, the cluster catalogue does not cover the very centre of the galaxy. NGC 1566 shows the clearest spiral structure. All galaxies show some spiral structure in the very young clusters ($<10$ Myr, top panels) and NGC 1566 also shows spiral structure in the 10-100 Myr age clusters (centre panel). There are very few older clusters though so it is not possible to determine whether any structure is present.}\label{fig:obsclusters}
\end{figure*}

We carried out similar analysis for the distribution of clusters in observed galaxies. We show the positions of clusters of different ages in Figure~\ref{fig:obsclusters} for the galaxies NGC 628, M83 and NGC 1566. The age ranges of the clusters are $<10$ Myr, 10-100 Myr and 100-500 Myr, but there are so few older clusters that the spatial distribution is largely independent of the maximum age (e.g. 200, 500 or 1000 Myr) in the third bin. This small number of older clusters is due to two competing effects. Firstly as clusters age, they have a higher chance to be disrupted. Secondly, as the stellar population of the clusters ages, the clusters become fainter and eventually fall below the detection limits of our datasets. Spiral structure is generally only clear in the very youngest clusters. The exception is the galaxy NGC 1566 which has a very clear spiral pattern, and still shows clear spiral arms in 10-100 Myr clusters. These results contrast with the distribution of star particles in both the simulated isolated galaxies and cosmological simulations, which still show spiral structure in the older ($>$ 100 Myr) stars. There is however greater similarity to the `clusters' in the isolated galaxy simulation, which do have the ability to disperse.
For the 100-500 Myr range in the observations, and the oldest `clusters' in the isolated galaxy simulations, there tend to be very few clusters, although there is a tendency for the clusters to have moved away from the spiral arms. 
Similar to the results from LEGUS, Chandar et al. 2016 in prep. find that spiral structure is not clear in older ($>100$ Myr) clusters in M51.

The Panchromatic Hubble Andromeda Treasury (PHAT) \citep{Dalcanton2012} survey provides high resolution maps of the star formation history in M31 \citep{LewisA2015}, with similar style maps to those presented here in their Figures~5 and 7. The structure of M31 is dominated by a ring, although given the inclination of M31, this structure could plausibly be a spiral arm \citep{Gordon2006}. The PHAT data show a clearest defined ring in the youngest stars. Even in the older stars though, this ring structure is still present. \citet{LewisA2015} argue that this is evidence that the ring is a stationary structure. We see similar behaviour for spiral arms in both our isolated galaxy and cosmological simulations though, and in the case of the cosmological simulations, the spiral arms are not steady. 

The differences between the cluster observations, PHAT survey and simulations may result from a number of factors that may determine whether clusters or stars adopt a spiral pattern. Firstly resolution, or the number of clusters may be an issue. Secondly, whether we are indeed using i) stars or simulated clusters which do not evolve, or ii) observed or simulated clusters which can evolve.
Galactic structure may also affect the cluster or stellar distribution. As we discussed in the last section, the cosmological simulations with weak spiral arms show little spiral structure in the stars, whilst the isolated galaxy simulations have quite strong spiral structure. In NGC 628, the galaxy with the weakest spiral structure, there is little structure in the distribution of intermediate age star clusters. \citet{Grasha2015} estimate that clustering of stars in clusters in NGC 628 decreases after around 40 Myr. Cluster dispersal timescales, and/or the length of time clusters spend in spiral arms, may be longer in strongly barred galaxies (such as NGC 1566) or galaxies with stronger arms (e.g. M83) and indeed in these two examples we see clearer structure in the intermediate age stars. \citet{Elmegreen2011} suggests that hierarchical clustering may be important and thus clusters formed in denser spiral arms may be more bound and long-lived. \citet{Dobbs2009} also predicted clearer patterns in galaxies with strong long-lived spiral arms, where the gas and stars tend to spend longer in the spiral arms.

\subsection{Mass distributions of clusters and comparison with observations}
We also investigate the mass distributions of `clusters' in the isolated galaxy simulations. We caution again that although we can group stars into clusters, they may not all necessarily be like real clusters. In particular they may be more like stellar associations rather than clusters of stars that have formed in the same molecular cloud at the same time (we investigate this specifically in Section 4.4). Thus comparing the absolute value of the slope of the mass distributions of the clusters to observations may not be particularly useful, although we can make quantitative comparisons. In this section, unlike the results in Figure~\ref{fig:isoclusters}, we do not make any attempt to remove clusters with large age spreads. Using just the clusters from Figure~\ref{fig:isoclusters}, i.e. with better defined ages, we obtain a mass distribution with a slightly steeper slope (and normalised to lower numbers of clusters) than those we present here.

\begin{figure}
\centerline{\includegraphics[scale=0.43]{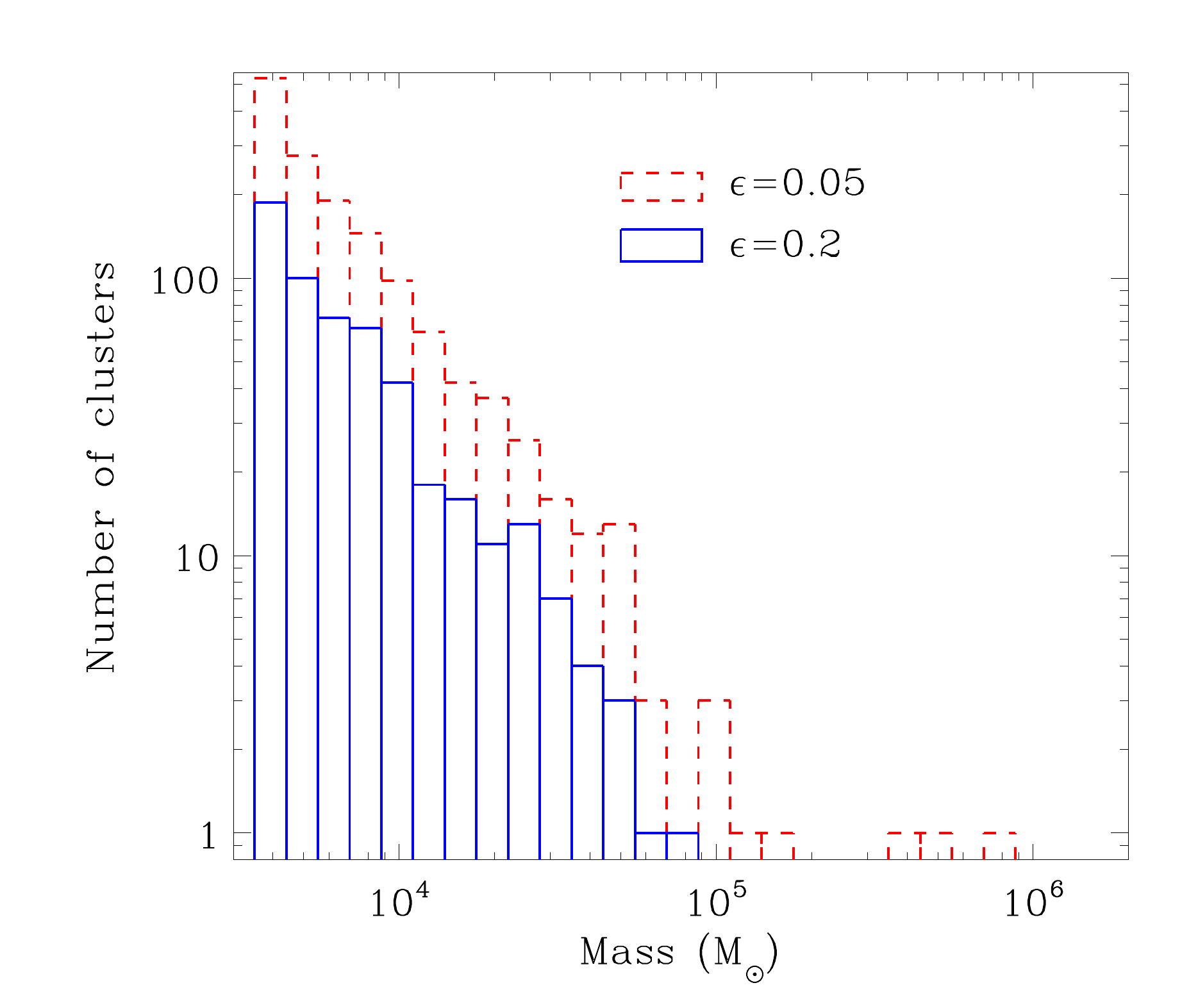}}
\centerline{\includegraphics[scale=0.43]{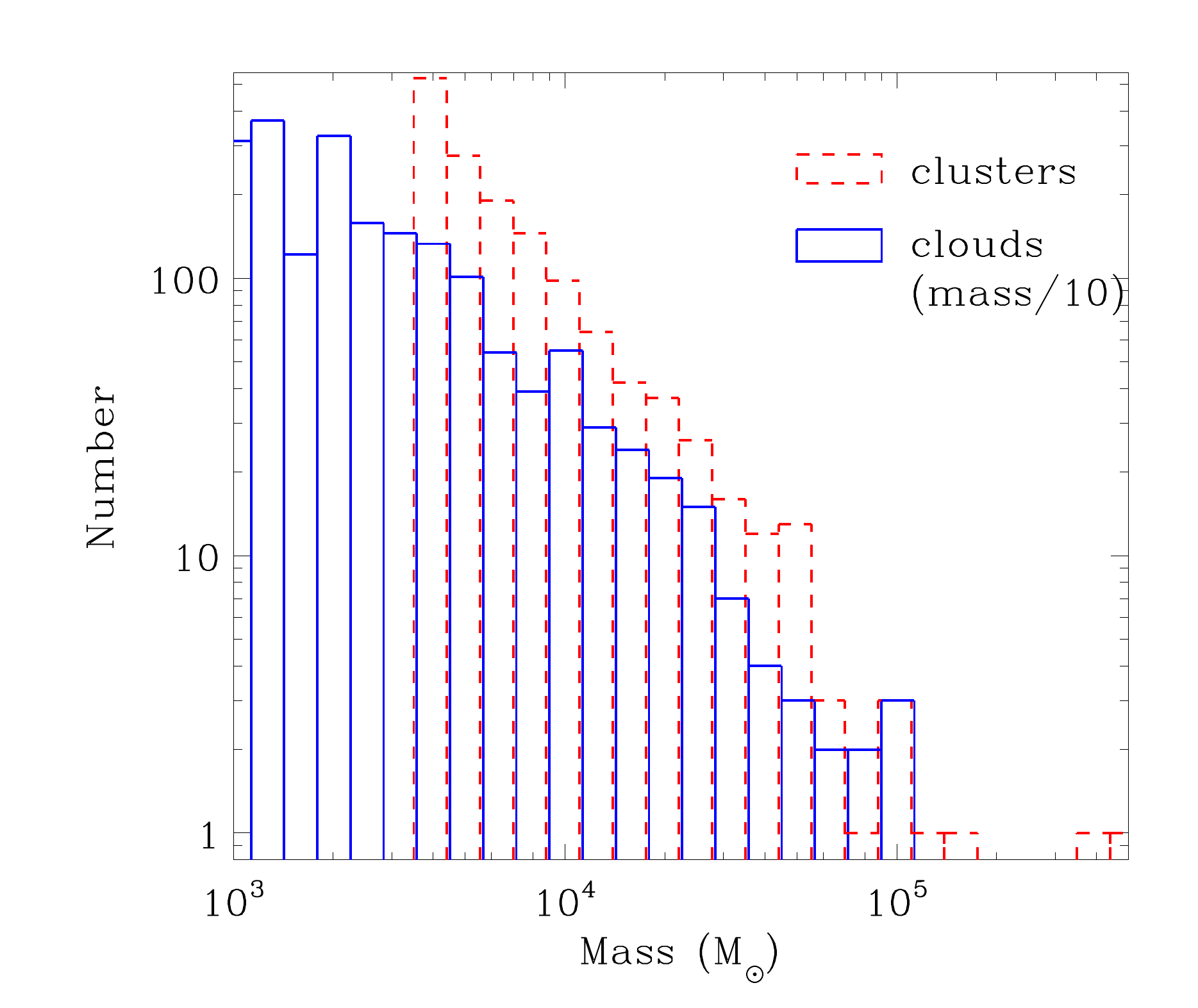}}
\caption{The mass distributions are shown for the clusters in the isolated galaxy simulations with different feedback (top panel). With lower feedback, the clusters extend to higher masses. The mass distributions for clouds and clusters for the low feedback model are shown in the lower panel (the distribution for the clouds has been rescaled to fit on the same axes). The distribution for the clusters is slightly steeper. The higher feedback simulation shows similar behaviour.}\label{fig:mass}
\end{figure}

We show mass distributions for the isolated galaxy simulations with different levels of feedback in the top panel of Figure~\ref{fig:mass}. The slope of the mass distribution appears similar regardless of the level of feedback, rather the masses of the clusters are simply shifted to lower masses with higher feedback. This is similar to the effect of feedback on cloud mass distributions as seen in \citet{Dobbs2011new}. The exponent of the distributions is around $-1.6$. This tends to be a little low compared to observed mass distributions, where the exponent is around $-2$ \citep{Adamo2015b, Chandar2014, Chandar2015}. Our findings of consistent distributions, but with different normalisations, resembles the results of \citet{Chandar2015}, who find different maximum cluster masses for different galaxies. They find a large star forming galaxy such as M83 or M51 has clusters of a few $10^5$ M$_{\odot}$ (similar to our low feedback model), whilst smaller galaxies have maximum cluster masses of only a few $10^4$ M$_{\odot}$. Although our results vary with feedback, rather than galaxy mass, in \citet{Dobbs2011new} we found that the same efficiency feedback was less effective in a more massive galaxy (in particular we showed an example with double the surface density). Although high mass clouds and clusters may generate more stellar feedback, they also tend to be more bound in the simulations and are thus less easily disrupted by  feedback.

In the lower panel of Figure~\ref{fig:mass}, we compare the mass distributions for clouds and clusters. The mass distribution for the clusters appears slightly steeper than that of the clouds. This difference was stable to changes to the friends of friends algorithm to locate the clouds and clusters. 
Steeper distributions are also found for observed stellar clusters compared to observed molecular clouds, as the exponent for molecular clouds tends to be $-1.5$ to $-2.3$ (e.g. \citealt{Solomon1987,Heyer2001,Roman2010,Gratier2012}).

\subsection{Evolution of star clusters in simulations - are the clusters real?}
In this section we study the evolution of `clusters' in the simulations. The resolution, and number of star particles, in the isolated galaxy simulations is sufficiently high that star particles are clearly grouped together into dense concentrations, which we pick out as `clusters' using our friends of friends algorithm. The resolution in the cosmological simulations is such that the mass of one particle is equivalent to a 2.3$\times10^4$ or 2.3$\times 10^5$ M$_{\odot}$ star cluster, and the number of star particles too low to produce strong concentrations. Thus from hereon we only consider the isolated galaxy simulations, predominantly the model with low feedback. Unlike observations, we can use the simulations to trace the time evolution of clusters, and the individual star particles within them. One question is whether, as typically supposed for clusters, the stars were all formed together, or whether they have come together over time, or whether most of the stars were formed together, but a few random stars are present in the cluster either due to chance or because they have been captured somehow. To study the evolution of star clusters, we first consider the backwards time evolution, i.e. we take a sample of clusters near the end of our simulation and examine how the star particles in those clusters originated. In Section 4.4.3 we instead study the forwards evolution of clusters, to determine the outcome for clusters selected at an intermediate point in our simulation. In Section 4.4.4 we compare the apparent evolution of clusters in the simulations with observations and theory.
\begin{figure}
\centerline{\includegraphics[scale=0.24, bb=50 0 500 500]{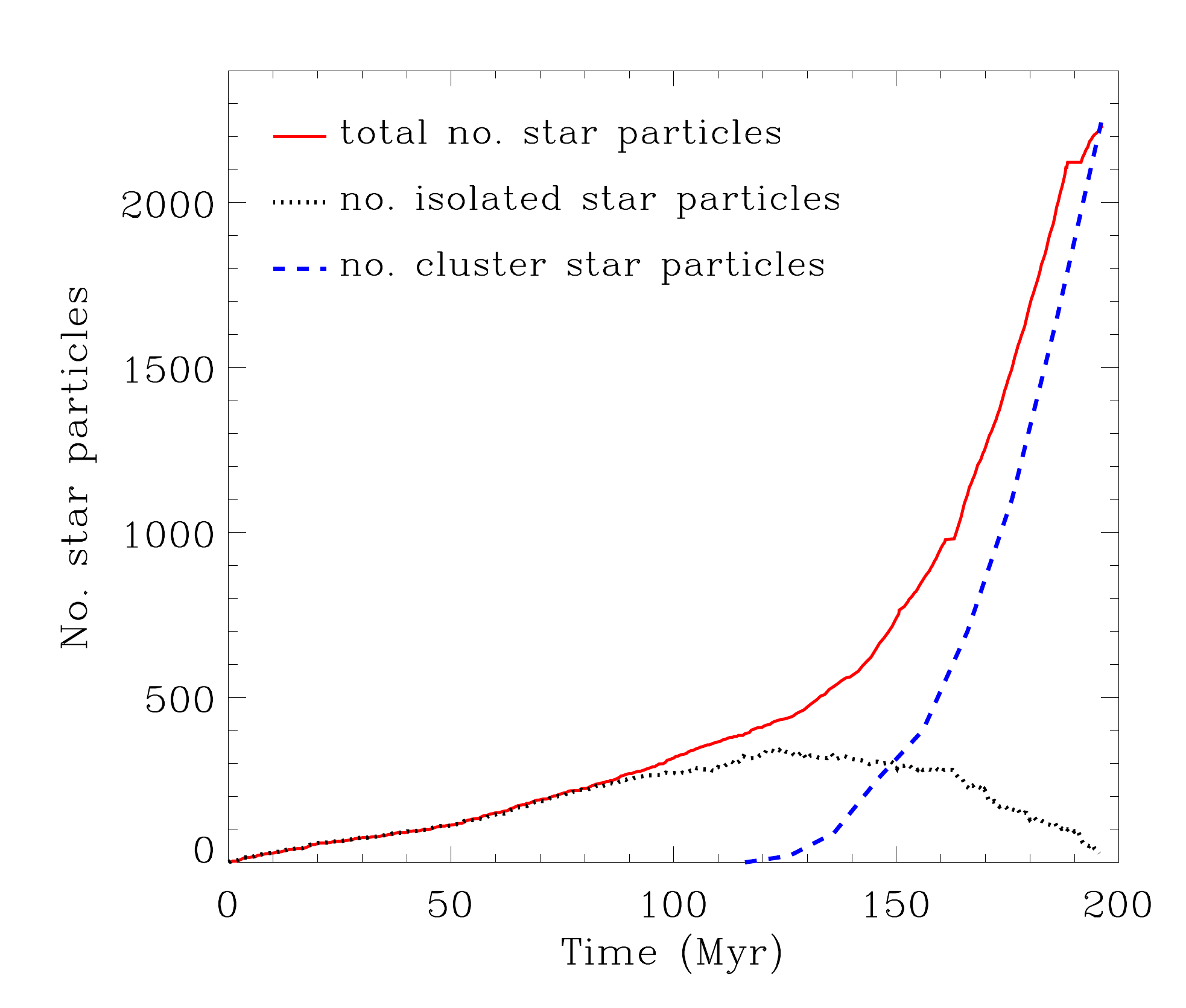}
\includegraphics[scale=0.24, bb=0 0 500 500]{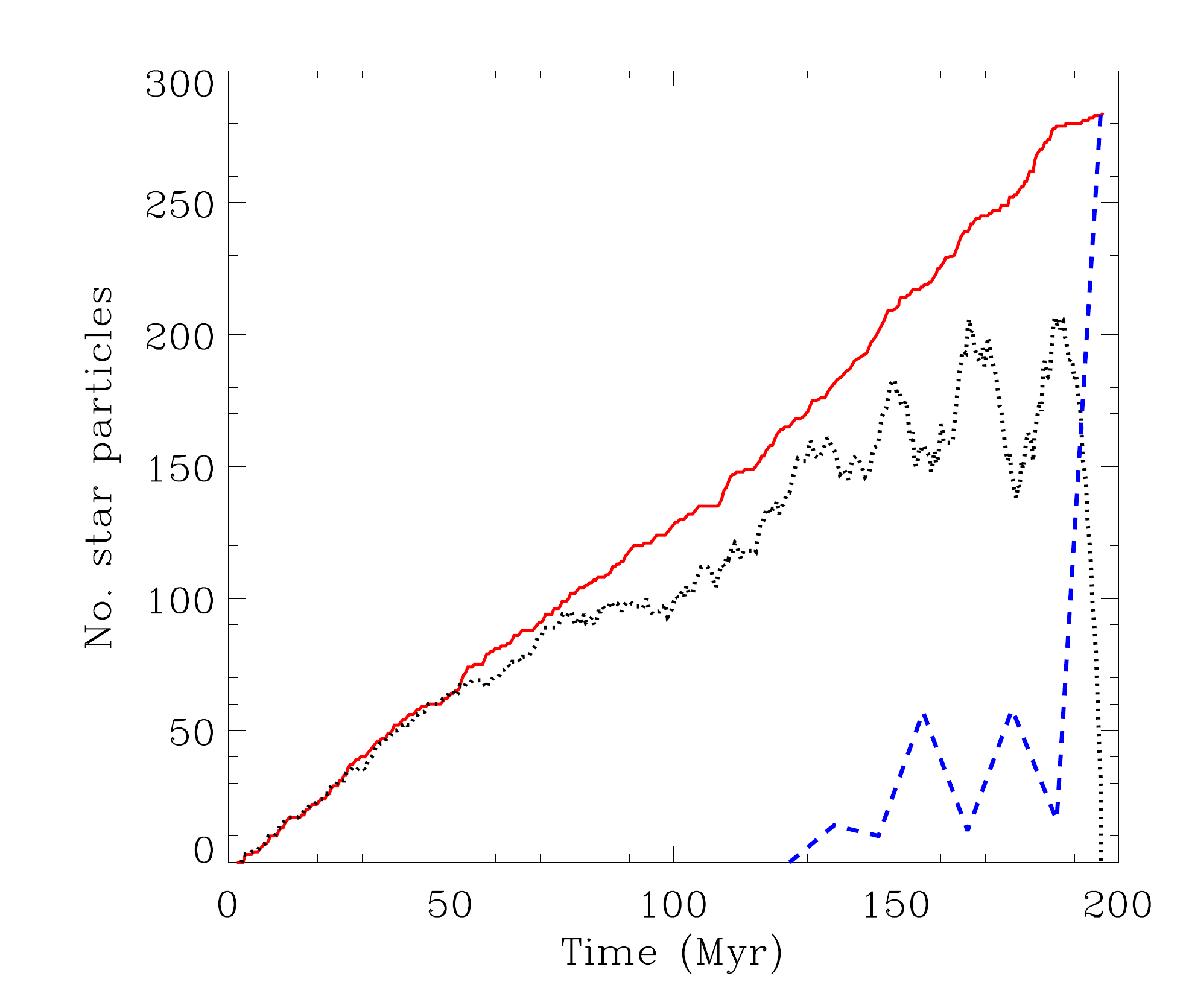}}
\caption{The total number of star particles, the number of isolated particles (separated by at least 20 pc from all other star particles), and the largest number of star particles in a cluster as determined by the clumpfinding algorithm are shown versus time for two example clusters. The clumpfinding algorithm is only applied in 10 Myr intervals rather than at each timestep. The cluster in the left hand panel is characterised by a large number of young stars which recently formed together rather than in isolation. These stars come from one main cluster, as indicated by the blue dashed line, three smaller clusters, and some stars not associated with a cluster.  The cluster in the right hand panel is characterised by gradual addition of star particles over time, from different locations, and thus does not resemble a `bona fide' cluster but instead would be just a stellar group. In this case, until 200 Myr, most star particles are not picked out as a cluster, but as isolated particles.}\label{fig:isoexample}
\end{figure}

\subsubsection{Backwards evolution of clusters}
\begin{figure}
\centerline{\includegraphics[scale=0.6, bb=0 0 350 350]{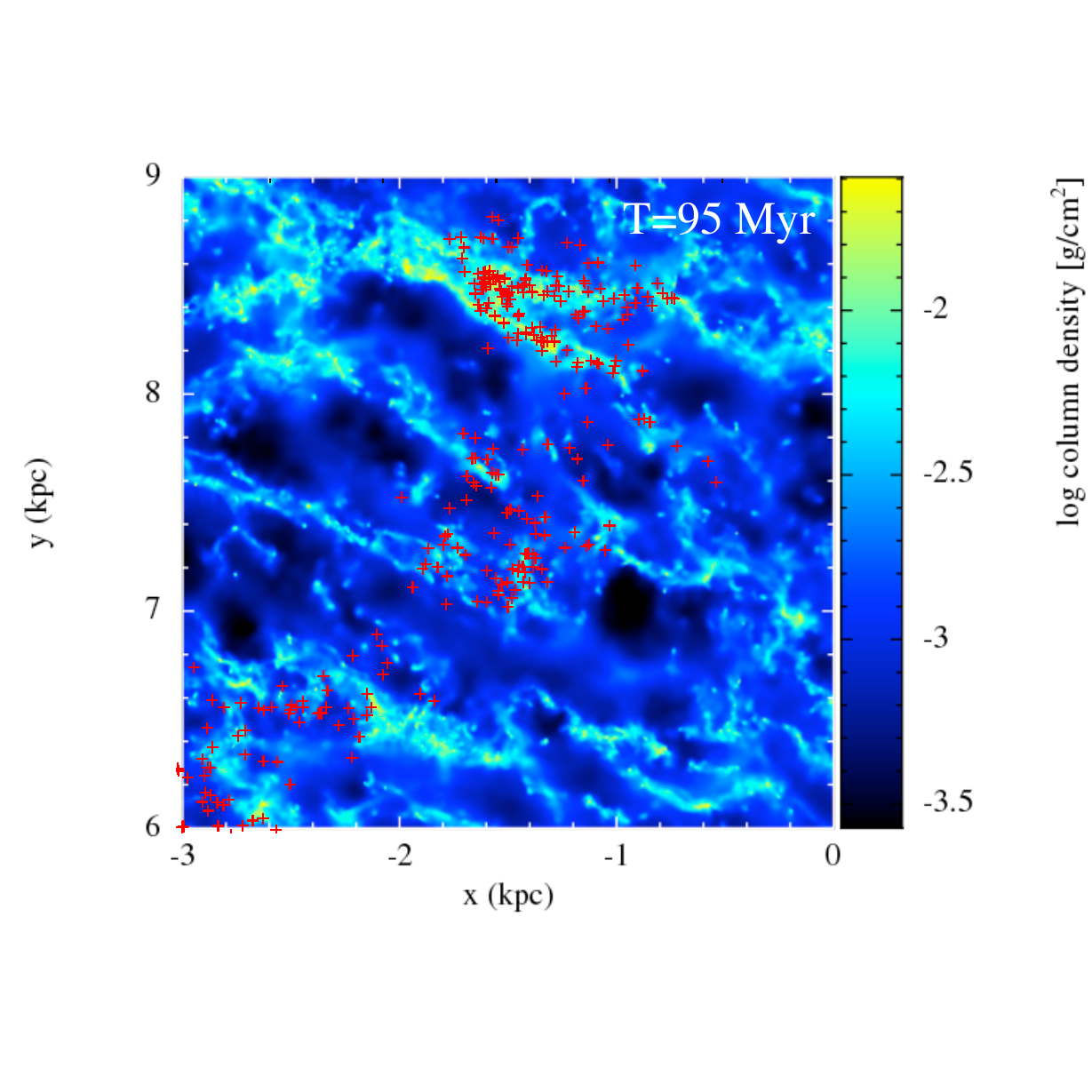}}
\centerline{\includegraphics[scale=0.47, bb=200 0 250 380]{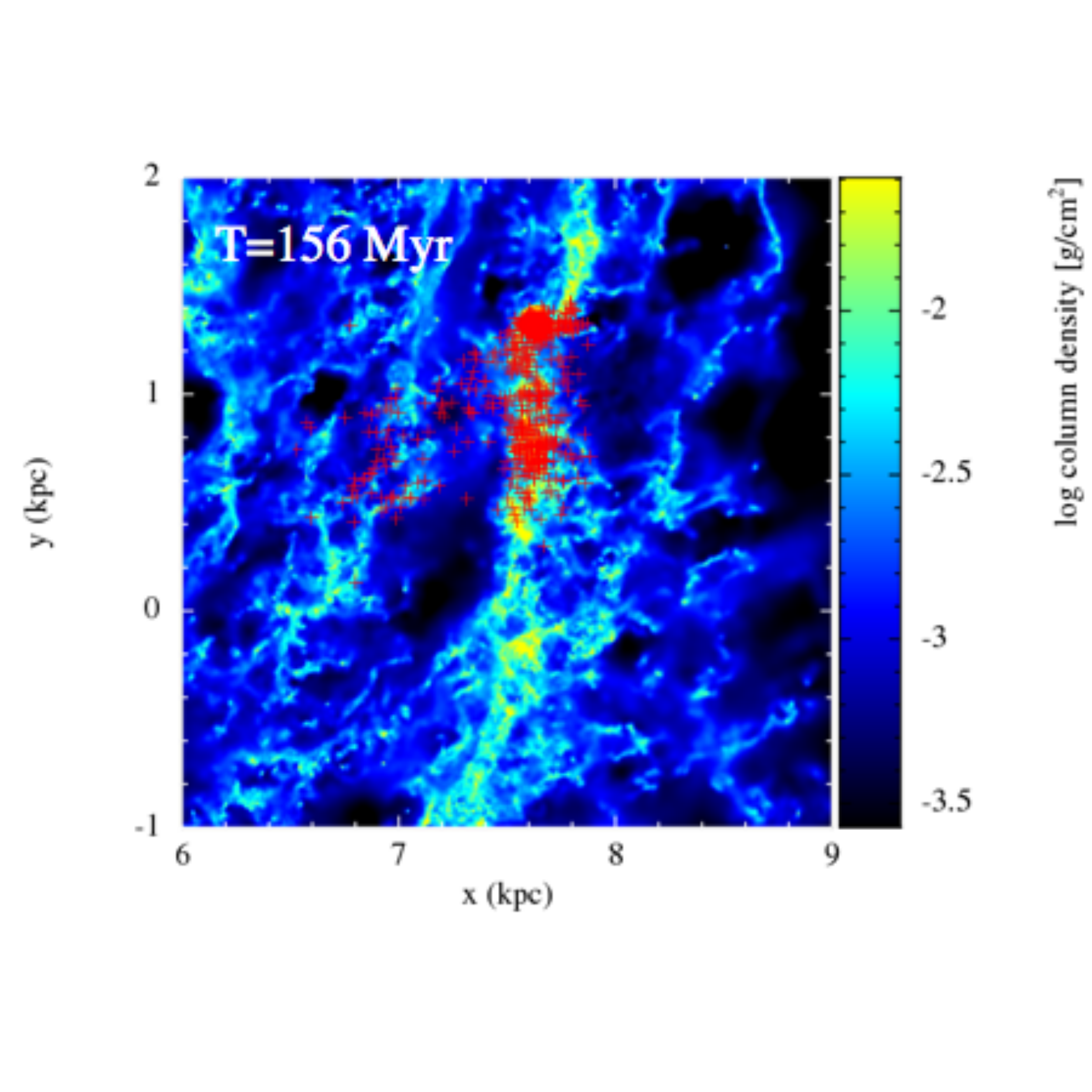}}
\centerline{\includegraphics[scale=0.6, bb=0 30 350 280]{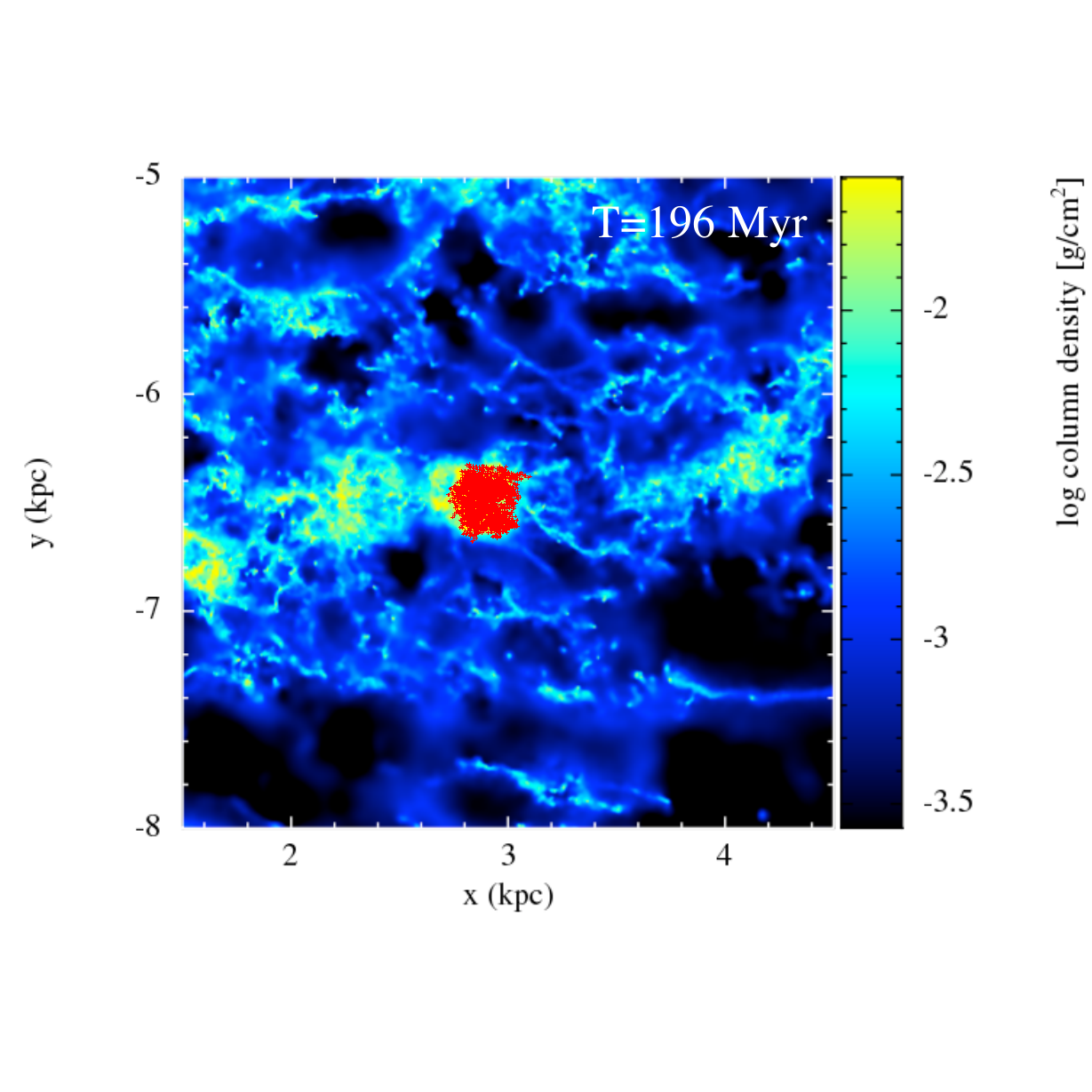}}
\caption{The evolution of an example cluster (the time the cluster is selected is in the lower panel) is shown. Star particles, shown in red, are over plotted on the column density. The evolution is shown over a 100 Myr time period. By 40 Myr before the cluster is selected (middle panel), there is a clear core of stars (associated with a dense cloud of gas), and a second group of stars in an adjacent cloud. Many of the stars of the cluster also form in the final 40 Myr of evolution, presumably also in these clouds of gas.}\label{fig:example1}
\end{figure}

To study how the clusters have originated, we look at when the star particles in a cluster were formed, and if they have been spatially coincident with the other stars in the cluster. We need to study the backwards evolution here since the forwards evolution assumes clusters already exist, and would not pick out cases where isolated star particles are accreted by a star cluster. We show two contrasting examples of cluster evolution in Figures~\ref{fig:isoexample}, \ref{fig:example1} and 8 from the low feedback model. We have preferentially selected larger clusters (giving better resolution), our examples containing 2243 and 280 particles (7$\times10^5$ M$_{\odot}$ and 8.9$\times10^4$ M$_{\odot}$). In Figure~\ref{fig:isoexample}, we show the total number of star particles versus time, and the number of isolated star particles. The isolated star particles are considered to be all those that are at least 20 pc from any other star particles in the cluster (so effectively these star particles would not be selected if applying our friends of friends algorithm at those times). This does not provide a complete indication of the number of star particles formed which are not part of the main cluster, as if multiple concentrations of stars form in different places, these are not picked up as isolated star particles. But there is nevertheless an indication of the number of random star particles that are forming part of the cluster. Changing this distance produces the same trends, the number of isolated particles is simply higher or lower.

We can see from Figure~\ref{fig:isoexample} that the two example clusters behave quite differently. For the first example (left), there is a rapid increase in the number of stars formed in the last 50 Myr or so. For this cluster, half the cluster members have been formed in the last 40 Myr. At early times, many (in fact all or nearly all) of the star particles form in locations quite separate from each other. However over time the number of isolated particles decreases, suggesting that the majority of star particles, which are formed at later times, are truly formed within the cluster. This picture is supported by Figure~\ref{fig:example1}, which shows the spatial distribution of star particles at different times. At the earliest time (96 Myr, top), the star particles are scattered randomly and not in a cluster. By 156 Myr (centre panel) a much clearer, concentrated cluster has developed (at $x=7.6$ kpc, $y=1.3$ kpc) where most of the star particles reside. Other star particles are still scattered away from the main cluster though. By the final time (196 Myr, lower panel) the star particles are clearly all together in the cluster, albeit this is not surprising as this is when the friends of friends algorithm is applied.

Star particles which are not initially associated with the cluster as observed at 196 Myr tend to start out at disparate distances from other stars, and then evolve with very similar coordinates for the last so many Myrs of their evolution round the galaxy. Hence in Figure~\ref{fig:isoexample} (left), the number of isolated star particles only declines at later times. This suggests that these star particles become bound to the main cluster, rather than randomly passing through. This is because this cluster, and its associated GMC (the stars are still coincident with a large amount of gas), are strongly bound, and the velocities of the star particles are small enough to become bound. It is also worth noting that additional star particles are not necessarily random, but often may be associated with molecular clouds which are merging with the GMC (or cloud) containing the main cluster. In Figure~\ref{fig:example1} middle panel, we see that  there are quite a number of star particles associated with a dense region of gas slightly below the main cluster (at $x=7.6$ kpc, $y=0.7$ kpc). Analysis using our clumpfinding algorithm (Figure~\ref{fig:isoexample}) finds that four clusters merge together over 100 Myr, albeit the main cluster is substantially more massive than the others. As shown in \citet{Dobbs2015}, cloud-cloud collisions are quite frequent, thus we may expect multiple populations to reside in longer lived clouds, from where clouds have collided. 

\begin{figure}
\centerline{\includegraphics[scale=0.6, bb=0 0 350 350]{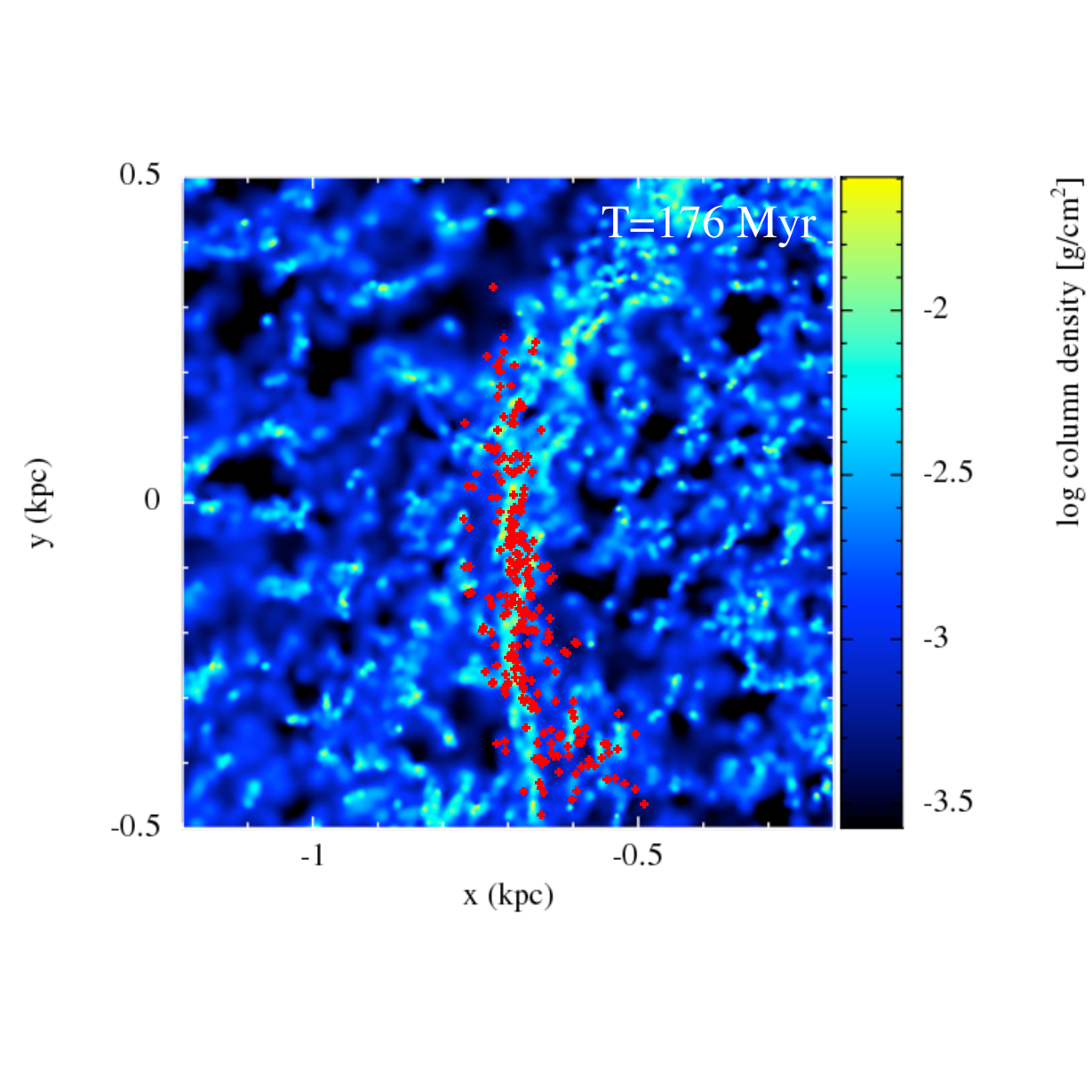}}
\centerline{\includegraphics[scale=0.6, bb=0 30 350 280]{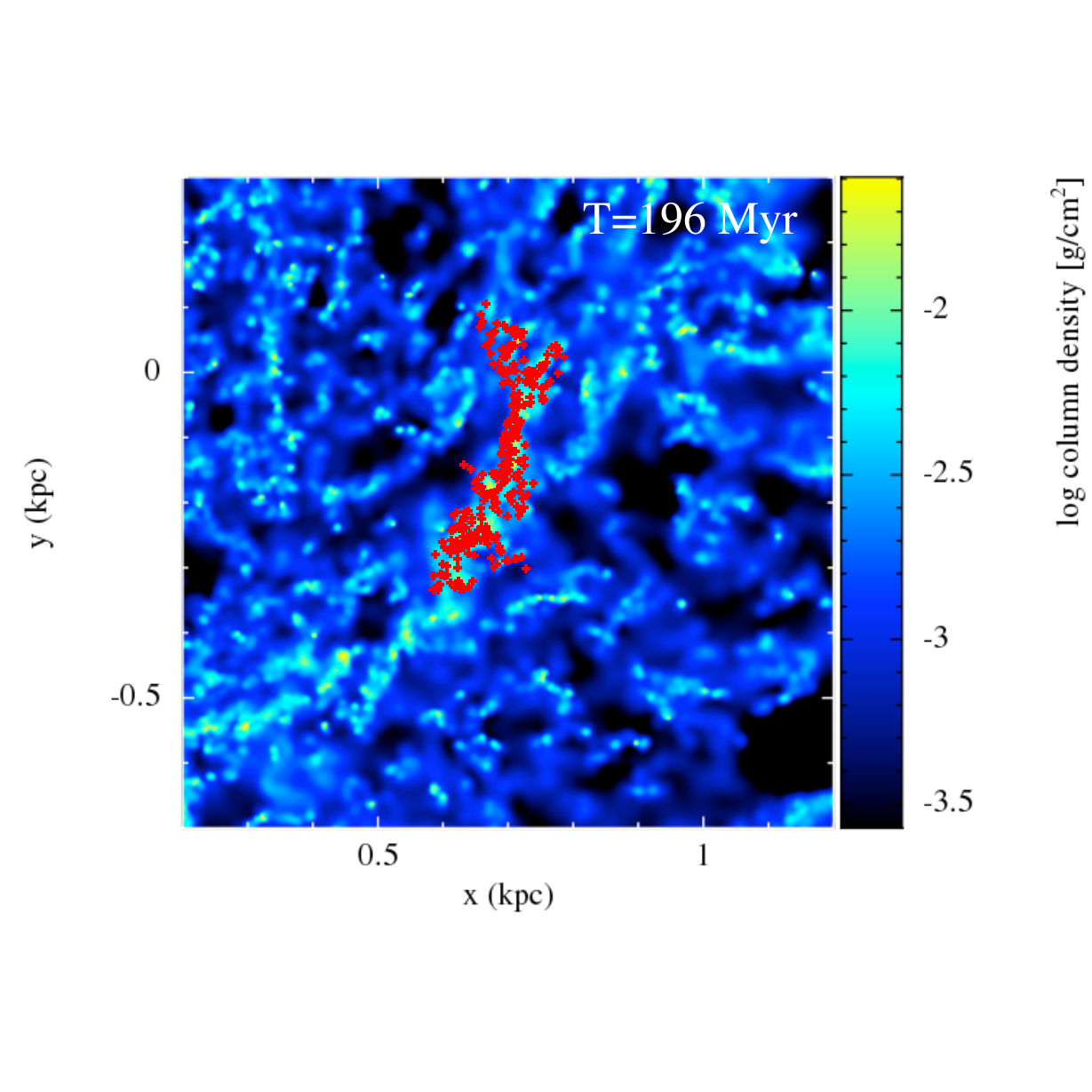}}
\caption{The evolution of an example cluster (the time the cluster is selected is in lower panel) is shown. Star particles, shown in red, are over plotted on the column density, over 20 Myr of time evolution. In this example, the `cluster' originates from disparate star particles (connected neither in time or space) so this is not a bona fide cluster as termed by observers.}\label{fig:example2}
\end{figure}

Our second example shows somewhat different behaviour. Here there appears to be a continuous addition of stars to the cluster over time (Figure~\ref{fig:isoexample}, right). Two stages of the evolution of this cluster are shown in Figure~\ref{fig:example2}, at 176 Myr and 196 Myr (when the cluster is selected). Unlike our first example, where there is a clear single cloud of gas associated with the main cluster, the star particles are much more spread out, and are not associated with any particular large clumps of gas. This particular cluster is located close to the centre of the galaxy, where because of the strong tidal forces, large GMCs do not form - typically large clusters away from the centre are more like that shown in example one. Clusters such as that in our second example are also much less concentrated than real clusters (although given our resolution it is difficult to obtain very concentrated clusters except for those which are more massive). Our second example is then not truly a cluster in the sense that the star particles have formed at different times and simply come together by chance. The assembly of the star particles, from quite elongated regions, into a more concentrated area, is also easier towards the centre of the galaxy where dynamical times, and orbital distances are quite short.

Overall, about half of all the clusters are more like our first example, when stars have preferentially formed around a particular time (these clusters correspond to those shown in Figure~\ref{fig:isoclusters}). The others show quite large age spreads (these clusters are absent from Figure~~\ref{fig:isoclusters}). Clusters with small age spreads tend to contain younger stars, indicating that either these clusters disperse on short timescales (except for those that are very massive), or they become contaminated with other stars. In Section 4.4.3 we consider the forwards evolution of clusters where we can determine how clusters evolve.

\subsubsection{Relation of star clusters to gas}
As typified by our examples in Figures~\ref{fig:example1} and \ref{fig:example2}, clusters in the simulations tend to be associated with gas. This suggests that either our clusters are typically young, or we are missing a population of clusters which have dispersed from the gas, but have not yet dispersed to field stars. 

In our low feedback model, we see a number of massive clusters that are still associated with gas, that have seen ongoing star formation for 40 Myr or more, as typified by our Example 1. Clusters can disassociate from gas simply by the change from gas mass to stellar mass, and the injection of feedback to expel remaining gas. In our Example 1, about 5 \% of the mass is in stars, so there is not yet a large fraction of the mass in stellar form, which may not yet be enough to see the expulsion of gas from the cluster. There is some indication of the star formation rate slowing, but only at the very end of the simulation. Unlike previous simulations of isolated clusters (e.g. \citealt{Kroupa2001,Boily2003,Proszkow2009,Moeckel2010}), transformation of the total mass of the GMC from gas into stars is hampered by accretion of gas onto the GMC at the same time (see also \citealt{Goldbaum2011,Zamora2012}). In this example, stellar feedback appears also too weak to lead to gas expulsion from the GMC. However for those cases where stellar feedback does disperse the cloud, the remnant stellar clusters do not appear to be strongly bound to stay together. This could well be a consequence of resolution, and / or limitations with our feedback scheme. Firstly we will underestimate the relaxation time of the clusters. Secondly, as an individual cloud disperses, it may contain multiple bound star clusters which diverge. In our simulations though, if this occurs, then each separate cluster would contain only a small number of star particles. Thirdly it is difficult to form strongly bound clusters in the first instance.
\begin{figure}
\centerline{\includegraphics[scale=0.48, bb=30 0 500 500]{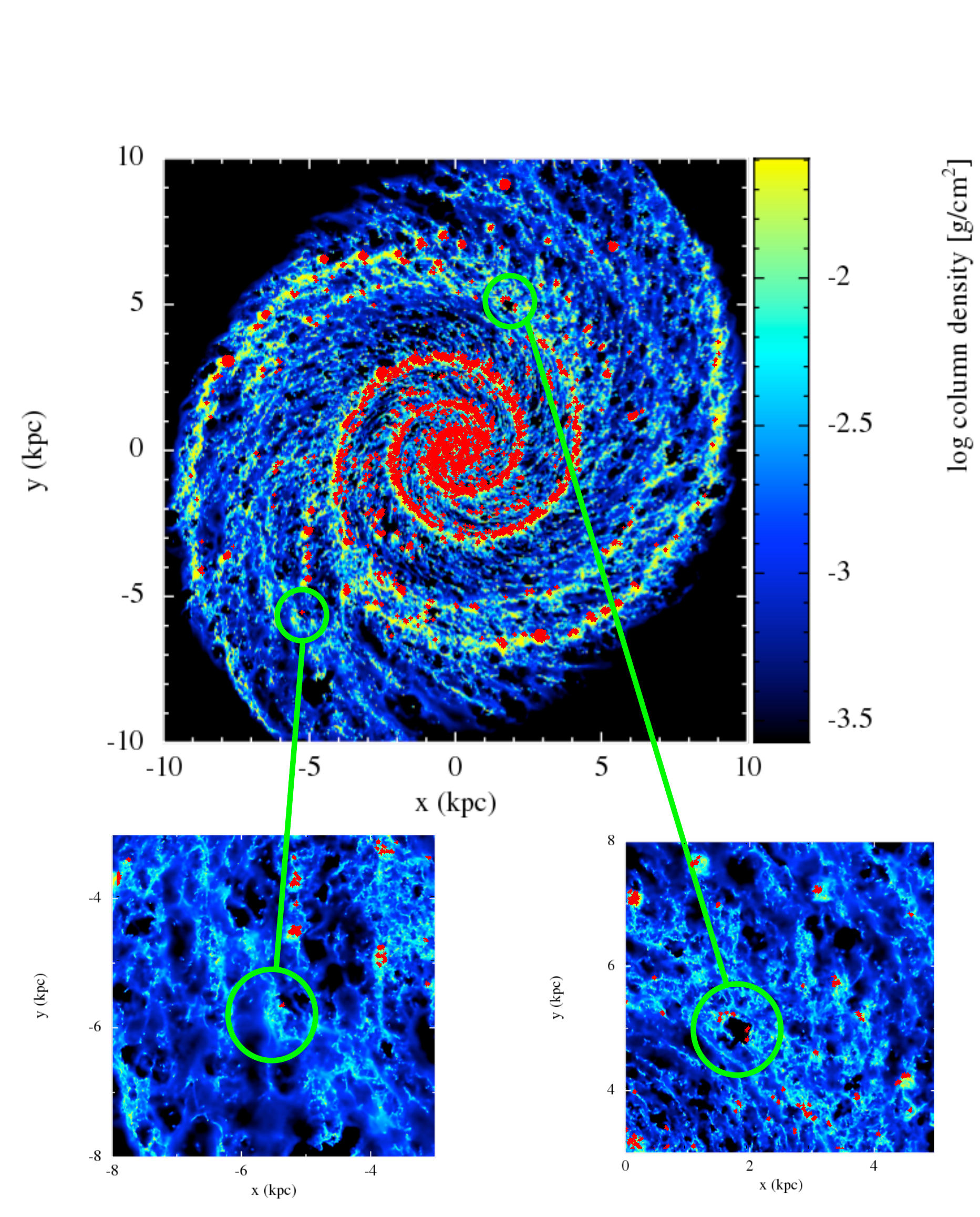}}
\caption{The column density of the gas for the low feedback model is shown, with the stellar clusters over plotted in red. The vast majority of clusters are associated with dense gas. The green circles highlight some clusters of particular interest which are discussed in the text.}\label{fig:gasandstars}
\end{figure} 

In Figure~\ref{fig:gasandstars} we show clusters overplotted on the gas column density for our low feedback simulation. The great majority of clusters appear coincident with dense gas, although given that the densities will tend to be high in the spiral arms, it is not necessarily clear that the clusters correspond to peaks along the spiral arms, at least for the inner, busier part of the galaxy. Even in the inter arm regions, the clusters tend to be associated with dense gas, although this is perhaps similar to clusters associated with spurs seen in M51. For the high feedback model (not shown), similarly most clusters are associated with dense gas.

We do however find a few clusters that are not coincident with dense gas, although they may tend to be close to dense clouds. We highlight a couple of interesting clusters, or groups of clusters in green circles in Figure~\ref{fig:gasandstars}. In the lower left we highlight a cluster that is not coincident with a dense gas cloud. This cluster was in a cloud about 40 Myr ago, and has remained intact since then. It is next to a filamentary dense feature, although the cluster originated from a different cloud of gas that has since dispersed. This cluster is the end cloud of a chain of clusters in a spur like feature, so perhaps only this one has been in existence long enough to dissociate from the gas. Other inter arm clusters (e.g. in the top part of the figure) have no indication of disassociating from their associated gas clouds. The other green circle, in the upper right quadrant of Figure~\ref{fig:gasandstars} highlights four star clusters. These clusters were chosen because they appear to border a hole, or low density region, in the disc. Hence although these clusters are now spatially distinct, this suggests they could have been initially part of the same cloud, and have been dispersed by a (or multiple) feedback event(s). We traced these clusters back over time, and find indeed that 40 Myr ago, they were all part of the same cloud, that has now dispersed. In all these examples though the clusters only contain a very small number of particles (10--20). This suggests that only the smaller clusters in our simulations are seen to dissociate from the gas, and thus we may have trouble picking them out, or they may disperse because we do not model them with enough star particles rather than the cluster is not bound.   

To conclude, it is clear that we see larger clusters, but these tend to still be associated with dense gas. Likely if we were able to run this simulation longer, the clouds would eventually reach a stage where accretion is minimal, less gas remains, and it is expelled from the cluster. However we would probably expect such clusters to have quite a large spread in ages (our largest clusters already have a $\sim40$ Myr spread in ages). In those cases where the gas is more readily dispersed, the clusters are often quite small. Similarly in our higher feedback model, most of the clusters are quite small, and they probably simply go below our cluster-finding criteria in the time taken for gas to dissociate. Massive clusters without gas, and small age spreads, tend not to be seen in our models, so if they are common in observations then that would be something that these simulations do not readily account for. 

\subsubsection{Forward evolution of clusters}
\begin{figure}
\centerline{\includegraphics[scale=0.36, bb=50 120 550 650]{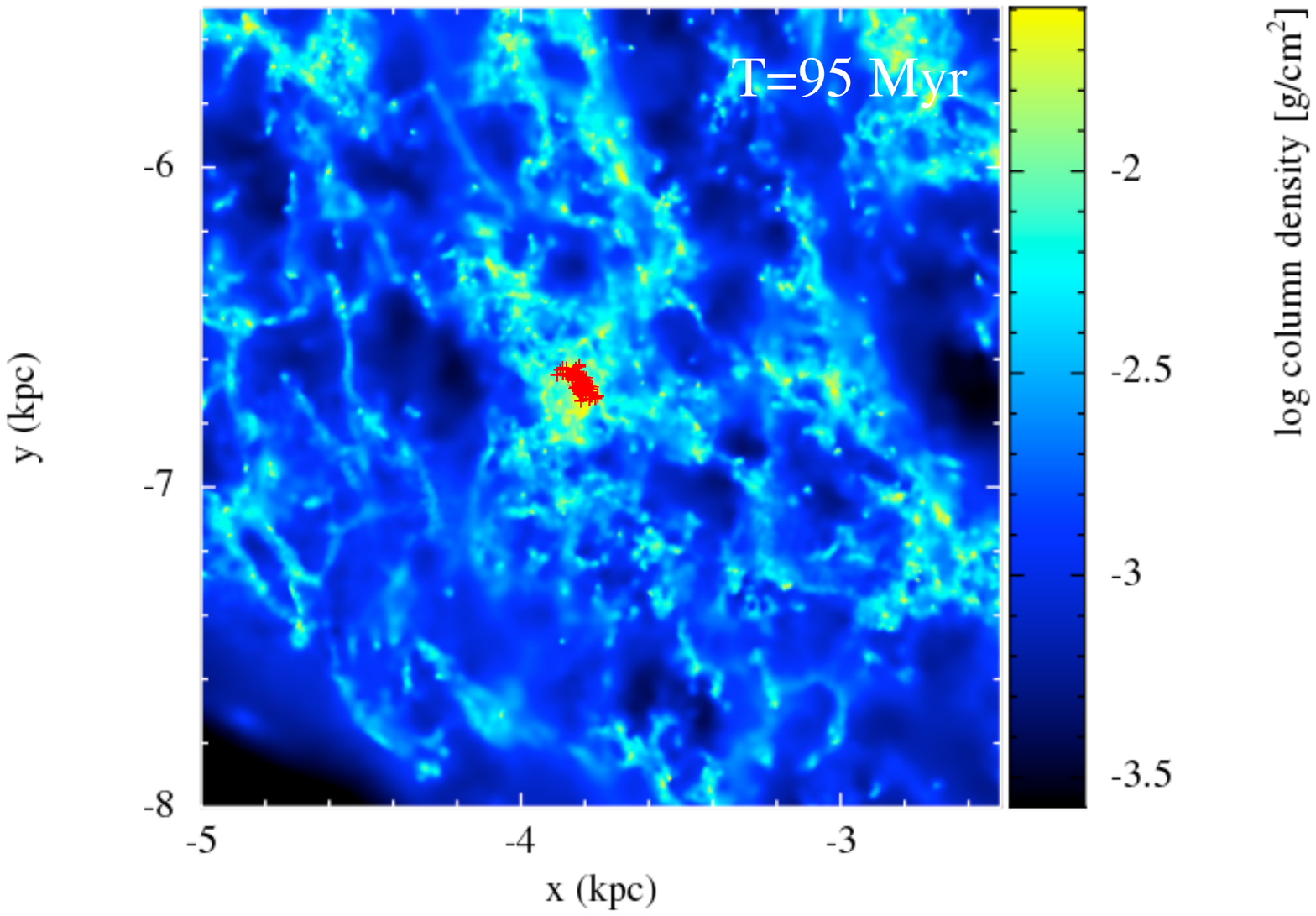}}
\centerline{\includegraphics[scale=0.6, bb=0 30 350 280]{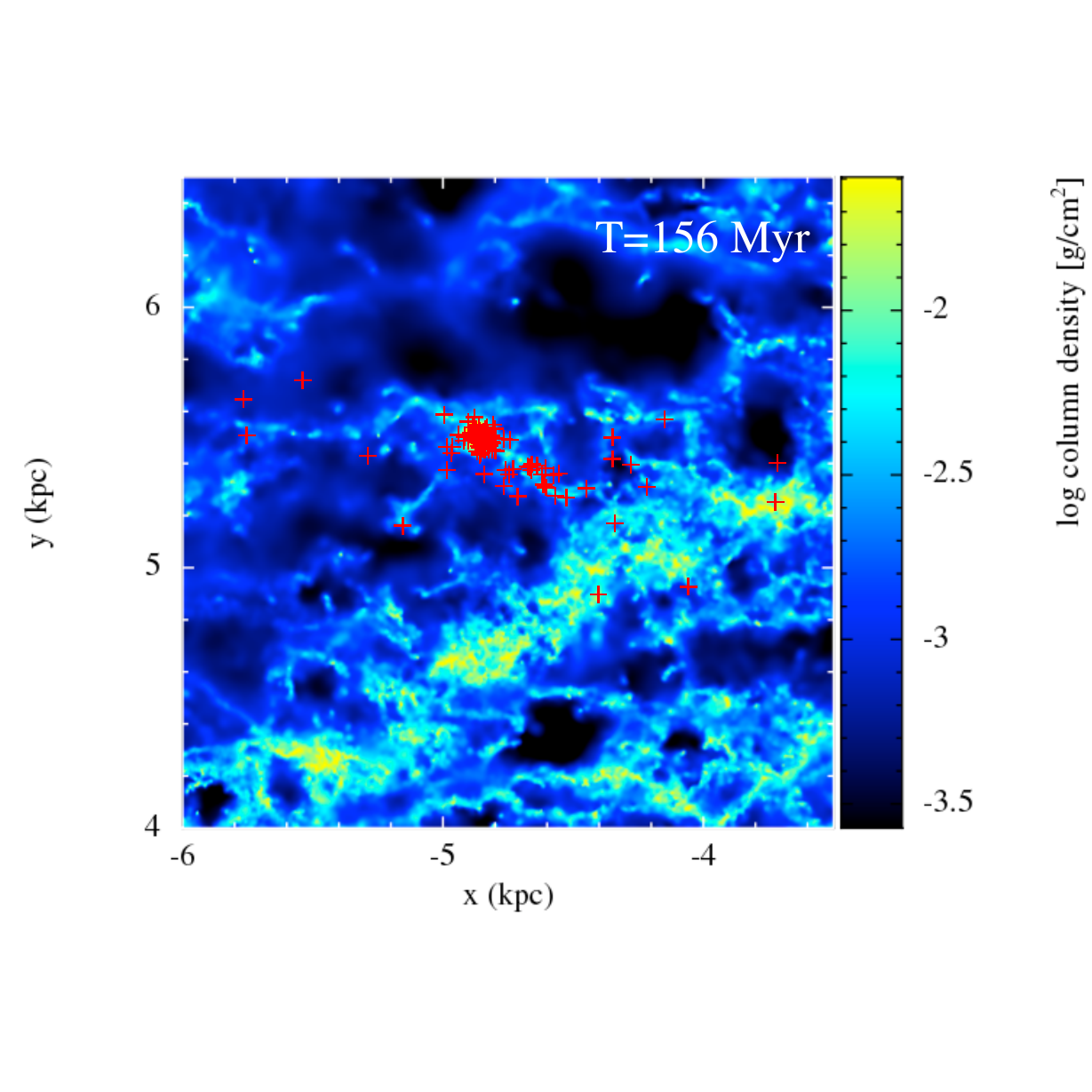}}
\caption{The evolution of a cluster (the time the cluster is selected is in the top panel) is shown, which initially lies in a spiral arm. Star particles (red) are over plotted on the gas column density. The time in the lower panel shows the evolution of the cluster after 60 Myr. After 60 Myr, there is still a dense cluster present (\textbf{at $x\sim -4.9$ kpc, $y\sim5.5$  kpc}), which is the main cluster picked out by the clumpfinding algorithm at this time. The cluster now lies in an inter-arm spur. Other star particles, constituting about half the mass of the original cluster, have dispersed.}\label{fig:forward1}
\end{figure}

\begin{figure}
\centerline{\includegraphics[scale=0.58, bb=0 0 350 350]{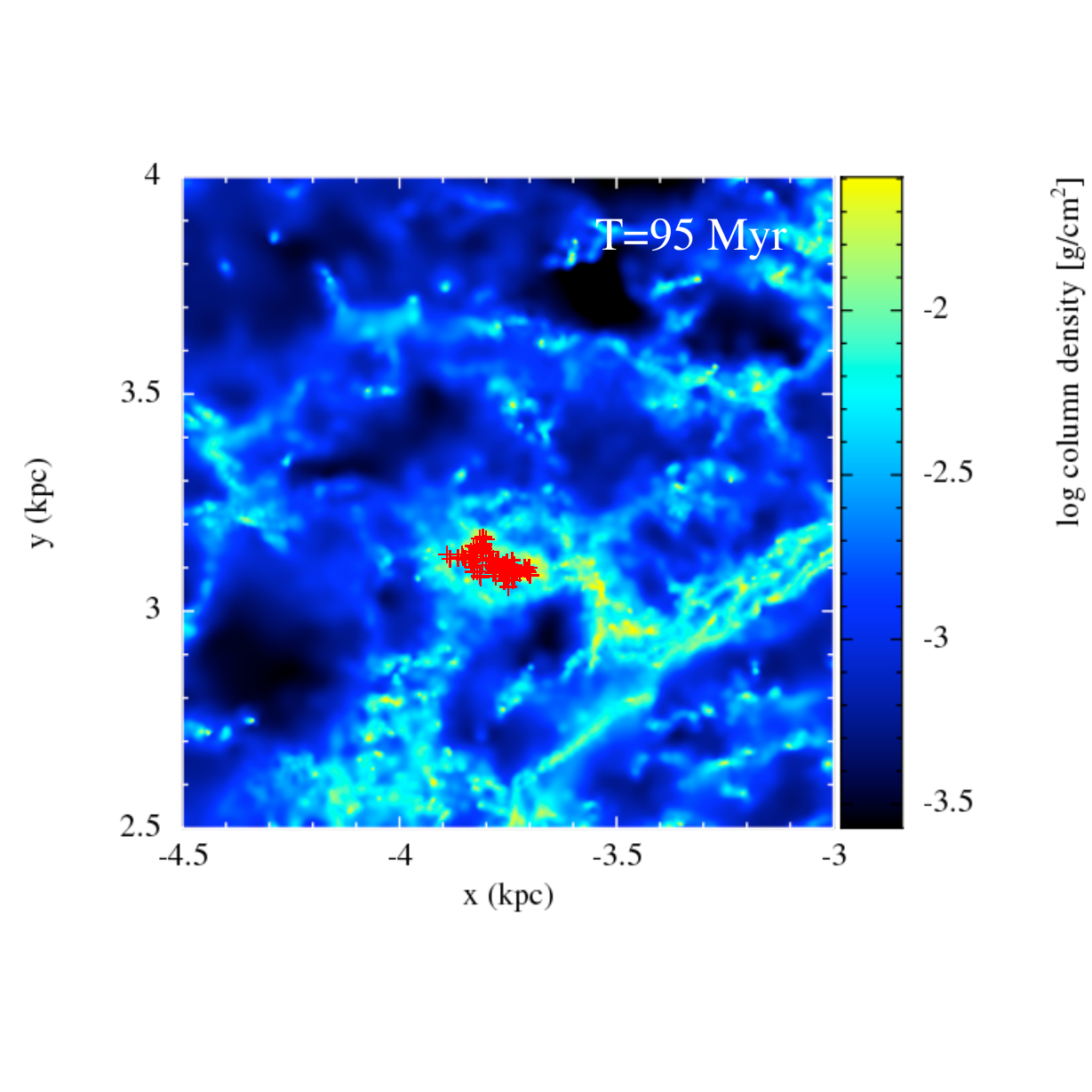}}
\centerline{\includegraphics[scale=0.6, bb=0 30 350 280]{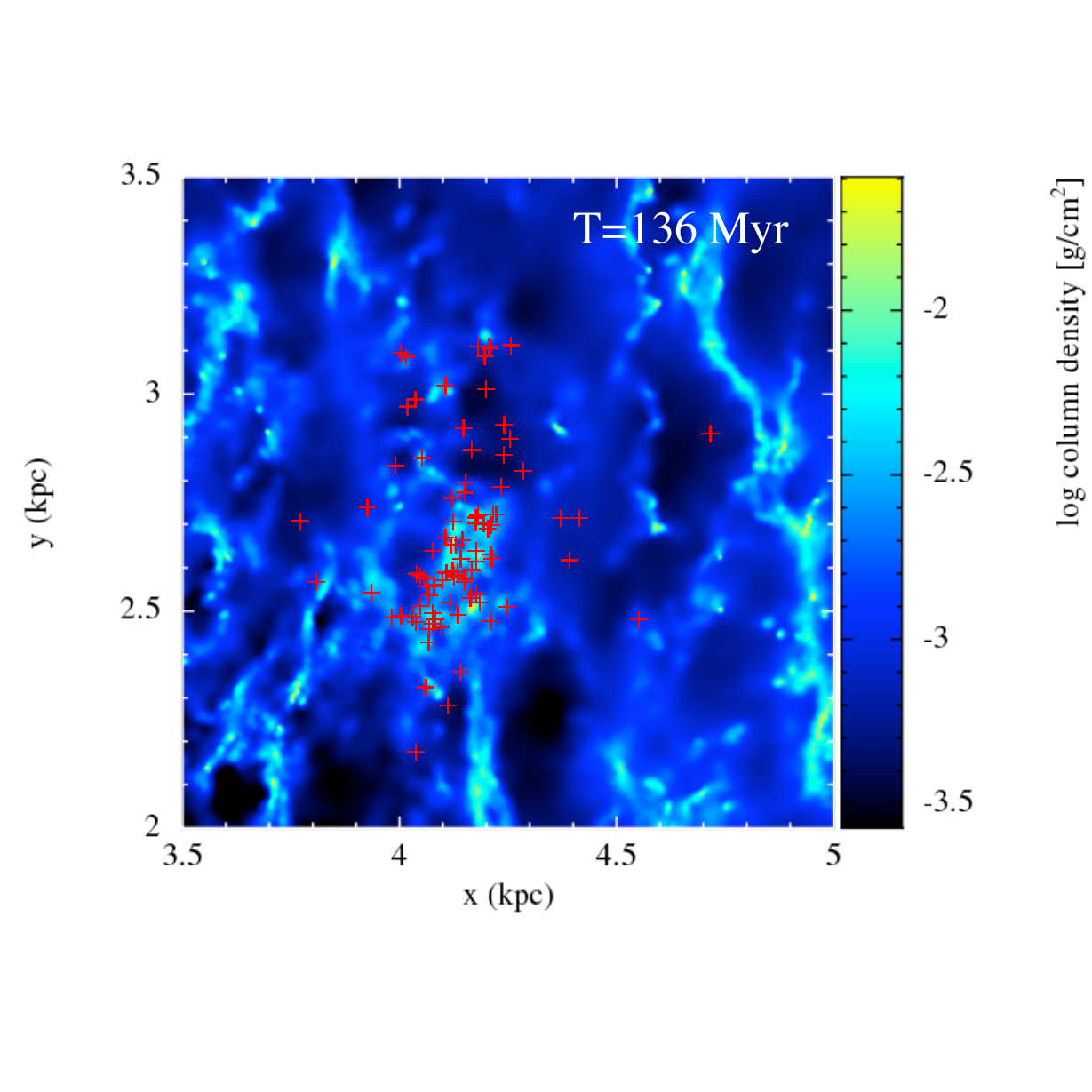}}
\caption{The evolution of a cluster (the time the cluster is selected is in the top panel) is shown. The cluster lies initially in a cloud which is just leaving the arm. Star particles (red) are over plotted on the gas column density. The time in the lower panel shows the evolution of the cluster after 40 Myr. The star particles (and thus the cluster) have dispersed after 40 Myr (no cluster is identified with the clumpfinding algorithm) and are no longer associated with a dense gas cloud. Likewise the cloud seen in the top panel has also largely dispersed.}\label{fig:forward2}
\end{figure}

So far we have only considered the backwards evolution of clusters. However this does not tell us how clusters found in our simulations will evolve. Taking the low feedback model, we select clusters at a time of 96 Myr (roughly halfway through our simulation) and determine how they evolve over a period of up to 100 Myr. We show two examples of different cluster behaviour in Figures~\ref{fig:forward1} and \ref{fig:forward2}. Both clusters are initially fairly similar, comprising of about 100 star particles, and masses of around 3$\times10^4$ M$_{\odot}$. However they evolve quite differently. In our first example (Figure~\ref{fig:forward1}), the cluster remains largely intact, so must be both bound and not disrupted by tidal forces. The lower panel of Figure~\ref{fig:forward1} shows the star particles 60 Myr after the cluster was selected. There is a compact group of star particles that has remained together over the 60 Myr. However some of the star particles (around one third) have dispersed away from the main cluster. In the top panel, the cluster is situated in an outer spiral arm (the arm is not that well defined), and then has moved into the inter arm region by 60 Myr. The main cluster is still associated with a dense cloud of gas, and likely new star particles have formed in this region, making this cluster probably similar to  the example shown in Figure~\ref{fig:example1}, though there is no indication that a collision has occurred like in Figure~\ref{fig:example1}. 

In our second example, shown in Figure~\ref{fig:forward2}, the stars do not remain together but disperse. The lower panel shows the star particles at a time of 40 Myr later. The star particles are much more widespread compared to the example shown in Figure~\ref{fig:forward1}, and there is no particular central concentration of star particles. Also, unlike the example in Figure~\ref{fig:forward1}, the star particles are no longer associated with a GMC. Instead the gas associated with the star particles has largely dispersed, or at least is now of lower density.  In this example, the cluster is in a GMC just leaving the spiral arm initially, then the star particles lie in a nondescript inter-arm region. In Figure~\ref{fig:forward1} the cluster remains in an inter-arm spur. Clusters situated along spurs are a common feature in M51. In Figure~\ref{fig:forward2}, the cluster is initially in a spur feature close to the arm, but the spur has largely dispersed after 40 Myr. Although the two examples look similar initially, partly because the symbol size is large so that the clusters are easily visible, the star particles in the cluster in Figure~\ref{fig:forward1} are situated in a region about half the dimensions of that of Figure~\ref{fig:forward2}. The stellar density is around 0.06 M$_{\odot}$ pc$^{-3}$ for the example in Figure~\ref{fig:forward1}, marginally less than that needed for the cluster to be tidally bound due to the potential, but there is gas present as well. The stellar density is around 0.02 M$_{\odot}$ pc$^{-3}$ for the example of Figure~\ref{fig:forward2}. 

As well as the background potential, tidal effects due to the gas are expected to be important, as cloud densities typically exceed the tidal density from the background potential \citep{Elmegreen2010}. Gas densities in the clouds and surrounding ISM are generally higher than the stellar densities, around 0.1-10 M$_{\odot}$ pc$^{-3}$ (if the star formation efficiency is around 10\%, then roughly every 10th particle will be converted to a star particle so naturally the gas densities will tend to be higher). Thereby tidal disruption from the clouds themselves will be more relevant and it is perhaps no surprise that the gas clouds and clusters appear to evolve largely together.

The example in Figure~\ref{fig:forward2} shows behaviour that we generally did not see when studying the backwards evolution of the clusters. This example shows that the gas clouds where clusters are born do get dispersed (in this case in around 30 Myr), but when we take a sample of clusters we observe a selection effect in that we do not pick out star particles from recently dispersed clouds. This is again likely a result of the limited resolution of our models. If some stars are still grouped together into a cluster, or multiple clusters, following the evolution of the example in Figure~\ref{fig:forward2}, then they likely are not resolved either by our unresolved Nbody dynamics, our stellar mass resolution, or both. 

\begin{figure}
\centerline{\includegraphics[scale=0.4]{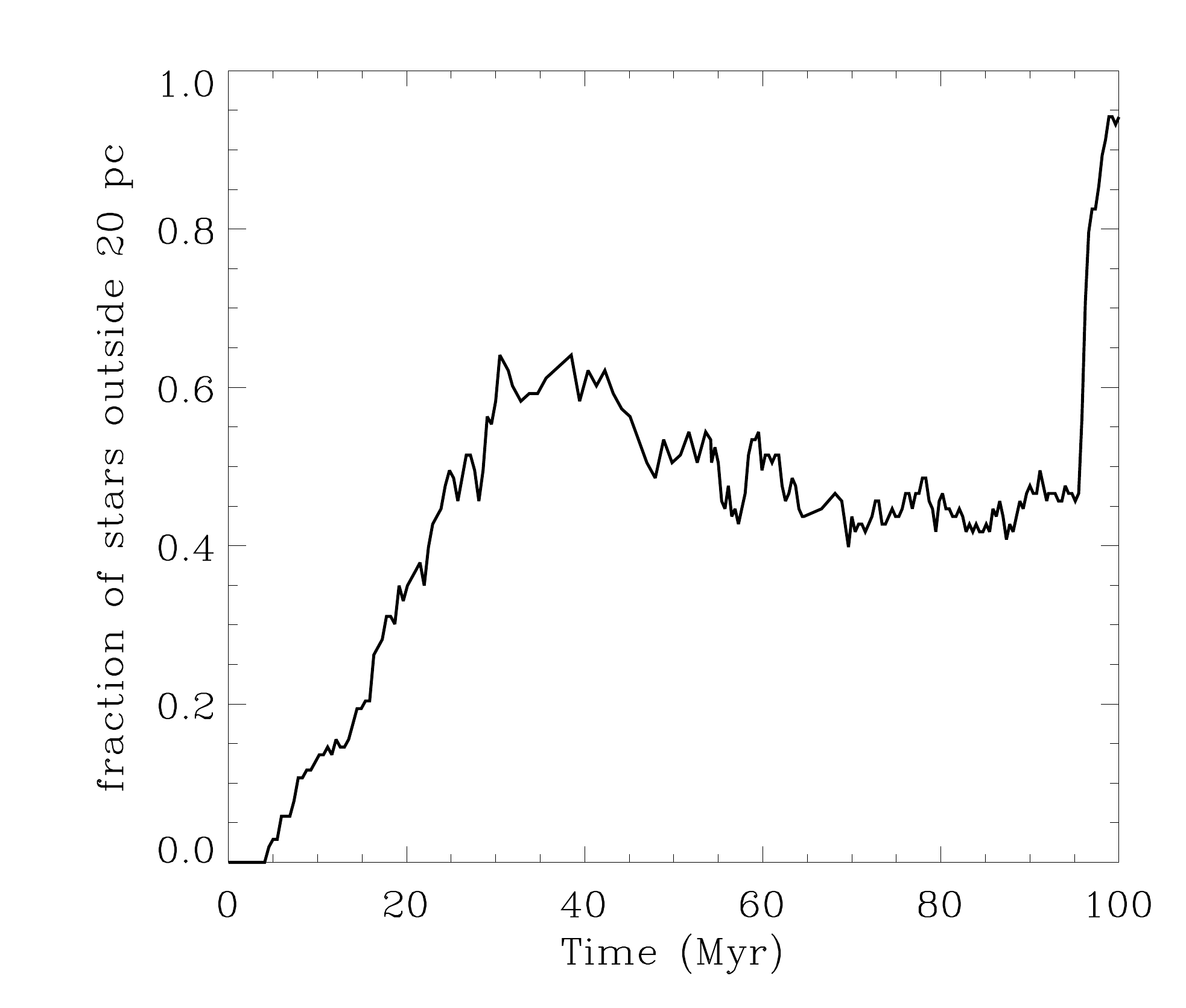}}
\caption{The fraction of isolated stars \textbf{is shown versus time} for the example shown in Figure~\ref{fig:forward1}. Here isolated means stars that are at least 20 pc from any neighbours. About half the stars are dispersed after about 40 Myr, then for the next 50 Myr, the cluster appears to undergo little evolution before dispersing. We also examined how the stars were grouped into clusters using the clumpfinding algorithm over this timescale (not shown). The fraction of stars found to be in a cluster decreases to between 15 and 60 per cent until 90 Myr, but with one main cluster splitting into two clusters after around 60 Myr. There are no clusters present according to the clumpfinding algorithm at 100 Myr, in agreement with the indication from the fraction of isolated stars that the cluster has completely dispersed. The time scale denotes the time since the cluster was initially selected (0 Myr on the above scale, 96 Myr in terms of the timescale of the simulation).}\label{fig:isoforward1}
\end{figure}

We can again look quantitatively at how long clusters appear to survive. We again use the separation between the stars in a cluster as a measure of how the cluster evolves. In Figure~12 we show the fraction of stars which are at least 20 pc away from another star, versus time, for our example from Figure~\ref{fig:forward1}. As we described above, by eye, some of the star particles in this example clearly stay together in a tight group for at least 60 Myr. Figure~\ref{fig:isoforward1} shows more generally how this cluster behaves with time. For 30 Myr or so, star particles become dispersed from the main, most concentrated group of stars, and the number of stars in the cluster thus decreases. This is seen by the more dispersed star particles seen in Figure~\ref{fig:forward1}, lower panel. Between 30 and around 95 Myr, the fraction of isolated stars stays fairly constant. During this time, the central main cluster stays fairly similar. As indicated by Figure~\ref{fig:isoforward1}, about half the star particles are close to other star particles, likely in the main cluster, and half isolated. At about 95 Myr, there is a substantial change as the number of isolated stars dramatically increases. Actually when viewing the cluster at 100 Myr, the stars have not suddenly dispersed, rather for the last 10 Myr or so they are slowly dispersing, and after about 95 Myr, their typical separations are larger than 20 pc. This evolution of the cluster is also roughly similar to that of the host cloud of the main cluster, which is also dispersing after 60 Myr.  

\begin{figure}
\centerline{\includegraphics[scale=0.4]{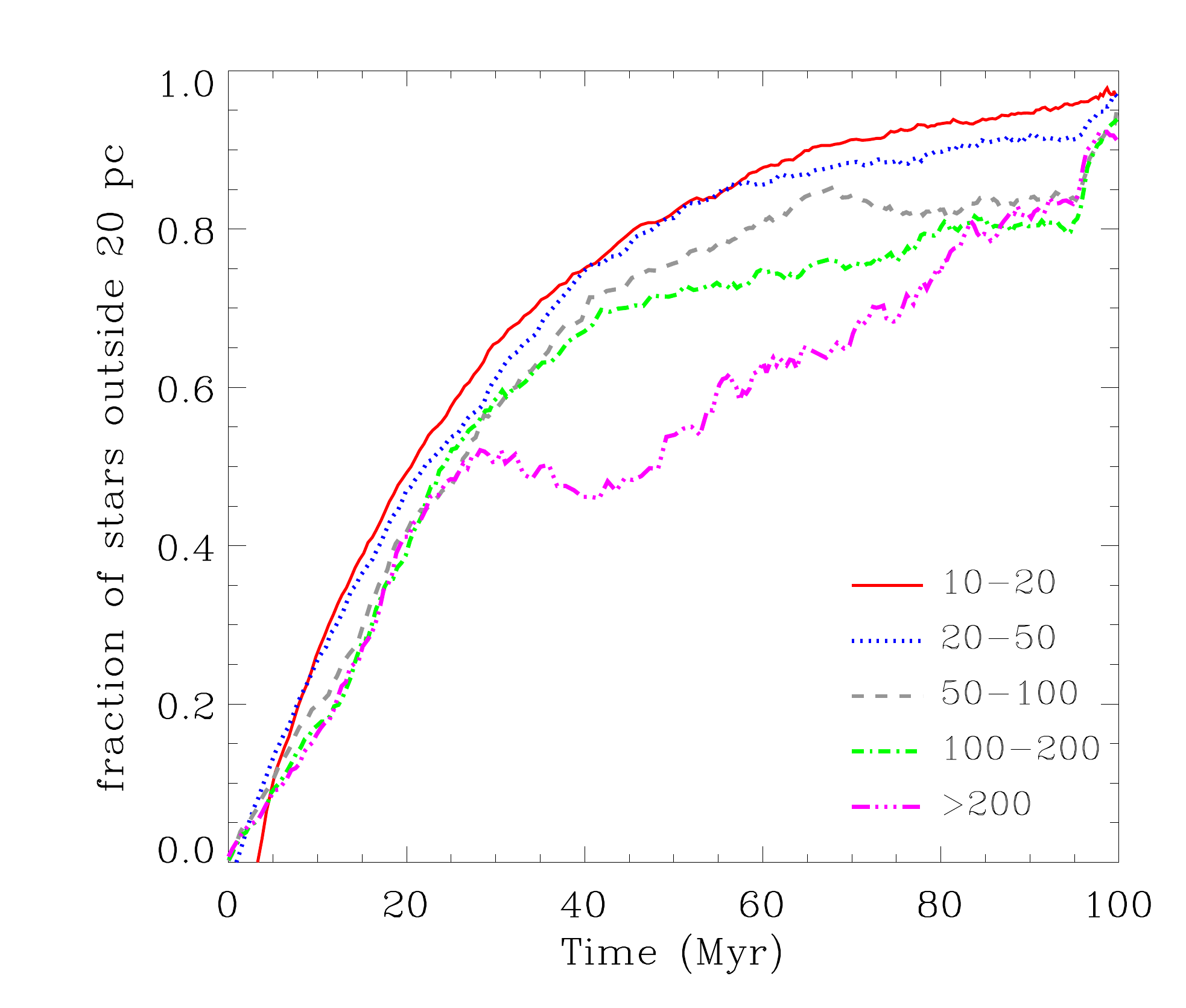}}
\caption{The fraction of isolated stars is shown versus time for clusters with different numbers of particles. Clusters typically appear to disperse over timescales of 10s of Myrs, the most massive showing a slight delay compared to the lower mass clusters. The time scale denotes the time since the cluster was initially selected (0 Myr on the above scale, 96 Myr in terms of the timescale of the simulation).}\label{fig:isoall}
\end{figure}

In Figure~\ref{fig:isoall} we extend our analysis to average over all clusters as they evolve during this 100 Myr period. We first bin the clusters according to how many star particles they contain. For each time we then compute the average number of isolated star particles over all the clusters in a bin with a given range of number of star particles. The behaviour is fairly similar regardless of the size of the cluster, and indicates that the typical behaviour of a cluster is to disperse over a fairly short time frame, a few 10s of Myrs. There is some tendency for the stars in the most massive star clusters to remain together for longer, particularly for the most massive simulated clusters, which have $>200$ star particles, or  several $10^4$ M$_{\odot}$ Observations of M31 and M83 find clusters disperse on a timescale of $\sim$100 Myr \citep{Fouesneau2014, Silva-Villa2014}. We note that Figure~\ref{fig:isoall} shows the \textit{typical} evolution of star clusters, but some will have different evolution, such as the example in Figure~\ref{fig:forward1} which takes longer to disperse, whilst some clusters may disperse very quickly, e.g. over 10-20 Myr. Overall though, the number of clusters which survive for a long time are small. In some cases, some stars may be dispersed whilst a core may remain and form the basis for the strongly bound clusters and clouds seen in Section 4.4.1.  

\begin{figure}
\centerline{\includegraphics[scale=0.4]{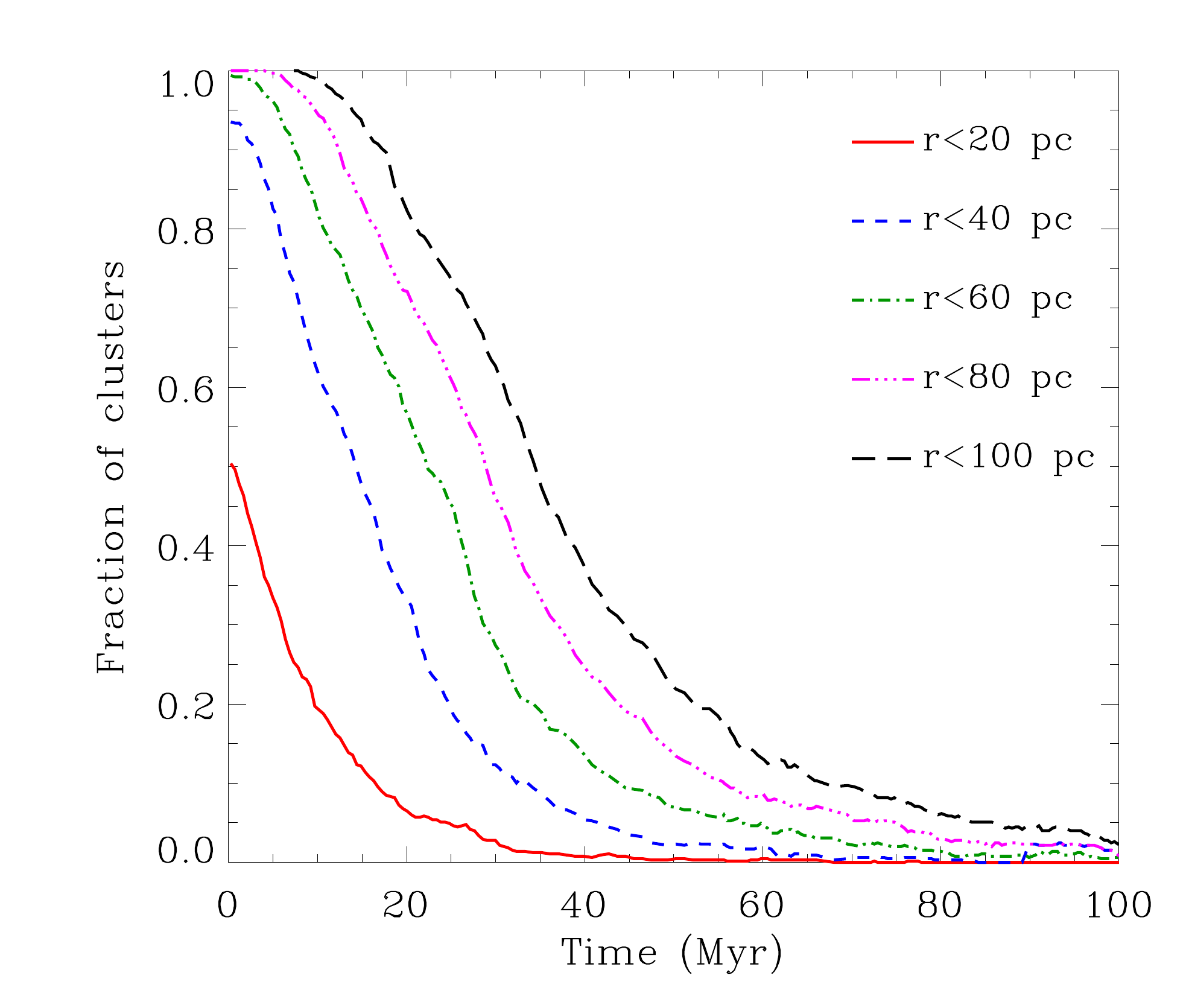}}
\caption{The fraction of clusters with radii below given values are shown versus time. The radius is the (3D) half mass radius. Cluster radii appear to increase over timescales of 10s of Myrs, again suggesting that they dissolve on these timescales.}\label{fig:radius}
\end{figure}

Alternatively to considering the number of isolated stars, we could also consider a property of the clusters such as density or radius (which is also more directly comparable to observations). Given that the simulated clusters are non spherical, and may contain both dense regions, or a dense core, and more dispersed stars, calculating radii and densities is not so easy in the calculations. Here we simply compute the radius which contains half the mass of the cluster, or the half mass radius. We compute the radius in 3D - using the 2D radius results in slightly lower (e.g. by 10 or 20 \%) values.  In Figure~\ref{fig:radius} we show the evolution of the fraction of clusters with half mass radii within given values versus time. Although Figure~\ref{fig:radius} does not indicate the evolution of individual clusters, just the distributions, the radius of a given cluster will generally increase with time. Similarly to when considering isolated star particles, Figure~\ref{fig:radius} indicates that the simulated clusters typically evolve over 10s of Myr timescales. The average radii of the simulated clusters when selected are around 20 pc, suggesting that many are more like observed unbound clusters rather than compact clusters \citep{Mackey2003a,Mackey2003b,Goul2003}. As they evolve over time, their radii increase to several 10s of pcs, which makes them more analagous to observed stellar associations \citep{Goul2011}.  

Overall, our analysis indicates that most clusters probably disperse fairly quickly, often so that we cannot readily detect them with the resolution of our simulations. Some clusters appear to be an exception though, particularly for our low feedback model, where a cluster, or some central core survives to produce massive strongly bound clusters. The behaviour of the clusters in our models seems at least partly tied with the behaviour of the natal molecular clouds. Molecular clouds that survive for several 10s of Myrs or more tend to be the hosts of more bound, likely more massive clusters, whereas clouds that are relatively quickly formed and dispersed host smaller clusters which similarly disperse relatively quickly. 

\subsubsection{Comparison of cluster dispersal with observations and theoretical interpretations}
To further quantify cluster evolution and better compare with observations, we show in Figure~15 plots of the distribution of cluster ages, even though we can only crudely assign clusters to a given age bin. We tend to use quite large logarithmic bins, so even simply identifying clusters as predominantly containing young, very young or older stars tends to cover our age bins. In Figure~15, we show three different arrangements of clusters into age bins. In the simplest case, we just assign an age bin according to the bin which has the highest frequency (taking into account the width of the bin) of stars (red dashed lines). In this case, clusters which have stars of all ages are still included. In our second case, we take the same approach as used for Figure~2, we only take clusters where there is an overdensity of ages clusters of at least 50\% for a given bin (blue solid lines). This will tend to remove clusters that are effectively just stellar groups, like shown in Figure~8 (preferentially older clusters). Thirdly we take an even more restrictive approach, where we require that a quarter of star particles lie within a 10 Myr bin,  and also that there are at least five star particles within that 10 Myr bin (magenta dotted lines). This third approach effectively tends to remove any clusters with a small number of stars from the sample and again older clusters with large age spreads.    

\begin{figure*}
\centerline{\includegraphics[scale=0.4]{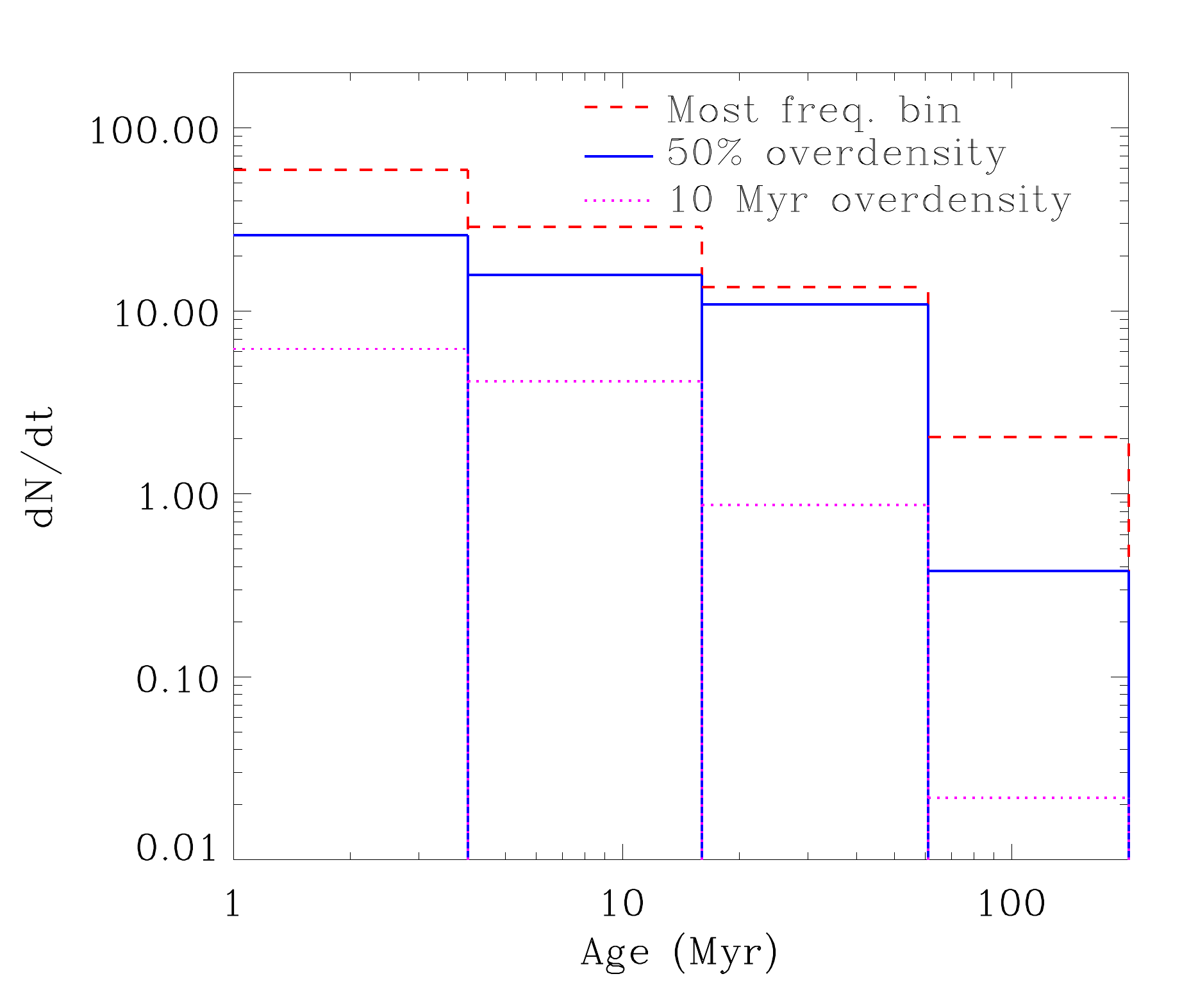}
\includegraphics[scale=0.4]{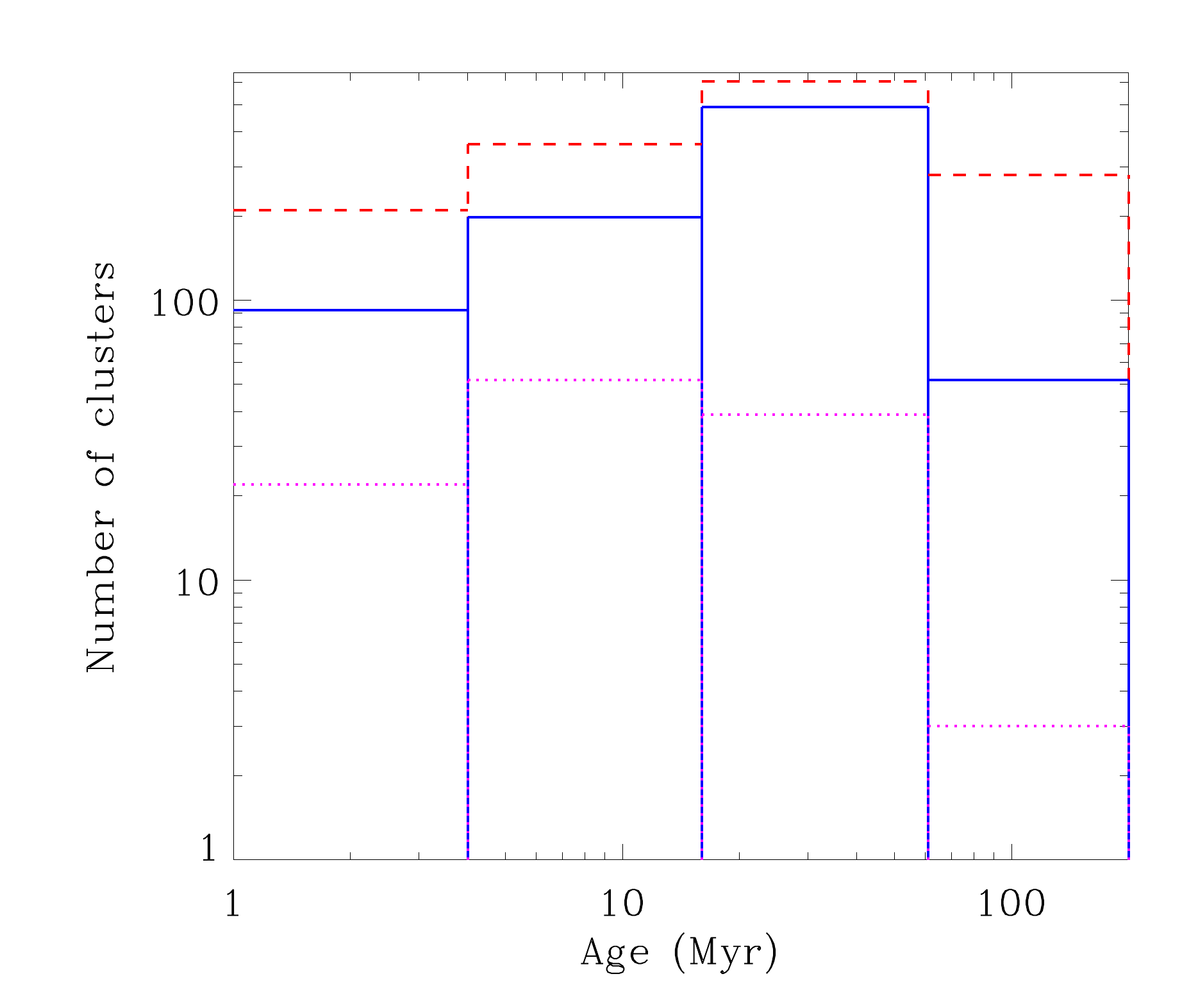}}
\caption{The number of clusters of different ages, divided by age bin is shown on the left, and simply the number of clusters of different ages is shown on the right. The clusters are defined in 3 different ways, with no restriction for clusters with large age spreads (red dashed lines) and with stricter definitions which discount clusters with large age spreads (blue solid and magenta dotted lines, see text).}\label{fig:ages}
\end{figure*}

In the left panel of Figure~\ref{fig:ages} we show the number of clusters divided by the size of the age bin, versus age. This representation can be compared to observed clusters. Our results for the less strict cluster definitions give a slope (typically denoted $\zeta$) in agreement with M83 clusters \citep{Silva-Villa2014}, where the slope is fairly shallow, $\zeta>-0.6$, and several other nearby galaxies \citep{Gieles2005,Gieles2008,Silva-Villa2011,Konstan2013,Baumgardt2013}. Similar to the observations, the simulations tend to show a relatively constant profile followed by a distinct downturn. The downturn tends to be at slightly lower ages in the simulations though (50--100 Myr), in agreement with those found in the previous section, but lower compared to observations (100-200 Myr). The results for the strictest cluster definition indicates a sharply declining slope, indicative of rapid cluster dispersal, after an initial period of 10s of Myrs.
Although there is some dependence on the cluster definition, it is at least encouraging that we find similar agreement with observations, and show similar patterns of a period of slow dispersal followed by more rapid dispersal, without particular refinement of the cluster definition. It is also unsurprising that the stricter we make our cluster definition, the more rapidly they appear to disperse. 

In the right panel of Figure~\ref{fig:ages} we show simply the number of clusters vs age. Here we see the same basic behaviour for all cluster definitions; the number of clusters increases and then decreases with age, although again there is a sharper decrease for the stricter cluster definition (and a peak at lower ages). We can compare these results with theoretical predictions for different models  of cluster evolution by \citet{Elmegreen2010}. The simpler models by \citet{Elmegreen2010} investigate instant versus continuous dispersal, with a standard power law and Schechter function mass distribution functions. The models also depend on a parameter $\chi$, which is a measure of the cluster disruption rate. Our results appear to be consistent with the cluster evolution models where $\chi \geq1$, which corresponds to moderate or rapid cluster disruption, but there is not a distinction between the other variations. This is consistent with the behaviour seen in the simulations. We see both instant and continuous dispersal, but the relatively short timescales in our models are more consistent with higher ($\geq1$) values of $\chi$. Our results are not consistent with the model of cluster dispersal by cloud collisions, but then we do not see any evidence that cloud-cloud collisions disperse clusters in our simulations (in fact during the one cloud merger we show, the clusters appear to merge together). 

\section{Discussion: Cluster evolution}
In this work we have made a simple attempt to identify clusters, study their spatial distribution and follow their evolution. The novelty of our approach lies in that at least for the isolated galaxy simulations, the resolution is such that multiple star particles constitute one star cluster, and thus we can consistently follow clusters as a series of N-body particles. However we are very far removed from the situation where we can model individual stars, and thus correctly account for effects such as two-body relaxation, or mass loss of individual stars. Because of our resolution, our clusters also do not tend to be as concentrated as observed clusters. Our massive clusters also often tend to have quite large age spreads (e.g. 40 Myr) whereas YMCs tend to have very small age spreads \citep{Longmore2014}. These large age spreads can be due to ongoing star formation, accretion of other stars, or both. 

Despite these caveats, we find some broad agreement between the evolution of clusters in the simulations and observations. In particular we see similar sorts of age distributions in the simulations and observations. 
 The detailed behaviour of the clusters depends somewhat on our definition. When restricting our sample to clusters with better defined ages (which is probably closest to observed clusters which tend to assume a single age population), we see more rapid dispersal, but better agreement with the spatial distribution of clusters compared to observations, since we tend to preferentially remove older clusters.  

The evolution of the simulated clusters is probably in strongest agreement with scenarios where cluster dispersal coincides with gas loss (or dynamics) of the natal molecular cloud. This is shown by the tendency of clusters to be associated with gas clouds. Because the stellar densities are comparable to, or lower than gas densities, the cluster evolution tends to be governed by the gas, with tidal effects from the clouds likely dominating disruption \citep{Elmegreen2010}. This behaviour tends to mean that clusters disperse on shorter timescales of 10s of Myrs, following more closely the lifetimes of GMCs, than the 100-200 Myr suggested by observations. Our results may also contradict observations that stellar clusters are no longer associated with gas after relatively short timescales (e.g. \citealt{Whitmore2014} and \citealt{Bash1977}). 

The discrepancies between our clusters and YMCs, and the association of gas with the clusters in our simulations, may arise for a number of reasons. With regards to the age spreads, we could be simply considering more distributed and larger clusters than YMCs.  There may also be a more fundamental problem of trying to gather a large mass of gas together in a short period of time in order to form stars. For cluster dispersal, there may be a selection effect that dispersal into smaller clusters is not seen again due to lack of resolution. Another factor affecting the age spreads, cluster lifetimes and presence of gas is the effectiveness of stellar feedback in the simulations. For the smaller clouds feedback likely readily disperses the cloud, but not the more massive clouds (see also \citealt{Calura2015} and \citealt{Krause2016}). It could be that the feedback is not strong enough, and the escape velocities too small. However the velocities tend to be at least a few km s$^{-1}$. Another, perhaps more important factor, is that because of our resolution, star formation tends to occur relatively uniformly in space and time (and the feedback effectively prevents star particles occurring very close together). In reality, if star formation is concentrated in a massive cluster of stars formed at a similar time, then both the stellar density will be higher, and the feedback will be much more concentrated in the clouds than our simulations. Ineffective feedback (and continued accretion of stars) could however be relevant to globular cluster formation where larger stellar age spreads are apparent.   

Finally, we note that although we cannot follow the evolution of clusters in the cosmological simulations, the spatial distributions highlight that there are too many older clusters compared to the observations, and that in reality many of the older clusters will have dispersed. Consequently the older clusters (and to some extent the star particles in the isolated galaxy simulations, although these are smaller masses) will be incorrectly modelled as a single particle or source of emission.

\section{Conclusions}
We have examined stellar clusters in isolated galaxy and cosmological galaxy simulations. We first considered the spatial distributions of star particles and clusters in these simulations and compared them with observational surveys. Most notably, the simulated clusters (both in the case of the isolated galaxies and cosmological galaxies) display clearer spiral structure in older clusters compared to the observations. We identified a couple of  possible reasons for this difference. Firstly there are few older clusters present in the observations compared to the simulations. Related to this is the fact that in the observations, clusters disperse, whereas in the cosmological simulation there is no cluster evolution, so the number of clusters does not decrease with age. Likewise the same occurs for the star particles in the isolated galaxy simulations, although if grouping stars into clusters, there are smaller numbers of older clusters and the distribution of clusters better resembles the observations. A second effect may be the global galactic structure. The simulations show particularly clear spiral structure, and likewise the observations of NGC 1566 with the strongest spiral arms show the clearest spiral structure in the older stars. 

We then considered the mass distributions of clusters for the isolated galaxy simulations, grouping stars together into clusters using a friends of friends algorithm. With different levels of feedback, the distributions show the same slope, but are shifted up and down. This is similar to the behaviour seen for clouds \citep{Dobbs2011new}. We also saw simply from the spatial distribution of star particles that the lower feedback model contained more massive clusters, since the feedback is less able to disperse the clouds and prevent continuing star formation and further cluster growth. The cluster distributions are slightly steeper than that for clouds, and a little shallower compared to observations.  

We then studied the evolution of clusters in the isolated galaxy simulations, again grouping star particles into clusters using a friends of friends algorithm. Some clusters are shown to resemble what we typically think of as clusters, i.e. groups of stars that form mostly together (although we also see indications that clusters may merge during cloud-cloud collisions) and those that are simply associations of stars that happen to be spatially coincident. We perform the analysis with different restrictions on the age spreads to try to deselect the latter examples. We compared age distributions from the clusters in the simulations with results both from observations and the theoretical models of \citet{Elmegreen2010}. We find a gradient of slope $\zeta \sim  -1$ to $-0.5$, indicative of moderate to rapid dispersal, dependent on the strictness of the cluster definition. Values at the less steep end of this range are in good agreement with observations, though correspond to our least strict definition of clusters, suggesting that probably our clusters disperse faster than observed clusters. 

The evolution of the star clusters in the simulations appears to largely follow the evolution of the GMCs themselves -- most dispersing relatively quickly, some surviving longer. Thus the timescales for cluster dispersal (10s of Myrs) we find are not dissimilar to cloud lifetimes. They are longer than the lifetimes typically found in \citet{Dobbs2013} but there we used a fairly strict definition of lifetime that at least half the gas contained by the cloud must be the same over its lifetime. However our resolution is such that we may well be underestimating the time clusters survive after the gas disperses. This may be reflected in the timescales we find compared to observed clusters. Achieving such resolution to study clusters fully (and in particular achieve high cluster densities) is not yet viable in galaxy simulations. A second point is that massive clusters with small age spreads tend not to occur in the simulations, as massive clouds tend to build up over longer timescales (i.e. 10s of Myrs). In the simulations, continuous gas accretion onto the clouds hinders gas dispersal and cluster dissociation compared to simulations or analysis of isolated clouds. Again though, resolution may be an issue, particularly in regards to how feedback is modelled and how effective stellar feedback is in dispersing the gas.

In our discussion section, we have mentioned a number of further caveats and uncertainties in our interpretation of the simulations. In particular, we have only used a very simple approach to identifying clusters, and due to our resolution they tend to be less concentrated or dense than real clusters. Again obtaining the densities of real clusters would require higher resolution that is higher than typically feasible in these type of simulations at present. We also make no attempt to follow processes such as mass loss of individual stars, or two-body effects of individual stars in the simulations which in reality may also effect cluster evolution.

\section{Acknowledgments}
We thank the referee for a useful report.
The calculations for this paper were performed primarily on the DiRAC machine `Complexity', as well as the supercomputer at Exeter, which is jointly funded by STFC, the Large Facilities Capital Fund of BIS, and the University of Exeter. 
We would like to thank Michele Fumagalli for work putting together the LEGUS cluster catalogues. CLD and CGF acknowledge funding from the European Research Council for the 
FP7 ERC starting grant project LOCALSTAR. CGF thanks Ben Thompson for performing data reduction. D.G. kindly acknowledges financial support by the German Research Foundation (DFG) through grant GO\,1659/3-2. Figures in this paper were produced using \textsc{splash} \citep{splash2007}.
\bibliographystyle{mn2e}
\bibliography{Dobbs}

\bsp
\label{lastpage}
\end{document}